%% file: article.tex
\documentstyle[aps,12pt,epsf]{revtex}    
   
\begin{document}   
   
\begin{center}   
{\Large \bf  Lightest Neutral Higgs Pair Production in Photon-photon Collisions   
 in the Minimal Supersymmetric Standard Model   
 \\}   
  \vskip1cm {\large Shou Hua Zhu,    
  Chong Sheng Li and    
  Chong Shou Gao\footnote{e-mail: huald@ccastb.ccast.ac.cn, csli@pku.edu.cn, gaochsh@pku.edu.cn}}   
  \vskip0.2cm   
  Department of Physics,   
  Peking University, Beijing, 100871, P. R. China \\   
  \vskip0.2cm    
  China Center of Advanced Science and Technology(World Laboratory),   
  P.O.\ Box 8730, Beijing, 100080, P. R. China \footnote{Mailing address} \\   
  \vskip 0.5cm   
  {\large\bf Abstract \\[10pt]} \parbox[t]{\textwidth}{   
  The cross sections for the lightest neutral Higgs pair production at the one-loop level in photon-photon   
collisions are calculated in the Minimal Supersymmetric Standard Model(MSSM). We find that the contributions to the lightest Higgs boson pair production cross sections  arising from the  genuine supersymmetric virtual particles dominate over that arising from the virtual particles in the two-Higgs-doublet model (2HDM) if there are mixing between the right- and left-handed stops, otherwise, the contributions from ones in the 2HDM  are dominant. We present the detailed numerical results of the cross sections of the process $e^+e^-\rightarrow \gamma\gamma\rightarrow h_0h_0$ in both beamstrahlung and laser back scattering photon modes at the Next Linear Collider (NLC). The cross sections are typically in the range of $10^{-3}fb $ to $10^5 fb$ depending on the choice of the mass of the lightest neutral Higgs boson, $\tan\beta$ and  
photon collision modes, especially whether there are mixing between the stops. 
 }   
  \vskip 0.5cm

\end{center}   
   
PACS number(s): 12.15.Lk, 12.60.Jv, 14.80.Cp 
   
\newpage   
\renewcommand{\thefootnote}{\arabic{footnote}}   
\setcounter{footnote}{0}   
   
   
\section{Introduction}   
\setcounter{equation}{0}   
   
The Higgs boson is the missing piece and also the least known one of   
the standard model(SM). Experimental data set a lower bound of $65.5 GeV$ at    
the $95\% $ confidence   
level(CL)\cite{Janot}. On the other hand, exploiting the sensitivity   
to the Higgs boson through quantum loops, a global fit to the latest   
electronweak precision data predicts $m_{H}=149_{-82}^{+148} GeV $ together   
with a $95\%$ CL upper bound at $550 GeV$ in the SM \cite{Boutemeur}.   
   
Beyond the SM, the most intensively studied class of the theories as a possible candidate  for new physics  is the supersymmetry(SUSY), especially the minimal supersymmetric extension of the standard model(MSSM)\cite{1} in which two Higgs doublets are necessary, giving masses separately to up- and down-type fermions and assuring cancellation of anomalies, which has the same Higgs sectors with 
the two-Higgs-doublet model (2HDM), i.e. so-called Model II, except the restrictions imposed by SUSY. In contrary to the SM in which there is just a single neutral Higgs boson, in the MSSM, there are three neutral and two charged Higgs bosons,    
$h_0$, $H$,   
$A$ and  $H^\pm$ of which $h_0$ and $H$ are $CP-$even and $A$ is $CP-$odd.   
The masses and coupling of these Higgs bosons are controlled by two    
parameters, e.g., $m_A$ and $\tan\beta$, at tree level. The upper bound   
for the mass \cite{4}   
 for the lightest $CP-$ even $h_0$ in MSSM is $m_{h0}<m_Z$ at   
tree level and $ m_{h0}<m_Z+\epsilon (m_t,\tilde{m})$ when including radiative   
corrections($\tilde{m}$ is the SUSY mass scale). The pursuit of the Higgs   
bosons predicted by the SM and the MSSM is one of the primary goals of the present   
and next generation of colliders.   
   
The main processes of the neutral Higgs boson production at the LEP2 and LHC are $e^+e^-\rightarrow Z h_0$\cite{5} and $gg \rightarrow (h_0, H, A)$\cite{6}, respectively. However, the Next Linear Collider(NLC) operating at a center-of-mass energy of $   
500-2000 GeV$ with the luminosity of the order of $10^{33} cm^{-2} s^{-1}$   
can also provide an ideal place to search for the Higgs boson, especially, it may produce  a   lightest neutral Higgs boson pair with an observable production   
rate, since the events would be much cleaner than in the LHC   
and the parameters of the  Higgs boson would be easier to   
extracted.  There are mainly two options for the photon sources at the NLC: laser back-scattering and beamstrahlung photons. These two kind of photon sources are the options of turning the electron-positron collider into a laser photon collider with   
 high energy and high luminosity which are expected to be   
comparable to that of the primary $e^+e^-$ collisions\cite{gammagama,Douglas}.

The primary mechanism of the neutral Higgs production in photon-photon collisions is $\gamma\gamma \rightarrow (h_0, H, A)$,   
but in order to study the triple and quadruple Higgs couplings at future high energy colliders,   
it is necessary to explore the Higgs boson pair production process.    
In the SM, the cross section for the neutral Higgs boson pair production in photon-photon collisions  has been calculated by   
 J.V. Jikia\cite{Jikia}, and in the 2HDM, the process $\gamma\gamma \rightarrow H_0 H_0$ has also been computed for a special case in Ref.\cite{Sun}. In this paper,   
 we present the complete calculations of the lightest neutral Higgs boson pair production cross section at one-loop level in the MSSM. In the MSSM, besides the loop diagrams arising from the SM and the 2HDM particles, there are also hundreds of additional loop diagrams arising from  genuine SUSY particles, which  make the calculations much more complicated than the case of the SM and the 2HDM.    
   
The structure of this paper is as follows. In Sec. II we give the analytical results in terms of the well-known standard notation of the one-loop integrals. In Sec. III we present some numerical examples with discussions. The tedious expressions of the form factors in the amplitude are summarized in Appendix A, B and C, respectively.

\section{Calculations}   
\setcounter{equation}{0}   
   
   
The process of   
\begin{eqnarray}   
\gamma (p_1,\lambda_1)+\gamma (p_2,\lambda_2)\longrightarrow h_0(k_1)+h_0(k_2)   
\end{eqnarray}   
is forbidden at the tree level but it can be induced through one-loop diagrams which are shown in  Fig.1-Fig.5 (all relevant cross diagrams are not explicitly shown), where $\lambda_{1,2}$ denote the helicities of photons. In the center-of-mass system(CMS) the momenta read in terms of the beam energy $E$ of the incoming photons and the scattering angle $\theta$   
\begin{eqnarray}   
p_1^\mu&=&E(1,0,0,-1)\nonumber \\   
p_2^\mu&=&E(1,0,0,1) \nonumber \\   
k_1^\mu&=&E(1,-\beta sin\theta ,0,-\beta cos\theta )\nonumber \\   
k_2^\mu&=&E(1,\beta sin\theta ,0,\beta cos\theta )   
\end{eqnarray}   
where $\beta =\sqrt{1-4m_{h0}^2/{\hat{s}}} $ is the velocity of the Higgs in the CMS.   
We define the Mandelstam variables as   
\begin{eqnarray}   
\hat{s}&=&(p_1+p_2)^2=(k_1+k_2)^2\nonumber \\   
\hat{t}&=&(p_1-k_1)^2=(p_2-k_2)^2 \nonumber \\   
\hat{u}&=&(p_1-k_2)^2=(p_2-k_1)^2    
\end{eqnarray}   
   
In order to calculate polarization cross section, we introduce the explicit    
polarization vectors of the helicities($\lambda_1\lambda_2$) for photons as follows   
\begin{eqnarray}   
\epsilon_1^\mu (p_1, \lambda_1=\pm 1)&=&-\frac{1}{\sqrt{2}}(0,1,\mp i ,0)  \nonumber \\   
\epsilon_2^\mu (p_2, \lambda_2=\pm 1)&=&\frac{1}{\sqrt{2}}(0,1,\pm i ,0)   
\end{eqnarray}   
This choice of polarization vectors for the photons implies   
\begin{eqnarray}   
\epsilon_i.p_j =0,\ \ \ \ \ (i,j=1,2)   
\end{eqnarray}   
and by the momentum conservation,   
\begin{eqnarray}   
\epsilon_i.k_1 =-\epsilon_i.k_2,\ \ \ \ \ ( i=1,2).   
\end{eqnarray}   
   
We perform the calculation in the 't Hooft-Feynman gauge, the relevant Feynman rules can be found in Ref\cite{1}. Because this process can only be induced through loop diagrams, we do not need  to consider the renormalization at one-loop level, and the ultraviolet divergence should be   
canceled automatically if the calculation is right. The virtual particles which enter loops are the third family (top and bottom) quarks, $W$ bosons, charged ghosts, charged Goldstone bosons, charged Higgs bosons, charginos, squarks and sleptons. As we will see below, the transversality to photon momentum has been obviously kept. In order to ensure the correctness of our calculations, as a check, we have compared our results with Ref. \cite{Jikia}, and found that our results in the case of the SM are agreement with theirs.    
   
Because of the transversality of photons and our choices of the photons helicities, the general amplitude for the process $\gamma\gamma\rightarrow h_0h_0$  can be simply written as   
\begin{equation}   
\begin{array}{ll}   
M&=M_1 \epsilon_1.\epsilon_2+   
M_2 \epsilon_1.k_1 \epsilon_2.k_1 +eps(\epsilon_1,   
\epsilon_2, p_1, p_2) M_3    
\end{array}   
\end{equation}   
with    
\begin{eqnarray}   
 eps(\epsilon_1,   
\epsilon_2, p_1, p_2)=\epsilon_{\mu\nu\rho\sigma}\epsilon_1^{\mu}\epsilon_2^{\nu}   
p_1^{\rho} p_2^{\sigma},  
\end{eqnarray}   
where $\epsilon_{\mu\nu\rho\sigma}$ is the total anti-symmetry tensor, and    
the $M_i (i=1, 2, 3)$,  which correspond to the factors of the $\epsilon_1.\epsilon_2$,   
$\epsilon_1.k_1 \epsilon_2.k_1$ and $eps(\epsilon_1,   
\epsilon_2, p_1, p_2)$, respectively,   are given by   
\begin{eqnarray}   
M_i=\sum_{j=1}^{12} f_i^{(j)} + (f_i^{(1)}+ f_i^{(6)}+ f_i^{(10)}+   
f_i^{(11)} )\left[k_1\leftrightarrow k_2,  \hat{u} \leftrightarrow \hat{t}\right].   
\end{eqnarray}   
Here, the form factor $f_i^{(j)}$ represents the contributions arising from Feynman diagrams with the indices of $j$, and the $ f_i^{(j)}\left[k_1\leftrightarrow k_2,  \hat{u} \leftrightarrow \hat{t}\right]$ stands for the contributions from the cross diagrams  of the corresponding diagrams with the indices of $j$. Among these form factors,   
\begin{eqnarray}   
f_2^{(k) }=0, \mbox{for $ (k=3, 4, 5...9)$}   
\end{eqnarray}   
and   
\begin{eqnarray}   
f_3^{(k)}=0, \mbox{for $(k=3, 4, 6, 7, 8...12)$}   
\end{eqnarray}   
the explicit expressions of the other form factors are given in Appendix A, B and C, respectively. Note that the amplitude in Eqs. 7 does not have 
gauge-invariance due to dropping terms that vanish for our specific choice of polarization vectors. But we find that the amplitude has indeed the  
gauge-invariance after adding the dropping terms, which confirms again that our 
calculations are correct.  
   
The polarization cross section and the cross section of   
subprocess $\gamma\gamma\rightarrow h_0h_0$ are given by    
\begin{eqnarray}   
\hat{\sigma}(\lambda_1 \lambda_2)& =&\int_{\hat{t}_{min}}^{\hat{t}_{max} } 
{1\over 32\pi \hat{s }^2} \left|  M(\lambda_1 \lambda_2)\right|^2 d \hat{t},  
\nonumber \\ 
\hat{\sigma} &=&\int_{\hat{t}_{min}}^{\hat{t}_{max} }{1\over 32\pi \hat{s }^2}\bar{ \sum_{ spins} }\left|  M\right|^2 d \hat{t},  
\end{eqnarray}   
respectively, 
where the bar over the summation recalls average over initial spins, and   
\begin{eqnarray}   
\hat{t}_{min}=(m_{h_0}^2-{\hat{s}\over 2})-{\beta \hat{s}\over 2}\nonumber \\   
\hat{t}_{max}=(m_{h_0}^2-{\hat{s}\over 2})+{\beta \hat{s}\over 2}.   
\end{eqnarray}   
   
The total cross section of $e^+e^- \rightarrow \gamma\gamma \rightarrow h_0h_0$    
can be obtained by folding the $\hat{\sigma}$, the cross section   
of the subprocess $\gamma\gamma\rightarrow h_0h_0$, with the   
photon luminosity   
$$   
\sigma (s)=\int^{x_{max}}_{2 m_{h0}/\sqrt{s}} dz   
{dL_{\gamma\gamma}\over dz} \hat{\sigma} (\gamma\gamma\rightarrow h_0h_0 \mbox{at   
$\hat{s}=z^2 s$}),  
$$   
where $\sqrt{s}$ and $\sqrt{\hat{s}}$ is the CMS energy of $e^+e^-$ and   
$\gamma\gamma$ respectively and $dL_{\gamma\gamma}/dz$ is the photon    
luminosity, defined as   
$$   
{dL_{\gamma\gamma}\over dz}=2 z \int_{z^2\over x_{max}}^{x_{max}}{dx\over x}   
f_{\gamma /e}(x) f_{\gamma /e}(z^2/x),  
$$   
$f_{\gamma /e}(x)$ is the photon structure function of the electron   
beam \cite{Halzen}. For a TeV collider with $\sigma_x/\sigma_y=25.5$, the beamstrahlung   
 photon structure function can be represented as \cite{Blankenbecler}   
$$    
f_{\gamma /e}(x)=\left\{   
\begin{array}{ll}   
\left( 2.25-\sqrt{x\over 0.166}\right)\left( {1-x\over x}\right)^{2/3} & \mbox{ for $x<0.84$, }\\   
0 &\mbox{ for $x>0.84,$}   
\end{array}   
\right.   
$$   
where $x$ is the relative momentum of the radiated photon and the parent electron.   
   
If we operate NLC as a mother machine of photon-photon collider in Compton back-scattering photon fusion mode with unpolarization initial electrons and   
laser, the energy spectrum of the photons is   
given by \cite{Bohm}   
$$   
f_{\gamma /e}(x)=\left\{   
\begin{array}{ll}   
{1\over 1.8397}\left(1-x+{1\over 1-x}-{4x\over x_i(1-x)}+   
{4 x^2\over x_i^2(1-x)^2}\right) & \mbox{ for $x<0.83$, $x_i=2(1+\sqrt{2}),$}\\   
0 &\mbox{ for $x>0.83.$}   
\end{array}   
\right.   
$$

\section{Numerical examples and conclusions}   
In the following we present some numerical results of a lightest neutral Higgs boson pair production    
cross section in the process of $e^+e^-\rightarrow \gamma\gamma\rightarrow h_0h_0$. In our numerical calculations, for the SM parameters, we choose $m_w=80.33 GeV$, $m_z=91.187 GeV$, $m_t=176 GeV$, $m_b=4.5 GeV$ and $\alpha={1\over 128}$.    
Other parameters are determined as follows   
   
(i) The Higgs boson masses $m_{h0}$, $m_H$, $m_A$ and $m_{H^\pm}$  are given by \cite{Espinosa}   
\begin{eqnarray}   
m_{h0}^2&=&{1\over 2}\left[m_A^2+m_z^2+\epsilon-\right.\nonumber\\   
&&\left.\sqrt{(m_A^2+m_z^2+\epsilon)^2-   
4 m_A^2 m_z^2 cos^22\beta- 4\epsilon (m_A^2sin^2\beta+m_z^2cos^2\beta)}\right],   
\end{eqnarray}   
\begin{eqnarray}   
m_H^2=m_A^2+m_z^2-m_{h0}^2+\epsilon,    
\end{eqnarray}   
and   
\begin{eqnarray}   
m_{H^\pm}^2=m_A^2+m_w^2   
\end{eqnarray}   
with    
\begin{eqnarray}   
\epsilon =\frac{3 G_F}{\sqrt{2}\pi^2}\frac{m_t^4}{\sin^2\beta} \log (1+ \frac{m_S^2}{m_t^2} ).   
\end{eqnarray}   
Here we take  $m_S=m_{\tilde{Q}}=m_{\tilde{U}}=m_{\tilde{D}}$, the definitions of which are given below.  
The mixing angle $\alpha$ is fixed by $\tan\beta$ and the Higgs boson  
mass $m_A$,   
\begin{eqnarray}   
\tan 2\alpha=\tan 2\beta{m_A^2+m_z^2 \over m_A^2-m_z^2 +\epsilon/cos 2\beta},   
\end{eqnarray}   
where   
$ -{\pi \over 2}<\alpha <0$.

(ii) In the MSSM the mass eigenstates $\tilde{q}_1$ and $\tilde{q}_2$ of the squarks are related to the current eigenstates $\tilde{q}_L$ and $\tilde{q}_R$ by   
\begin{eqnarray}   
\left(\begin{array}{c}   
\tilde{q}_1 \\ \tilde{q}_2\end{array}\right)=   
R^{\tilde{q}}\left(\begin{array}{c}   
\tilde{q}_L \\ \tilde{q}_R\end{array}\right)\ \ \ \ \mbox{with}\ \ \ \   
R^{\tilde{q}}=\left(\begin{array}{cc}   
                   \cos\theta_{\tilde{q}} & \sin\theta_{\tilde{q}}\\   
                   -\sin\theta_{\tilde{q}} & \cos\theta_{\tilde{q}}    
                   \end{array}   
                   \right).   
\label{eq1}   
\end{eqnarray}   
For the squarks, the mixing angle $\theta_{\tilde{q}}$ and the masses $m_{\tilde{q}_{1,2}}$ can be calculated by diagonizing the following mass   
matrices \cite{1}   
\begin{eqnarray}   
M^2_{\tilde{q}}=\left(\begin{array}{cc}   
          M_{LL}^2 & m_q M_{LR}\\   
           m_q M_{RL} & M_{RR}^2   
           \end{array} \right), \nonumber \\   
M_{LL}^2=m_{\tilde{Q}}^2+m_q^2+m_{z}^2\cos 2\beta (I_q^{3L}-e_q\sin^2\theta_w), \nonumber \\   
M_{RR}^2= m_{\tilde{U},\tilde{D}}^2 +m_q^2+m_{z}^2\cos 2\beta e_q\sin^2\theta_w , \nonumber \\   
M_{LR}= M_{RL}=\left\{ \begin{array}{ll}   
                A_t-\mu \cot \beta & (\tilde{q}= \tilde{t})\\   
                A_b-\mu \tan \beta & (\tilde{q}= \tilde{b}),   
                \end{array}   
                \right.   
\label{eq2}   
\end{eqnarray}   
where $ m_{\tilde{Q}}^2$, $ m_{\tilde{U},\tilde{D}}^2$ are soft SUSY breaking mass terms of the left- and right-handed squark, respectively. Also, $\mu$ is the coefficient of the $H_1H_2$   
mixing term in the superpotential, $A_t$ and $A_b$ are the coefficient of the dimension-three trilinear soft SUSY-breaking term. $I_q^{3L}, e_q$ are the weak isospin and electric charge of the squark $\tilde{q}$.    
From Eqs. \ref{eq1} and \ref{eq2}, $m_{\tilde{t}_{1,2}}$ and $\theta_{\tilde{t}}$ can be derived as   
\begin{eqnarray}   
m^2_{\tilde{t}_{1,2}}&=&{1\over 2}\left[ M^2_{LL}+   
M^2_{RR}\mp \sqrt{   
(M^2_{LL}-M^2_{RR})^2+4 m^2_t M^2_{LR}}\right], \nonumber \\   
\tan\theta_{\tilde{t}}&=&{m^2_{\tilde{t}_1}-M^2_{LL} \over m_t M_{LR}}.   
\end{eqnarray}   
Because of the small mass of the bottom quark, we will neglect the mixing between left- and right-sbottoms, thus $\theta_{\tilde{b}}=0$. Similarly, we will neglect the mixing in the other light squarks. For simplicity, we shall assume the masses of all the light squarks (the superpartners of the first and the second family quarks)  and all the sleptons (the superpartners of the three family leptons) are degenerate.   
For simplicity, we don't show here the explicit expressions of the masses 
of mass-eigenstates of the sbottoms and the sleptons, which can be found 
in Ref. \cite{1}. In our numerical calculations, we choose $m_{\tilde{Q}}^2= m_{\tilde{U}}^2= m_{\tilde{D}}^2=(1 TeV)^2$. 
  
(iii) For the parameters $M$, $\tan\beta$ and $\mu$ in the chargino  
matrix \cite{1}, we put  $M=100 GeV$ and $\mu=-300 GeV$ unless otherwise stated, and $\tan\beta$ remains a variable.    
  
Some typical numerical calculations of the cross sections of the processes  
$\gamma\gamma\rightarrow h_0h_0$ and $e^+e^-\rightarrow \gamma\gamma   
\rightarrow h_0h_0$ are given in Figure \ref{fig6}-\ref{fig10} and   
Figure \ref{fig11}-\ref{fig14}, respectively.  
   
Figure \ref{fig6} shows the cross sections for  the subprocess    
$\gamma\gamma\rightarrow h_0h_0 $ as   
a function  of the Higgs boson mass $m_{h0}$ for the opposite photon helicities    
$\lambda_1=-\lambda_2=1$ in the MSSM, assuming  $\tan\beta=4$ and    
$\sqrt{\hat{s}}=0.5$, $1$ and $1.5 TeV$.    
In order to compare the contributions arising from different virtual particles, we also present the results of the cross section of the process in the 2HDM, 
of which the parameters space used in our numerical calculations only is the subset of the general 2HDM (Model II), where there are the relations between the Higgs boson masses (Eqs. 14-16) required by the MSSM.  
The figure shows that the cross section can increase from several $fb$ to the order of $10^4 fb$ when $m_{h0}$ increases from $60 GeV$ to $120 GeV$ in the the mixing case of the stops. But it is only of the order of $10^{-1} fb$ in the no-mixing case of stops. The cross sections in the 2HDM are almost the  
same as that in the MSSM for no-mixing case. 
   
Figure \ref{fig7} and Figure \ref{fig8} give the  cross section of the subprocess as a function of the $m_{h0}$ for the equal photon  helicities $\lambda_1=\lambda_2=+1$  and   
 $\lambda_1=\lambda_2=-1$, respectively. From Figure \ref{fig7} and Figure \ref{fig8}, one sees that the cross sections are smaller than that in the case of opposite photons helicities in the 2HDM and in the MSSM with the  
no-mixing case. And the cross sections for the mixing case of the stops are much larger than that for the no-mixing case of the stops and in the 2HDM. 
Figure \ref{fig7} shows that the cross 
sections of the no-mixing case in the MSSM for the photon  helicities $\lambda_1=\lambda_2=+1$ are almost the same as that in the 2HDM, but Figure \ref{fig8} shows the cross sections for the photon  helicities $\lambda_1=\lambda_2=-1$ are always bigger than that in the 2HDM. 
Such difference between the cross sections of the process  for the photons helicities $\lambda_1=\lambda_2=+1$ and $\lambda_1=\lambda_2=-1$ is relative to the different contributions to the amplitude from the charginos.

In Figure \ref{fig9} and \ref{fig10}, we present the cross sections of the  subprocess $ \gamma\gamma \rightarrow h_0h_0 $ for the opposite and equal photon helicities as a function of the Higgs boson mass $m_{h0}$, respectively, assuming $\tan\beta=40$. 
Because the difference between  the cross sections of the process  for the    
photons helicities  $\lambda_1=\lambda_2=+1$  and  $\lambda_1=\lambda_2=-1$  is negligibly small, we only present the results for the photons helicities  $\lambda_1=\lambda_2=+1$.  
From Figure \ref{fig9}, we can see that the cross sections in the MSSM for the mixing case of stops are larger than that in the MSSM for the no-mixing case and in the 2HDM, especially for $m_{h0}> 90 GeV$, and the former can vary from $\sim 1 fb$ to $10^4 fb$ when the Higgs mass is in the range of  $60 GeV$ to $130 GeV$, otherwise, the cross sections are only $\sim 0.5 fb $ and insensitive to the Higgs boson mass. 
Figure \ref{fig10} shows that the cross sections in the MSSM for the mixing case of stops are much larger than that in the MSSM for the no-mixing and in the 2HDM for $\sqrt{\hat{s}}= 500 GeV$, while for $\sqrt{\hat{s}}= 1000 GeV$ and $1500 GeV$, the mixing effects of stops are only important for  $m_{h0} >120 GeV$.

Figure \ref{fig11} and  Figure \ref{fig12} give the total cross section for the process $e^+e^- \rightarrow \gamma\gamma \rightarrow h_0h_0$ as the function of the Higgs boson mass $m_{h0}$ in beamstrahlung    
photon mode for $\tan\beta=4$ and $40$, respectively.  
From our calculations, we find that the contributions from the low energy cross section of the subprocess $\gamma\gamma \rightarrow h_0h_0$ are very important. 
In general, the total cross sections in the MSSM are greater than that in the 2HDM. We can see from Figure \ref{fig11} that the total cross sections in the mixing case of stops are greatly enhanced, which are greater than $10^4 fb$ for all the Higgs boson masses in the range of $60$ to $120 GeV$. But the total cross sections in the no-mixing case of stops and in the 2HDM are smaller 
than $1 fb$ due to decoupling effects of the heavy squarks. 
From Figure \ref{fig12}, one sees that the cross sections for the mixing case of stops in the MSSM increase from order of $10 fb$ to order of $10^4 fb$ with the increment of the mass of the Higgs boson, on the contrary, the total cross sections in the MSSM for the no-mixing case and in the 2HDM decrease. 
 
In Figure \ref{fig13} and Figure \ref{fig14}, we present the total cross sections of the  process $e^+ e^- \gamma\gamma \rightarrow h_0h_0$ in    
laser back-scattering photons mode as function of the Higgs boson mass $m_{h0}$   
 at $e^+e^-$ CMS energy $500, 1000$ and $1500 GeV$ for $\tan\beta=4$ and $40$, respectively.  
The results given by these figures are similar to Figure \ref{fig11} and \ref{fig12} except that the  
total cross sections in the MSSM for the mixing case of stops decrease with  
 increasing $e^+ e^-$ CMS energy.

To summarize, we have calculated the total cross sections of the process $e^+e^- \rightarrow \gamma\gamma\rightarrow   
h_0h_0$ at one-loop level in the MSSM in both photon-photon collision mode of the laser back scattering and beamstrahlung.  The results of numerical calculations for several typical parameter values show that the contributions to the total cross sections arising from the virtual particles in 2HDM play an important role   
when there are not mixing between the stops, otherwise, the genuine supersymmetric contributions to the total cross sections dominate over that arising from the ones in the 2HDM.    
The total cross sections of the process in the MSSM with the mixing case of the stops are much larger than that for the no-mixing case and in the 2HDM for both of $\tan\beta =4$ and $40$.    
Note that our numerical calculations indicate that the total cross sections are 
insensitive to the parameter $\mu$, and so we do not show the corresponding curves versus $\mu$.  
We conclude that the total cross section of the process $e^+e^-\rightarrow \gamma\gamma \rightarrow   
h_0h_0$ varies from $10^{-3} fb$ to $ 10^5 fb$ at $e^+e^-$ center-of-mass energy $\sqrt{s}=500, 1000$ and $1500 GeV$, mainly depending on the choice of the mass of the lightest neutral Higgs boson, $\tan\beta$ and photon collision modes, especially whether there are mixing between the stops. 
   
\section*{Acknowledgments}   
This work was supported in part by the National Natural Science Foundation of China and a grant from the State Commission of Science and Technology of China.   
  
\input{formula}

\newpage   
\input feyn.tex   
\input fig.tex

\end{document}

%% file: formula.tex
\section*{Appendix A}  
In this paper, all form factors are decomposed to two parts of contributions
from the virtual particles in the 2HDM and the genuine SUSY, in which
the contributions from the 2HDM include ones of the third generation
quarks and the bosons in the SM as well as the additional bosons in the
2HDM.

The form factor $f_1^{(1)}$ is given by  
\begin{eqnarray}  
f_{1}^{(1)}&=&f_1^{(1, 2HDM)}+f_{1}^{(1, charginos)}+f_1^{(1,sfermions)},  
\end{eqnarray}  
with 
\begin{eqnarray}  
f_1^{(1, 2HDM)}&=&f_1^{(1, fermions)}+f_1^{(1,bosons)}. 
\end{eqnarray}  
Here $f_1^{(1, fermions)}$, $f_1^{(1,bosons)}$,  $f_{1}^{(1, charginos)}$ and 
$f_1^{(1,sfermions)}$ are  
\begin{eqnarray}  
&&f_1^{(1,fermions)}=\nonumber \\  
&&{e^2 g^2 m_t^2 \cos (\alpha)^2 csc(\beta)^2\over 6 m_{w}^2 \pi^2}   
\left[ 2 B_0(0, m_t^2, m_t^2)  +   
 (m_{h0}^2 + 4 m_t^2 - \hat{t} - 2 \hat{u})    
      C_0^9 \right. \nonumber \\  
&+&   
   2 m_t^2 (\hat{s} + 2 \hat{t})    
      D_0^7   
+   
 (m_{h0}^2 - \hat{u})    
      C_1^9+   
   2 (m_{h0}^2 + 4 m_t^2) (m_{h0}^2 + \hat{t})    
      D_1^7   
    +2 \hat{s}    
C_2^9 \nonumber \\  
&+&   
 (m_{h0}^4 + 2 m_{h0}^2 \hat{t} + 8 m_t^2 \hat{t} - \hat{t} \hat{u})    
     D_2^7   
  + 4 m_{h0}^2   (\hat{t} + \hat{u})    
      D_{11}^7 \nonumber \\  
&+& 2 (4 m_{h0}^2 - \hat{s}) (m_{h0}^2 + \hat{t})   
      D_{12}^7  
\nonumber \\  
&+& \left.  
 (8 m_{h0}^4 - 4 m_{h0}^2 \hat{s} + \hat{s}^2)    
     D_{13}^7   
    + (m_{h0}^2 + \hat{t})^2    
      D_{22}^7   
  -   
   4 C_{00}^9   
   + 4 (-4 m_t^2 + \hat{t} + \hat{u})   
      D_{00}^7  
   \right]\nonumber \\  
&+&  
{e^2 g^2 m_b^2 \sin (\alpha)^2 sec(\beta)^2\over 24 m_{w}^2 \pi^2}\left[m_t\rightarrow m_b \right],  
\end{eqnarray}  
\begin{eqnarray}  
&&f_1^{(1,bosons)}=\nonumber \\  
&&{e^2 g^2 \cos (\alpha - \beta)^2\over 32 \pi^2}   
\left( -4 C_{00}^6 + 2 (-m_{h0}^2 + \hat{t}) (-m_{h0}^2 + \hat{u})   
      D_{11}^4  
     \right. \nonumber \\  
&+& 2 (-m_{h0}^2 + \hat{t}) (-\hat{t} + \hat{u})   
      D_{12}^4  
   + (-2 m_{h0}^4 + 2 m_{h0}^2 \hat{t} - \hat{t}^2 + 2 m_{h0}^2 \hat{u} - \hat{u}^2)   
      D_{13}^4\nonumber \\  
 &-&  (-m_{h0}^2 + \hat{t})^2   
      D_{22}^4  
    + 2 (3 m_{h0}^2 + 2 \hat{t} + \hat{u})   
      D_{001}^4 \nonumber \\  
     &+&\left. 8 (-4 m_{h0}^2 + \hat{s})   
     D_{001}^5 +   
   2 (m_{h0}^2 + 2 \hat{t})   
      D_{002}^4  
    - 8 (m_{h0}^2 + \hat{t})   
      D_{002}^5\right) \nonumber \\  
 &+&   
 {e^2 g^2 \sin (\alpha - \beta)^2\over 32 \pi^2}   
  \left( 2 (-m_{h0}^2 + \hat{t}) (-m_{h0}^2 + \hat{u}) D_{11}^1 +   
   2 (-m_{h0}^2 + \hat{t}) (-\hat{t} + \hat{u}) D_{12}^1 \right. \nonumber \\  
&+&   
   (-2 m_{h0}^4 + 2 m_{h0}^2 \hat{t} - \hat{t}^2 + 2 m_{h0}^2 \hat{u} - \hat{u}^2)   
      D_{13}^1  
 \nonumber \\  
&-&   
   (-m_{h0}^2 + \hat{t})^2 D_{22}^1 +   
   2 (-13 m_{h0}^2 + 4 \hat{s} + 2 \hat{t} + \hat{u})   
D_{001}^1 \nonumber \\  
& - &\left.   
   2 (3 m_{h0}^2 + 2 \hat{t}) D_{002}^1\right)  
 +   
   {e^2 g^2\over 32 \pi^2}   
\left( C_{00}^1   
      (\cos (\alpha - \beta)^2 - 3 \sin (\alpha - \beta)^2) \right. \nonumber \\  
& - & \left.  
   B_0(0, m_{w}^2, m_{w}^2) +   
(m_{h0}^2 - \hat{u}) C_1^1  
       -   
   \hat{s} C_2^1\right)  
      +   
{e^2 g^2\over 32 \cos\theta_w^2 \pi^2}   
   \left( \right. \nonumber \\  
&& 4 D_{00}^6   
      (-2 \cos\theta_w m_{w} \sin (\alpha - \beta) + m_z \cos (2 \beta) \sin (\alpha + \beta))^2  
\nonumber \\  
&+& 2 D_2^1 \sin (\alpha - \beta)   
  (3 \cos\theta_w m_{h0}^2 m_{w}^2 \sin (\alpha - \beta) \nonumber \\  
&-& \cos\theta_w m_{h0}^2 \hat{t} \sin (\alpha - \beta) +   
        \cos\theta_w m_{w}^2 \hat{t} \sin (\alpha - \beta) + \cos\theta_w \hat{t}^2 \sin (\alpha - \beta)  
\nonumber \\  
& + &  
        m_{h0}^2 m_{w} m_{z} \cos (2 \beta) \sin (\alpha + \beta) -   
        m_{w} m_{z} \hat{t} \cos (2 \beta) \sin (\alpha + \beta)) \nonumber \\  
&+&   
   2 D_{1}^1   
      \sin (\alpha - \beta) (6 \cos\theta_w m_{h0}^2 m_{w}^2 \sin (\alpha - \beta) -   
        \cos\theta_w m_{w}^2 \hat{s} \sin (\alpha - \beta) \nonumber \\  
&+& \cos\theta_w \hat{t}^2 \sin (\alpha - \beta) +   
        2 \cos\theta_w m_{w}^2 \hat{u} \sin (\alpha - \beta) - \cos\theta_w \hat{t} \hat{u} \sin (\alpha - \beta) \nonumber \\  
& - &  
        m_{w} m_{z} \hat{t} \cos (2 \beta) \sin (\alpha + \beta) +   
        m_{w} m_{z} \hat{u} \cos (2 \beta) \sin (\alpha + \beta)) \nonumber \\  
&+ &  
   2 (-m_{h0}^2 + \hat{t}) \cos (\alpha - \beta)   
     D_2^4 (-(\cos\theta_w m_{H^\pm}^2 \cos (\alpha - \beta)) + \cos\theta_w m_{w}^2 \cos (\alpha - \beta) \nonumber \\  
&+ &  
        \cos\theta_w \hat{t} \cos (\alpha - \beta) - m_{w} m_{z} \sin (2 \beta) \sin (\alpha + \beta))  
    \nonumber \\  
& + &2 (-\hat{t} + \hat{u}) \cos (\alpha - \beta)   
     D_{1}^4   
      (\cos\theta_w m_{H^\pm}^2 \cos (\alpha - \beta) - \cos\theta_w m_{w}^2 \cos (\alpha - \beta) \nonumber \\  
&- &  
        \cos\theta_w \hat{t} \cos (\alpha - \beta) + m_{w} m_{z} \sin (2 \beta) \sin (\alpha + \beta))  
    \nonumber \\  
& - &D_0^4   
      (\cos\theta_w m_{H^\pm}^2 \cos (\alpha - \beta) - \cos\theta_w m_{w}^2 \cos (\alpha - \beta) \nonumber \\  
&-&   
         \cos\theta_w \hat{t} \cos (\alpha - \beta) + m_{w} m_{z} \sin (2 \beta) \sin (\alpha + \beta))^2  
    \nonumber \\  
&+ &C_0^1   
      (-(\cos\theta_w m_{H^\pm}^2 \cos (\alpha - \beta)^2) - \cos\theta_w m_{h0}^2 \cos (\alpha - \beta)^2 +   
        \cos\theta_w m_{w}^2 \cos (\alpha - \beta)^2\nonumber \\  
& + &3 \cos\theta_w \hat{t} \cos (\alpha - \beta)^2 -   
        \cos\theta_w \hat{u} \cos (\alpha - \beta)^2 - \cos\theta_w m_{h0}^2 \sin (\alpha - \beta)^2 \nonumber \\  
&+ &  
        3 \cos\theta_w m_{w}^2 \sin (\alpha - \beta)^2 + 3 \cos\theta_w \hat{t} \sin (\alpha - \beta)^2 -   
        \cos\theta_w \hat{u} \sin (\alpha - \beta)^2 \nonumber \\  
&- &  
        2 m_{w} m_{z} \cos (2 \beta) \sin (\alpha - \beta) \sin (\alpha + \beta) -   
        2 m_{w} m_{z} \cos (\alpha - \beta) \sin (2 \beta) \sin (\alpha + \beta))  
   \nonumber \\  
&+&D_{00}^1   
      (8 \cos\theta_w^2 m_{h0}^2 \sin (\alpha - \beta)^2 +   
        48 \cos\theta_w^2 m_{w}^2 \sin (\alpha - \beta)^2 - 17 \cos\theta_w^2 \hat{t} \sin (\alpha - \beta)^2\nonumber \\  
& - &  
        18 \cos\theta_w^2 \hat{u} \sin (\alpha - \beta)^2 +   
        6 \cos\theta_w m_{w} m_{z} \cos (2 \beta) \sin (\alpha - \beta) \sin (\alpha + \beta) \nonumber \\  
&+ &  
        4 m_{z}^2 \cos (2 \beta)^2 \sin (\alpha + \beta)^2)   
+   
  D_0^1   
      (-16 \cos\theta_w^2 m_{h0}^2 m_{w}^2 \sin (\alpha - \beta)^2 \nonumber \\  
& - &  
        4 \cos\theta_w^2 m_{w}^2 \hat{s} \sin (\alpha - \beta)^2   
+   
        10 \cos\theta_w^2 m_{w}^2 \hat{t} \sin (\alpha - \beta)^2\nonumber \\  
& - &\cos\theta_w^2 \hat{t}^2 \sin (\alpha - \beta)^2 +   
        12 \cos\theta_w^2 m_{w}^2 \hat{u} \sin (\alpha - \beta)^2 \nonumber \\  
&- &  
        2 \cos\theta_w m_{w}^3 m_{z} \cos (2 \beta) \sin (\alpha - \beta) \sin (\alpha + \beta) \nonumber \\  
&+&   
        2 \cos\theta_w m_{w} m_{z} \hat{t} \cos (2 \beta) \sin (\alpha - \beta) \sin (\alpha + \beta) \nonumber \\  
&- &  
        m_{w}^2 m_{z}^2 \cos (2 \beta)^2 \sin (\alpha + \beta)^2) +   
   4 D_{00}^5   
      (-2 \cos\theta_w^2 \hat{t} \cos (\alpha - \beta)^2 \nonumber \\  
&-& 2 \cos\theta_w^2 \hat{u} \cos (\alpha - \beta)^2 -   
        2 \cos\theta_w m_{w} m_{z} \cos (\alpha - \beta) \sin (2 \beta) \sin (\alpha + \beta) \nonumber \\  
&+&   
        m_{z}^2 \sin (2 \beta)^2 \sin (\alpha + \beta)^2)   
 +   
   D_{00}^4   
      (\cos\theta_w^2 m_{H^\pm}^2 \cos (\alpha - \beta)^2 + 8 \cos\theta_w^2 m_{h0}^2 \cos (\alpha - \beta)^2 \nonumber \\  
&- &  
        \cos\theta_w^2 m_{w}^2 \cos (\alpha - \beta)^2 - 9 \cos\theta_w^2 \hat{t} \cos (\alpha - \beta)^2 -   
        10 \cos\theta_w^2 \hat{u} \cos (\alpha - \beta)^2 \nonumber \\  
&- &\left.  
        2 \cos\theta_w m_{w} m_{z} \cos (\alpha - \beta) \sin (2 \beta) \sin (\alpha + \beta) +   
        4 m_{z}^2 \sin (2 \beta)^2 \sin (\alpha + \beta)^2)  
\right),   
\end{eqnarray}  
\begin{eqnarray}  
&&f_1^{(1,charginos)}= \nonumber \\
&&\left[{e^2\over 16 \pi^2}\left( 8 (-a_3^2 + b_{3}^2) B_0(0, m_{\tilde{\chi}_2^\pm}^2, m_{\tilde{\chi}_2^\pm}^2) +   
2 (-(a_3^2 m_{h0}^2) + b_{3}^2 m_{h0}^2 \right. \right. \nonumber \\  
&-& 2 a_3^2 m_{\tilde{\chi}_1^\pm}^2 + 2 b_{3}^2 m_{\tilde{\chi}_1^\pm}^2 -   
       2 a_3^2 m_{\tilde{\chi}_1^\pm} m_{\tilde{\chi}_2^\pm} - 2 b_{3}^2 m_{\tilde{\chi}_1^\pm} m_{\tilde{\chi}_2^\pm} + a_3^2 \hat{t} - b_{3}^2 \hat{t} + 2 a_3^2 \hat{u} -   
       2 b_{3}^2 \hat{u}) C_0^{11} \nonumber \\  
&+&  
2 m_{\tilde{\chi}_1^\pm} (a_3^2 m_{\tilde{\chi}_1^\pm}   
- b_{3}^2 m_{\tilde{\chi}_1^\pm} + a_3^2 m_{\tilde{\chi}_2^\pm} + b_{3}^2 m_{\tilde{\chi}_2^\pm})   
     (-2 m_{\tilde{\chi}_1^\pm}^2   
+ 2 m_{\tilde{\chi}_2^\pm}^2 - \hat{s} - 2 \hat{t}) D_0^{12} \nonumber \\  
&+&   
2 (a_3^2 - b_{3}^2) (-m_{h0}^2 + \hat{u}) C_1^{11} \nonumber \\  
&+&   
    4 (-(a_3^2 m_{h0}^4) + b_{3}^2 m_{h0}^4 - 3 a_3^2 m_{h0}^2 m_{\tilde{\chi}_1^\pm}^2 +   
       3 b_{3}^2 m_{h0}^2 m_{\tilde{\chi}_1^\pm}^2 - 2 a_3^2 m_{h0}^2 m_{\tilde{\chi}_1^\pm} m_{\tilde{\chi}_2^\pm} \nonumber \\  
&-& 2 b_{3}^2 m_{h0}^2 m_{\tilde{\chi}_1^\pm} m_{\tilde{\chi}_2^\pm} +   
       a_3^2 m_{h0}^2 m_{\tilde{\chi}_2^\pm}^2 - b_{3}^2 m_{h0}^2 m_{\tilde{\chi}_2^\pm}^2 - a_3^2 m_{h0}^2 \hat{t} + b_{3}^2 m_{h0}^2 \hat{t} \nonumber \\  
&-&   
       2 a_3^2 m_{\tilde{\chi}_1^\pm}^2 \hat{t} + 2 b_{3}^2 m_{\tilde{\chi}_1^\pm}^2 \hat{t} - 2 a_3^2 m_{\tilde{\chi}_1^\pm} m_{\tilde{\chi}_2^\pm} \hat{t} -   
       2 b_{3}^2 m_{\tilde{\chi}_1^\pm} m_{\tilde{\chi}_2^\pm} \hat{t} - a_3^2 m_{\tilde{\chi}_1^\pm}^2 \hat{u} \nonumber \\  
&+& b_{3}^2 m_{\tilde{\chi}_1^\pm}^2 \hat{u} + a_3^2 m_{\tilde{\chi}_2^\pm}^2 \hat{u} -   
       b_{3}^2 m_{\tilde{\chi}_2^\pm}^2 \hat{u}) D_1^{12}   
+   
4 (-a_3^2 + b_{3}^2) \hat{s} C_2^{11}   
+   
    2 (-(a_3^2 m_{h0}^4) + b_{3}^2 m_{h0}^4 \nonumber \\  
	&-& 2 a_3^2 m_{h0}^2 m_{\tilde{\chi}_1^\pm}^2 +   
       2 b_{3}^2 m_{h0}^2 m_{\tilde{\chi}_1^\pm}^2 + 2 a_3^2 m_{h0}^2 m_{\tilde{\chi}_2^\pm}^2 - 2 b_{3}^2 m_{h0}^2 m_{\tilde{\chi}_2^\pm}^2 -   
       2 a_3^2 m_{h0}^2 \hat{t}\nonumber \\   
&+& 2 b_{3}^2 m_{h0}^2 \hat{t} - 4 a_3^2 m_{\tilde{\chi}_1^\pm}^2 \hat{t}  
+ 4 b_{3}^2 m_{\tilde{\chi}_1^\pm}^2 \hat{t} -   
       4 a_3^2 m_{\tilde{\chi}_1^\pm} m_{\tilde{\chi}_2^\pm} \hat{t} - 4 b_{3}^2 m_{\tilde{\chi}_1^\pm} m_{\tilde{\chi}_2^\pm} \hat{t} + a_3^2 \hat{t} \hat{u} - b_{3}^2 \hat{t} \hat{u}) D_2^{12} \nonumber \\  
&+&   
    8 (a_3^2 - b_{3}^2) C_{00}^{11}+   
    8 (a_3^2 m_{\tilde{\chi}_1^\pm}^2 - b_{3}^2 m_{\tilde{\chi}_1^\pm}^2 + 2 a_3^2 m_{\tilde{\chi}_1^\pm} m_{\tilde{\chi}_2^\pm} + 2 b_{3}^2 m_{\tilde{\chi}_1^\pm} m_{\tilde{\chi}_2^\pm} +   
       a_3^2 m_{\tilde{\chi}_2^\pm}^2 - b_{3}^2 m_{\tilde{\chi}_2^\pm}^2 - a_3^2 \hat{t} \nonumber \\  
&+& b_{3}^2 \hat{t} - a_3^2 \hat{u} + b_{3}^2 \hat{u}) D_{00}^{12} +   
8 (-a_3^2 + b_{3}^2) m_{h0}^2 (\hat{t} + \hat{u}) D_{11}^{12}   
+   
4 (a_3^2 - b_{3}^2) (-4 m_{h0}^2 + \hat{s}) (m_{h0}^2 + \hat{t}) D_{12}^{12} \nonumber \\  
&+&\left. \left.  
    2 (a_3^2 - b_{3}^2) (-8 m_{h0}^4 + 4 m_{h0}^2 \hat{s} - \hat{s}^2) D_{13}^{12} +   
2 (-a_3^2 + b_{3}^2) (m_{h0}^2 + \hat{t})^2 D_{22}^{12}\right) \right]  
 \nonumber \\  
&+&  
\left[ a_3\rightarrow  a_2, b_{3}\rightarrow  b_{2}, m_{\tilde{\chi}_1^\pm}\rightarrow  m_{\tilde{\chi}_2^\pm}\right]  
+  
\left[a_3\rightarrow  a_1, b_{3}\rightarrow  b_{1}, m_{\tilde{\chi}_2^\pm}\rightarrow  m_{\tilde{\chi}_1^\pm}\right]\nonumber \\  
&+&  
\left[m_{\tilde{\chi}_1^\pm}\rightarrow  m_{\tilde{\chi}_2^\pm}, m_{\tilde{\chi}_2^\pm}\rightarrow  m_{\tilde{\chi}_1^\pm}\right],  
\label{appa1} 
\end{eqnarray}  
\begin{eqnarray}  
f_1^{(1,sfermions)}&=& 
-N_c \left[{e^2 e_t^2\over 2 \pi^2}\left(\xi_{\tilde{t}3}^2 D_{00}^{18}+  
\xi_{\tilde{t}2}^2 D_{00}^{15}+  
\xi_{\tilde{t}1}^2 D_{00}^{14}+  
\xi_{\tilde{t}3}^2 D_{00}^{19}\right) \right]\nonumber \\  
&-& N_c\left[e_t \rightarrow e_b, \xi_{\tilde{t}1} \rightarrow \xi_{\tilde{b}1}, 
\xi_{\tilde{t}2}\rightarrow\xi_{\tilde{b}2},\xi_{\tilde{t}3}\rightarrow\xi_{\tilde{b}3}, 
m_{\tilde{t}_1}\rightarrow m_{\tilde{b}_1}, m_{\tilde{t}_2}\rightarrow m_{\tilde{b}_2} 
 \right]\nonumber \\ 
&-&\left[e_t \rightarrow e_{\tau}, \xi_{\tilde{t}1} \rightarrow \xi_{\tilde{\tau}1}, 
\xi_{\tilde{t}2}\rightarrow\xi_{\tilde{\tau}2},\xi_{\tilde{t}3}\rightarrow\xi_{\tilde{\tau}3}, 
m_{\tilde{t}_1}\rightarrow m_{\tilde{\tau}_1}, m_{\tilde{t}_2}\rightarrow m_{\tilde{\tau}_2} 
\right],  
\label{appa2} 
\end{eqnarray}  
where $C_m^i$, $C_{mn}^i$ $\equiv C_m, C_{mn}(0, 0, \hat{s}, i)$ and $D_m^i$, $D_{mn}^i$, $ 
D_{mnl}^i $ $\equiv D_m$, $D_{mn}$, $D_{mnl}( m_{h0}^2, 0, 0, m_{h0}^2, \hat{t}, \hat{s}, i)$ are three-point and four-point Feynmann integrals,  
definitions for which can be found in Ref.\cite{denner}, and $i$ represent the  internal  
particles masses, which are given for $i =1-19$ by 
\begin{eqnarray}  
&&(m_{w}^2,m_{w}^2,m_{w}^2),(m_{H^\pm}^2,m_{w}^2,m_{w}^2),  
(m_{w}^2,m_{H^\pm}^2,m_{H^\pm}^2),  
(m_{w}^2,m_{w}^2,m_{H^\pm}^2),\nonumber \\  
&&(m_{H^\pm}^2,m_{H^\pm}^2,m_{w}^2),(m_{H^\pm}^2,m_{H^\pm}^2,m_{H^\pm}^2),  
(m_{H^\pm}^2,m_{w}^2,m_{H^\pm}^2),  
(m_{w}^2,m_{H^\pm}^2,m_{w}^2), \nonumber \\  
&&(m_t^2,m_t^2,m_t^2),  
(m_{\tilde{\chi}_1^\pm}^2, m_{\tilde{\chi}_1^\pm}^2, m_{\tilde{\chi}_1^\pm}^2),  
(m_{\tilde{\chi}_2^\pm}^2, m_{\tilde{\chi}_2^\pm}^2, m_{\tilde{\chi}_2^\pm}^2),  
(m_{\tilde{\chi}_1^\pm}^2, m_{\tilde{\chi}_1^\pm}^2, m_{\tilde{\chi}_2^\pm}^2), \nonumber\\  
&&(m_{\tilde{\chi}_2^\pm}^2, m_{\tilde{\chi}_2^\pm}^2, m_{\tilde{\chi}_1^\pm}^2),  
(m_{\tilde{\chi}_1^\pm}^2, m_{\tilde{\chi}_2^\pm}^2, m_{\tilde{\chi}_2^\pm}^2),  
(m_{\tilde{\chi}_2^\pm}^2, m_{\tilde{\chi}_1^\pm}^2, m_{\tilde{\chi}_1^\pm}^2),  
(m_{\tilde{t}_1 }^2, m_{\tilde{t}_1 }^2, m_{\tilde{t}_1 }^2), \nonumber \\  
&&(m_{\tilde{t}_2 }^2, m_{\tilde{t}_2 }^2, m_{\tilde{t}_2 }^2),  
(m_{\tilde{t}_1 }^2, m_{\tilde{t}_2 }^2, m_{\tilde{t}_1 }^2),  
(m_{\tilde{t}_2 }^2, m_{\tilde{t}_1 }^2, m_{\tilde{t}_2 }^2)  
\label{mass1}
\end{eqnarray}  
for $C_m^i, C_{mn}^i$, and  
\begin{eqnarray}  
&&(m_{w}^2, m_{w}^2, m_{w}^2, m_{w}^2), (m_{H^\pm}^2, m_{H^\pm}^2,m_{w}^2, m_{w}^2), (m_{w}^2, m_{w}^2,  m_{H^\pm}^2, m_{H^\pm}^2),\nonumber \\  
&&(m_{H^\pm}^2, m_{w}^2, m_{w}^2, m_{w}^2),   
(m_{w}^2, m_{H^\pm}^2, m_{H^\pm}^2, m_{H^\pm}^2), (m_{H^\pm}^2, m_{H^\pm}^2, m_{H^\pm}^2, m_{H^\pm}^2),\nonumber \\  
&&(m_t^2, m_t^2, m_t^2, m_t^2),  
(m_{\tilde{\chi}_1^\pm}^2, m_{\tilde{\chi}_1^\pm}^2, m_{\tilde{\chi}_1^\pm}^2, m_{\tilde{\chi}_1^\pm}^2),  
(m_{\tilde{\chi}_2^\pm}^2, m_{\tilde{\chi}_2^\pm}^2, m_{\tilde{\chi}_2^\pm}^2, m_{\tilde{\chi}_2^\pm}^2),\nonumber \\  
&&(m_{\tilde{\chi}_1^\pm}^2, m_{\tilde{\chi}_1^\pm}^2, m_{\tilde{\chi}_2^\pm}^2, m_{\tilde{\chi}_2^\pm}^2),  
(m_{\tilde{\chi}_2^\pm}^2, m_{\tilde{\chi}_2^\pm}^2, m_{\tilde{\chi}_1^\pm}^2, m_{\tilde{\chi}_1^\pm}^2),  
(m_{\tilde{\chi}_1^\pm}^2, m_{\tilde{\chi}_2^\pm}^2, m_{\tilde{\chi}_2^\pm}^2, m_{\tilde{\chi}_2^\pm}^2),\nonumber \\  
&&(m_{\tilde{\chi}_2^\pm}^2, m_{\tilde{\chi}_1^\pm}^2, m_{\tilde{\chi}_1^\pm}^2, m_{\tilde{\chi}_1^\pm}^2),  
(m_{\tilde{t}_1 }^2, m_{\tilde{t}_1 }^2, m_{\tilde{t}_1 }^2, m_{\tilde{t}_1 }^2),  
(m_{\tilde{t}_2 }^2, m_{\tilde{t}_2 }^2, m_{\tilde{t}_2 }^2, m_{\tilde{t}_2 }^2),\nonumber \\  
&&(m_{\tilde{t}_1 }^2, m_{\tilde{t}_1 }^2, m_{\tilde{t}_2 }^2, m_{\tilde{t}_2 }^2),  
(m_{\tilde{t}_2 }^2, m_{\tilde{t}_2 }^2, m_{\tilde{t}_1 }^2, m_{\tilde{t}_1 }^2),  
(m_{\tilde{t}_1 }^2, m_{\tilde{t}_2 }^2, m_{\tilde{t}_2 }^2, m_{\tilde{t}_2 }^2),   
(m_{\tilde{t}_2 }^2, m_{\tilde{t}_1 }^2, m_{\tilde{t}_1 }^2, m_{\tilde{t}_1 }^2)  
\label{mass2}
\end{eqnarray}  
for $D_m^i, D_{mn}^i, D_{mnl}^i$. 
And $a_i, b_i (i=1, 2, 3)$ in Eqs. \ref{appa1} are the coupling constants of the vertex  
$h_0-\chi^{\pm}_{1,2}-\chi^{\mp}_{1,2}$, which  are given by  
\begin{eqnarray} 
a_1&=&{i g \over 2}((Q^*_{11}+Q_{11}) \sin\alpha -(S^*_{11}+S_{11}) \cos\alpha)  
\nonumber \\ 
b_1&=&{i g \over 2}((-Q^*_{11}+Q_{11}) \sin\alpha +(S^*_{11}-S_{11}) \cos\alpha)  
\nonumber \\ 
a_2&=&{i g \over 2}((Q^*_{22}+Q_{22}) \sin\alpha -(S^*_{22}+S_{22}) \cos\alpha)  
\nonumber \\ 
b_2&=&{i g \over 2}((-Q^*_{22}+Q_{22}) \sin\alpha +(S^*_{22}-S_{22}) \cos\alpha)  
\nonumber \\ 
a_3&=&{i g \over 2}((Q^*_{21}+Q_{12}) \sin\alpha -(S^*_{21}+S_{12}) \cos\alpha)  
\nonumber \\ 
b_3&=&{i g \over 2}((-Q^*_{21}+Q_{12}) \sin\alpha +(S^*_{21}-S_{12}) \cos\alpha), 
\end{eqnarray} 
where the details of the matrix $Q$ and $S$ can be found in Ref. \cite{1}. 
In Eqs. \ref{appa2} we only write down the contributions from the third generation sfermions, and $N_c$ is the number of colors, $e_t, e_b, e_{\tau}$ are the electric charge of 
the top quark, the bottom quark and the tau, respectively, the $\xi_{\tilde{t}j}, \xi_{\tilde{b}j}, \xi_{\tilde{\tau} j}(j=1, 2, 3)$ are the  coupling constants 
 of the vertexes 
$ h_0-\tilde{t}-\tilde{t}$,  $ h_0-\tilde{b}-\tilde{b}$, $ h_0-\tilde{\tau}-\tilde{\tau}$, respectively, which are given by (cf. Ref. \cite{1} ) 
\begin{eqnarray}  
\left( \begin{array}{cc}  
\xi_{\tilde{t}1} & \xi_{\tilde{t}3} \\  
\xi_{\tilde{t}3} & \xi_{\tilde{t}2}  
\end{array}\right)=i g  R^{\tilde{t}} A_{\tilde{t}} (R^{\tilde{t}})^T 
\end{eqnarray} 
with 
\begin{eqnarray} 
A_{\tilde{t}}=\left(\begin{array}{cc}  
 {m_{z} ((1/2)-(2/3) \sin^2\theta_w) \sin ( \alpha +\beta  )\over \cos\theta_w} -  
 {m_t^2 \cos ( \alpha )\over m_{w} \sin ( \beta  )}  
&{m_t\over 2 m_{w} \sin\beta} (-A_t \cos\alpha-\mu \sin\alpha) \\  
 {m_t\over 2 m_{w} \sin\beta} (-A_t \cos\alpha-\mu \sin\alpha)&  
 {m_{z} (2/3) \sin^2\theta_w \sin (\alpha+\beta) \over \cos\theta_w}-  
{m_t^2 \cos \alpha \over m_{w} \sin\beta}  
 \end{array} \right),  
\end{eqnarray}  
\begin{eqnarray}  
\left( \begin{array}{cc}  
\xi_{\tilde{b}1} & \xi_{\tilde{b}3} \\  
\xi_{\tilde{b}3} & \xi_{\tilde{b}2}  
\end{array}\right)=i g  R^{\tilde{b}} A_{\tilde{b}} (R^{\tilde{b}})^T 
\end{eqnarray} 
with 
\begin{eqnarray} 
A_{\tilde{b}}=\left(\begin{array}{cc}  
 {-m_{z} ((1/2)-(1/3) \sin^2\theta_w) \sin ( \alpha +\beta  )\over \cos\theta_w} +  
 {m_b^2 \sin ( \alpha )\over m_{w} \cos ( \beta  )}  
&{m_b\over 2 m_{w} \cos\beta} (A_b \sin\alpha+\mu \cos\alpha) \\  
 {m_b\over 2 m_{w} \cos\beta} (A_b \sin\alpha+\mu \cos\alpha)&  
 {m_{z} (-1/3) \sin^2\theta_w \sin (\alpha+\beta) \over \cos\theta_w }+  
{m_b^2 \sin \alpha \over m_{w} \cos\beta}  
 \end{array} \right),  
\end{eqnarray}  
and 
\begin{eqnarray}  
\left( \begin{array}{cc}  
\xi_{\tilde{\tau}1} & \xi_{\tilde{\tau}3} \\  
\xi_{\tilde{\tau}3} & \xi_{\tilde{\tau}2}  
\end{array}\right)=({i g m_z\over \cos\theta_w}) 
\left(\begin{array}{cc}  
(-1/2+\sin^2\theta_w)\sin(\alpha+\beta)  
&0\\  
 0& -\sin^2\theta_w \sin(\alpha+\beta) 
 \end{array} \right). 
\end{eqnarray} 
  
The form factor $ f_{1}^{(2)}$ is given by 
\begin{eqnarray}  
f_{1}^{(2)}&=&f_1^{(2, 2HDM)}+f_{1}^{(2, charginos)}+f_1^{(2,sfermions)},  
\end{eqnarray}  
with 
\begin{eqnarray}  
f_1^{(2, 2HDM)}&=&f_1^{(2, fermions)}+f_1^{(2,bosons)}. 
\end{eqnarray} 
Here the  $f_1^{(2, fermions)}$, $f_1^{(2,bosons)}$,  $f_{1}^{(2, charginos)}$ and $f_1^{(2,sfermions)}$ are  
 \begin{eqnarray}  
&&f_1^{(2,fermions)}=\nonumber \\  
&&{e^2 g^2m_t^2 \cos (\alpha)^2 csc(\beta)^2 \over 12 m_{w}^2 \pi^2}   
\left[ -2 B_0(0, m_t^2, m_t^2) +   
   (m_{h0}^2 - 2 \hat{t} - \hat{u})   
      C_0^9 +   
   2 m_t^2 \hat{s}   
      D_0^7  \right. \nonumber \\  
&+&  
 (-4 m_{h0}^2 + \hat{s})   
     C_1^9 -   
   2 (m_{h0}^2 + \hat{t})   
      C_2^9  \nonumber \\  
&+&   
 (m_{h0}^4 - m_{h0}^2 \hat{u} + \hat{s} \hat{u})   
      D_2^7  
    + m_{h0}^2  (m_{h0}^2 - \hat{u})   
      D_3^7  
    + 2 (-m_{h0}^2 - 8 m_t^2 + 2 \hat{t} + \hat{u})   
     D_{00}^7 \nonumber \\  
&+&   
  (-m_{h0}^2 + \hat{t}) (-m_{h0}^2 + \hat{u})   
      D_{12}^7  
    - (-m_{h0}^2 + \hat{t})^2   
      D_{13}^7  
     + (m_{h0}^2 - \hat{u}) (m_{h0}^2 + \hat{u})   
     D_{22}^7 \nonumber \\  
&+&\left.  (m_{h0}^4 + m_{h0}^2 \hat{t} - 3 m_{h0}^2 \hat{u} + \hat{t} \hat{u})   
      D_{23}^7 +   
   2 m_{h0}^2  (-m_{h0}^2 + \hat{t})   
     D_{33}^7  
\right] \nonumber \\  
&+&   
{e^2 g^2 m_b^2 \sin (\alpha)^2 sec(\beta)^2\over 48 m_{w}^2 \pi^2}\left[m_t\rightarrow m_b   
\right],   
\end{eqnarray}  
\begin{eqnarray}  
&&f_1^{(2, bosons)}= \nonumber \\  
&&{e^2 g^2 \cos (\alpha - \beta)^2\over 16 \pi^2}   
\left( (-\hat{t} + \hat{u}) (D_{001}^2 -   
     D_{001}^3) +   
   2 (m_{h0}^2 + \hat{u}) (  
      D_{002}^2 -   
  D_{002}^3) \right. \nonumber \\  
&+& \left.  
 (4 m_{h0}^2 - \hat{s}) (  
     D_{003}^2-   
   D_{003}\right)  
 +    
  { e^2 g^2 \over 16 \cos\theta_w^2 \pi^2}   
\left( \right. \nonumber \\  
&&2 D_{00}^6   
      (-2 \cos\theta_w m_{w} \sin (\alpha - \beta) + m_{z} \cos (2 \beta) \sin (\alpha + \beta))^2  
     \nonumber \\  
&+& m_{w}^2 D_{0}^1 \sin (\alpha - \beta)   
      (-12 \cos\theta_w m_{h0}^2 \sin (\alpha - \beta) - \cos\theta_w \hat{s} \sin (\alpha - \beta)\nonumber \\   
&+&   
        7 \cos\theta_w \hat{t} \sin (\alpha - \beta) + 7 \cos\theta_w \hat{u} \sin (\alpha - \beta) -   
        2 m_{w} m_{z} \cos (2 \beta) \sin (\alpha + \beta)) \nonumber \\  
&+&   
  2 D_{00}^1   
      (2 \cos\theta_w^2 m_{h0}^2 \sin (\alpha - \beta)^2 +   
        12 \cos\theta_w^2 m_{w}^2 \sin (\alpha - \beta)^2 \nonumber \\  
&-& 4 \cos\theta_w^2 \hat{t} \sin (\alpha - \beta)^2   
-  
        4 \cos\theta_w^2 \hat{u} \sin (\alpha - \beta)^2 \nonumber \\  
& +&   
        2 \cos\theta_w m_{w} m_{z} \cos (2 \beta) \sin (\alpha - \beta) \sin (\alpha + \beta) +   
        m_{z}^2 \cos (2 \beta)^2 \sin (\alpha + \beta)^2) \nonumber \\  
&+&   
  D_{00}^3   
      (\cos\theta_w^2 m_{h0}^2 \cos (\alpha - \beta)^2 - 4 \cos\theta_w^2 \hat{t} \cos (\alpha - \beta)^2 -   
        5 \cos\theta_w^2 \hat{u} \cos (\alpha - \beta)^2 \nonumber \\  
&-&   
        2 \cos\theta_w m_{w} m_{z} \cos (\alpha - \beta) \sin (2 \beta) \sin (\alpha + \beta) +   
        2 m_{z}^2 \sin (2 \beta)^2 \sin (\alpha + \beta)^2) \nonumber \\  
&+&   
  D_{00}^2   
      (3 \cos\theta_w^2 m_{h0}^2 \cos (\alpha - \beta)^2 - 4 \cos\theta_w^2 \hat{t} \cos (\alpha - \beta)^2 -   
        3 \cos\theta_w^2 \hat{u} \cos (\alpha - \beta)^2 \nonumber \\  
&-&\left.   
        2 \cos\theta_w m_{w} m_{z} \cos (\alpha - \beta) \sin (2 \beta) \sin (\alpha + \beta) +   
        2 m_{z}^2 \sin (2 \beta)^2 \sin (\alpha + \beta)^2)  
\right),  
\end{eqnarray}  
\begin{eqnarray}  
&&f_1^{(2,charginos)}=\nonumber \\  
&&\left[ {e^2\over 16 \pi^2}\left(8 (a_3^2 - b_{3}^2) B_0(0, m_{\tilde{\chi}_2^\pm}^2, m_{\tilde{\chi}_2^\pm}^2) +   
2 (a_3^2 - b_{3}^2) (-m_{h0}^2 + 2 m_{\tilde{\chi}_1^\pm}^2 - 2 m_{\tilde{\chi}_2^\pm}^2 + 2 \hat{t} + \hat{u}) C_0^{13} \right. \right.\nonumber \\  
&+&   
2 m_{\tilde{\chi}_1^\pm} (-(a_3^2 m_{\tilde{\chi}_1^\pm}) + b_{3}^2 m_{\tilde{\chi}_1^\pm} - a_3^2 m_{\tilde{\chi}_2^\pm} - b_{3}^2 m_{\tilde{\chi}_2^\pm}) \hat{s} D_0^{10} \nonumber \\  
&+&   
    2 (a_3^2 - b_{3}^2) (4 m_{h0}^2 - \hat{s}) C_1^{14} +   
    4 (a_3^2 - b_{3}^2) (m_{h0}^2 + \hat{t}) C_2^{14} +   
2 (a_3^2 - b_{3}^2) (-m_{h0}^4 \nonumber \\  
&-&  
 m_{h0}^2 m_{\tilde{\chi}_1^\pm}^2 + m_{h0}^2 m_{\tilde{\chi}_2^\pm}^2 + m_{h0}^2 \hat{u} +   
       m_{\tilde{\chi}_1^\pm}^2 \hat{u} - m_{\tilde{\chi}_2^\pm}^2 \hat{u} - \hat{s} \hat{u}) D_2^{10} \nonumber \\  
&+&   
2 (a_3^2 - b_{3}^2) (-m_{h0}^4 + m_{h0}^2 m_{\tilde{\chi}_1^\pm}^2 - m_{h0}^2 m_{\tilde{\chi}_2^\pm}^2 - m_{\tilde{\chi}_1^\pm}^2 \hat{t} \nonumber \\  
&+&   
       m_{\tilde{\chi}_2^\pm}^2 \hat{t} + m_{h0}^2 \hat{u}) D_3^{10} +   
4 (a_3^2 m_{h0}^2 - b_{3}^2 m_{h0}^2 \nonumber \\  
&+& 2 a_3^2 m_{\tilde{\chi}_1^\pm}^2 - 2 b_{3}^2 m_{\tilde{\chi}_1^\pm}^2   
+   
       4 a_3^2 m_{\tilde{\chi}_1^\pm} m_{\tilde{\chi}_2^\pm} + 4 b_{3}^2 m_{\tilde{\chi}_1^\pm} m_{\tilde{\chi}_2^\pm} + 2 a_3^2 m_{\tilde{\chi}_2^\pm}^2 - 2 b_{3}^2 m_{\tilde{\chi}_2^\pm}^2 \nonumber \\  
&-&   
       2 a_3^2 \hat{t} + 2 b_{3}^2 \hat{t} - a_3^2 \hat{u} + b_{3}^2 \hat{u}) D_{00}^{10} +   
2 (a_3^2 - b_{3}^2) (m_{h0}^2 - \hat{t}) (-m_{h0}^2 + \hat{u}) D_{12}^{10}\nonumber \\  
& +&   
    2 (a_3^2 - b_{3}^2) (m_{h0}^2 - \hat{t})^2 D_{13}^{10}+   
2 (a_3^2 - b_{3}^2) (-m_{h0}^2 + \hat{u}) (m_{h0}^2 + \hat{u}) D_{22}^{10} \nonumber \\  
&+& \left.\left.  
    2 (a_3^2 - b_{3}^2) (-m_{h0}^4 - m_{h0}^2 \hat{t} + 3 m_{h0}^2 \hat{u} - \hat{t} \hat{u}) D_{23}^{10} +   
4 (a_3^2 - b_{3}^2) m_{h0}^2 (m_{h0}^2 - \hat{t}) D_{33}^{10}\right)\right] \nonumber \\  
&+&  
\left[a_3\rightarrow  a_2, b_{3}\rightarrow  b_{2}, m_{\tilde{\chi}_1^\pm}\rightarrow  m_{\tilde{\chi}_2^\pm}\right] +  
\left[a_3\rightarrow  a_1, b_{3}\rightarrow  b_{1}, m_{\tilde{\chi}_2^\pm}\rightarrow  m_{\tilde{\chi}_1^\pm}\right] \nonumber \\  
&+&  
\left[ m_{\tilde{\chi}_1^\pm}\rightarrow  m_{\tilde{\chi}_2^\pm}, m_{\tilde{\chi}_2^\pm}\rightarrow  m_{\tilde{\chi}_1^\pm}\right],  
\end{eqnarray}  
\begin{eqnarray}  
f_1^{(2, sfermions)}&=&-N_c \left[ {e^2 e_t^2\over 2 \pi^2}\left( 
\xi_{\tilde{t}3}^2 D_{00}^{16}+  
\xi_{\tilde{t}2}^2 D_{00}^{15}+  
\xi_{\tilde{t}1}^2 D_{00}^{14}+  
\xi_{\tilde{t}3}^2 D_{00}^{17}\right) \right]\nonumber \\  
&-& N_c\left[e_t \rightarrow e_b, \xi_{\tilde{t}1} \rightarrow \xi_{\tilde{b}1}, 
\xi_{\tilde{t}2}\rightarrow\xi_{\tilde{b}2},\xi_{\tilde{t}3}\rightarrow\xi_{\tilde{b}3}, 
m_{\tilde{t}_1}\rightarrow m_{\tilde{b}_1}, m_{\tilde{t}_2}\rightarrow m_{\tilde{b}_2} 
 \right]\nonumber \\ 
&-&\left[e_t \rightarrow e_{\tau}, \xi_{\tilde{t}1} \rightarrow \xi_{\tilde{\tau}1}, 
\xi_{\tilde{t}2}\rightarrow\xi_{\tilde{\tau}2},\xi_{\tilde{t}3}\rightarrow\xi_{\tilde{\tau}3}, 
m_{\tilde{t}_1}\rightarrow m_{\tilde{\tau}_1}, m_{\tilde{t}_2}\rightarrow m_{\tilde{\tau}_2} 
\right],  
\end{eqnarray}  
where $C_m^i, C_{mn}^i\equiv C_m, C_{mn}(m_{h0}^2, 0, \hat{t}, i)$,   
$C_0^i \equiv C_0(0,m_{h0}^2,\hat{t}, i)$ and 
$D_m^i, D_{mn}^i, D_{mnl}^i \equiv   D_m, D_{mn}, D_{mnl}(0, m_{h0}^2, 0, m_{h0}^2, \hat{u}, \hat{t}, i)$.
Here and below, the definition of $i$ is the same with the case of $f_1^{(1)}$,
and their explicit expressions are given by Eqs. \ref{mass1} and
Eqs. \ref{mass2}.
 
 The form factors $f_{1}^{(3)}$ and $f_{1}^{(4)}$ are given by  
\begin{eqnarray}  
&&f_{1}^{(3)}=f_{1}^{(4)} \nonumber \\  
&=&{e^2 g^2 m_{w} m_{z} B_0( 0, m_{w}^2, m_{w}^2)  \cos( \alpha - \beta) \over  
 32 \cos\theta_w \pi^2 (m_{H}^2 - \hat{s})}  
    \left( -\cos( 2 \alpha)  \cos( \alpha + \beta)  + 2 \sin( 2 \alpha)  \sin( \alpha + \beta) \right)   
  \nonumber \\  
&+&  
{3 e^2 g^2 m_{w} m_{z} B_0( 0, m_{w}^2, m_{w}^2)  \cos( 2 \alpha)  \sin( \alpha - \beta)    
    \sin( \alpha + \beta) \over 32 \cos\theta_w \pi^2 (-m_{h0}^2 + \hat{s})}.  
\end{eqnarray}  
  
The form factor $f_{1}^{(5)}$ is given by 
\begin{eqnarray}  
f_{1}^{(5)}=f_1^{(5, 2HDM)}+f_1^{(5, charginos)}+f_1^{(5, sfermions)},  
\end{eqnarray}  
with 
\begin{eqnarray}  
f_1^{(5, 2HDM)}=f_1^{(5, fermions)}+f_1^{(5, bosons)}. 
\end{eqnarray}  
Here $f_1^{(5, fermions)}$, $f_1^{(5, bosons)}$,  $f_1^{(5, charginos)}$ and  
$f_1^{(5, sfermions)}$ are 
\begin{eqnarray}  
&&f_1^{(5, fermions)}=\nonumber \\  
&&{(-\cos( 2 \alpha)  \cos( \alpha + \beta) +  
 2 \sin( 2 \alpha)  \sin( \alpha + \beta) ) (e^2 g^2 m_{z})\over   
48 \cos\theta_w m_{w} \pi^2 (m_{H}^2 - \hat{s})}\left( \right. \nonumber \\   
&&\left.  
4 m_{t}^2 csc( \beta)  \sin( \alpha) \left[ -2 B_0( 0, m_{t}^2, m_{t}^2)  -   
      \hat{s} C_0^9 - 2 \hat{s}C_2^9 + 8 C_{00}^9\right]    
+   m_b^2 sec( \beta)  \cos( \alpha) \left[m_t\rightarrow m_b\right] \right)  
\nonumber \\  
&+&  
 {e^2 g^2 m_{z} \cos( 2 \alpha)  \sin( \alpha + \beta)\over   
16 \cos\theta_w m_{w} \pi^2 (-m_{h0}^2 + \hat{s})}   
\left(  
4 m_{t}^2 \cos( \alpha)  csc( \beta)  
 \right. \nonumber \\  
&&\left.   
\left[ 2 B_0( 0, m_{t}^2, m_{t}^2)  +   
      \hat{s} C_0^9+   
      2 \hat{s} C_2^9 -   
      8 C_{00}^9\right]   
-  m_b^2 \sin(\alpha) sec(\beta) \left[  m_t\rightarrow m_b \right] \right),  
\end{eqnarray}  
\begin{eqnarray}  
&&f_1^{(5, bosons)}=\nonumber \\  
&&{(-\cos( 2 \alpha)  \cos( \alpha + \beta) +  
 2 \sin( 2 \alpha)  \sin( \alpha + \beta) ) (e^2 g^2 m_{z} m_{w})\over   
32 \cos\theta_w  \pi^2 (m_{H}^2 - \hat{s})}   
\left(   \cos( \alpha - \beta) ( 4 -   
      2  B_0( 0, m_{w}^2, m_{w}^2) \right. \nonumber \\  
& +& \left.  
     6 \hat{s} C_0^1 -   
     \hat{s} C_1^1 -   
      2 \hat{s} C_2^1 -   
      8  C_{00}^6 -   
      24 C_{00}^6 ) +   
 \cos( 2 \beta)  \cos( \alpha + \beta) (m_{w}^2 C_0^1- 4 C_{00}^1 +   
      4 C_{00}^6) \right)\nonumber \\  
&+&  
 {3 e^2 g^2 m_{z} m_{w}\cos( 2 \alpha)  \sin( \alpha + \beta)\over   
32 \cos\theta_w  \pi^2 (-m_{h0}^2 + \hat{s})}   
    \left(   
\sin( \alpha - \beta) (4 -   
      2  B_0( 0, m_{w}^2, m_{w}^2)  +   
     6  \hat{s} C_0^1 -   
      \hat{s} C_1^1 -   
      2 \hat{s} C_2^1    
     \right. \nonumber \\  
&-&8 C_{00}^6 - 24 C_{00}^1)    
+ \left.  
 \cos( 2 \beta)  \sin( \alpha + \beta) (m_{w}^2  C_0^1  
 +4 C_{00}^6 - 4 C_{00}^1)\right),  
\end{eqnarray}  
\begin{eqnarray}  
f_1^{(5, charginos)}&=&\left[ {e^2 \eta_{2}\over 4 \pi^2 (m_{H}^2 - \hat{s})}   
(2 a_5 m_{\tilde{\chi}_2^\pm} B_0(0, m_{\tilde{\chi}_2^\pm}^2, m_{\tilde{\chi}_2^\pm}^2) +   
       a_5 m_{\tilde{\chi}_2^\pm} \hat{s} C_0^{11}+  
       2 a_5 m_{\tilde{\chi}_2^\pm} \hat{s} C_2^{11}\right. \nonumber \\  
& -&\left.   
       8 a_5 m_{\tilde{\chi}_2^\pm} C_{00}^{11}) +(a_5\rightarrow  a_4, m_{\tilde{\chi}_2^\pm}\rightarrow  m_{\tilde{\chi}_1^\pm}) 
 \right]\nonumber \\  
&+&  
\left[ \eta_{2}\rightarrow  \eta_{1},a_5\rightarrow  a_2, a_4\rightarrow  a_1, m_{H}\rightarrow  m_{h0}\right],  
\end{eqnarray}  
\begin{eqnarray}  
f_1^{(5, sfermions)}&=&N_c \left[{2 e^2 e_t^2 \eta_1 \over 4 \pi^2 (m_{h0}^2 - \hat{s}) } \left( (\xi_{\tilde{t1}} C_{00}^{16}+ \xi_{\tilde{t2}}C_{00}^{17}\right)  
+  
{2  e_t^2 e^2 \eta_2 \over 4 \pi^2 (m_{H}^2 - \hat{s})}\left( \xi_{\tilde{t4}} C_{00}^{16}+\xi_{\tilde{t5}} C_{00}^{17} \right)\right] \nonumber \\  
&+&  
N_c\left[e_t \rightarrow e_b, \xi_{\tilde{t}1} \rightarrow \xi_{\tilde{b}1}, 
\xi_{\tilde{t}2}\rightarrow\xi_{\tilde{b}2},\xi_{\tilde{t}4}\rightarrow\xi_{\tilde{b}4},  
\xi_{\tilde{t}5}\rightarrow\xi_{\tilde{b}5}, 
m_{\tilde{t}_1}\rightarrow m_{\tilde{b}_1}, m_{\tilde{t}_2}\rightarrow m_{\tilde{b}_2} 
 \right]\nonumber \\ 
&+&\left[e_t \rightarrow e_{\tau}, \xi_{\tilde{t}1} \rightarrow \xi_{\tilde{\tau}1}, 
\xi_{\tilde{t}2}\rightarrow\xi_{\tilde{\tau}2},\xi_{\tilde{t}4}\rightarrow\xi_{\tilde{\tau}4}, 
\xi_{\tilde{t}5}\rightarrow\xi_{\tilde{\tau}5}, 
m_{\tilde{t}_1}\rightarrow m_{\tilde{\tau}_1}, m_{\tilde{t}_2}\rightarrow m_{\tilde{\tau}_2} 
\right],  
\end{eqnarray}  
where $C_m^i, C_{mn}^i\equiv    C_m, C_{mn}(0, 0, \hat{s}, i)$, and $\eta_1, \eta_2$ are coupling constants of  
the vertexes  
$h_0-h_0-h_0$ and 
$H-h_0-h_0$, respectively, which are given by 
\begin{eqnarray} 
\eta_1&=&-{3 i g m_z \over 2 \cos\theta_w} \cos(2\alpha) \sin(\alpha+\beta) \nonumber \\ 
\eta_2&=&-{ i g m_z \over 2 \cos\theta_w}(2 \sin(2 \alpha)\sin(\alpha+\beta)- 
\cos(2 \alpha) \cos (\alpha+\beta)), 
\end{eqnarray} 
and $a_4, a_5$ are the coupling constants of the vertex $H-\chi^{\pm}_{1,2}-\chi^{\mp}_{1,2}$, 
which are given by 
\begin{eqnarray} 
a_4&=&-{i g \over 2} ((Q^*_{11}+Q_{11}) \cos\alpha+ 
(S^*_{11}+S_{11}) \sin\alpha) \nonumber\\ 
a_5&=&-{i g \over 2} ((Q^*_{22}+Q_{22}) \cos\alpha+ 
(S^*_{22}+S_{22}) \sin\alpha). 
\end{eqnarray} 
And $\xi_{\tilde{t}i}, \xi_{\tilde{b}i}, \xi_{\tilde{\tau}i}, (i=4,5)$ are the coupling  
constants of the vertexes $H-\tilde{t}-\tilde{t}$,  $H-\tilde{b}-\tilde{b}$ and $H-\tilde{\tau}-\tilde{\tau}$, respectively, which are given by 
\begin{eqnarray}  
\left( \begin{array}{cc}  
\xi_{\tilde{t}4}& \xi_{\tilde{t}6}\\  
\xi_{\tilde{t}6}&\xi_{\tilde{t}5} 
\end{array}\right)=(-i g) R^{\tilde{t}} A_{\tilde{t}}  (R^{\tilde{t}})^T 
\end{eqnarray} 
with 
\begin{eqnarray} 
A_{\tilde{t}}=  
\left(\begin{array}{cc}  
 {m_{z} ((1/2)-(2/3) \sin^2\theta_w) \cos ( \alpha +\beta  )\over \cos\theta_w} +  
 {m_t^2 \sin ( \alpha )\over m_{w} \sin ( \beta  )}  
&{m_t\over 2 m_{w} \sin\beta} (A_t \sin\alpha-\mu \cos\alpha) \\  
 {m_t\over 2 m_{w} \sin\beta} (A_t \sin\alpha-\mu \cos\alpha)&  
 {m_{z} (2/3) \sin^2\theta_w \cos (\alpha+\beta)\over \cos\theta_w}+  
{m_t^2 \sin\alpha \over m_{w} \sin\beta}  
 \end{array} \right), 
\end{eqnarray}  
\begin{eqnarray}  
\left( \begin{array}{cc}  
\xi_{\tilde{b}4}& \xi_{\tilde{b}6}\\  
\xi_{\tilde{b}6}&\xi_{\tilde{b}5} 
\end{array}\right)=(i g) R^{\tilde{b}} A_{\tilde{b}}  (R^{\tilde{b}})^T 
\end{eqnarray} 
with 
\begin{eqnarray} 
A_{\tilde{b}}=  
\left(\begin{array}{cc}  
 {m_{z} ((1/2)-(1/3) \sin^2\theta_w) \cos ( \alpha +\beta  )\over \cos\theta_w} - 
 {m_b^2 \cos ( \alpha )\over m_{w} \cos ( \beta  )}  
&{m_b\over 2 m_{w} \cos\beta} (-A_b \cos\alpha+\mu \sin\alpha) \\  
{m_b\over 2 m_{w} \cos\beta} (-A_b \cos\alpha+\mu \sin\alpha) &  
 {m_{z} (1/3) \sin^2\theta_w \cos (\alpha+\beta)\over \cos\theta_w}-  
{m_b^2 \cos\alpha \over m_{w} \cos\beta}  
 \end{array} \right), 
\end{eqnarray}  
 and 
\begin{eqnarray}  
\left( \begin{array}{cc}  
\xi_{\tilde{\tau}4}& \xi_{\tilde{\tau}6}\\  
\xi_{\tilde{\tau}6}&\xi_{\tilde{\tau}5} 
\end{array}\right)=({i g m_z \cos(\alpha+\beta) \over \cos\theta_w})  
\left(\begin{array}{cc}  
 1/2-\sin^2\theta_w 
&0 \\  
0&  
 \sin^2\theta_w 
 \end{array} \right). 
\end{eqnarray}  
  
The form factor $ f_{1}^{(6)}$ is given by 
\begin{eqnarray}  
f_{1}^{(6)}={-e^2 g^2\over 32 \pi^2}\left( B_0(\hat{t}, m_{H^\pm}^2, m_{w}^2) \cos (\alpha - \beta)^2)+  
   B_0(\hat{t}, m_{w}^2, m_{w}^2) \sin (\alpha - \beta)^2\right).  
\end{eqnarray}  
  
The form factor $ f_{1}^{(7)}$ is given by 
\begin{eqnarray}  
f_{1}^{(7)}&=&f_1^{(7, 2HDM)}+f_1^{(7, sfermions)},  
\end{eqnarray}  
with 
\begin{eqnarray}  
f_1^{(7, 2HDM)}&=&e^2 g^2/(32 \cos\theta_w^2 \pi^2)   
(-4 \cos\theta_w^2    
+ 4 \cos\theta_w^2 B_0(0, m_{w}^2, m_{w}^2) \nonumber \\  
&+&   
   \cos\theta_w^2 \hat{s} C_1^1   
+   
   4 \cos\theta_w^2 \hat{s} C_2^1 +   
  4 C_{00}^6   
      (\cos\theta_w^2 \nonumber \\  
&+& \sin\theta_w^2 \cos (2 \alpha) \cos (2 \beta) - \cos\theta_w^2 \sin (2 \alpha) \sin (2 \beta))  
\nonumber \\     
&+& 4 C_{00}^1   
      (6 \cos\theta_w^2 - \sin\theta_w^2 \cos (2 \alpha) \cos (2 \beta) +   
        \cos\theta_w^2 \sin (2 \alpha) \sin (2 \beta)) \nonumber \\  
&+&   
   C_0^1   
      (\cos\theta_w^2 m_{w}^2 - 4 \cos\theta_w^2 \hat{s} + m_{w}^2 \sin\theta_w^2 \cos (2 \alpha) \cos (2 \beta) \nonumber \\  
&-&   
        \cos\theta_w^2 m_{w}^2 \sin (2 \alpha) \sin (2 \beta)) 
),  
\end{eqnarray}  
\begin{eqnarray}  
f_1^{(7, sfermions)}&=&N_c \left[{i  e^2 e_t^2\over 2 \pi^2}\left(\xi_{\tilde{t}2}^q C_{00}^{17}+  
   \xi_{\tilde{t}1}^q D_{00}^{16} \right)\right] \nonumber \\   
&+& N_c \left[e_t \rightarrow e_b, \xi_{\tilde{t}1}^q \rightarrow \xi_{\tilde{b}1}^q, 
\xi_{\tilde{t}2}^q \rightarrow \xi_{\tilde{b}2}^q, m_{\tilde{t}_1} \rightarrow m_{\tilde{b}_1}, 
 m_{\tilde{t}_2} \rightarrow m_{\tilde{b}_2} \right] \nonumber \\ 
&+& \left[e_t \rightarrow e_{\tau}, \xi_{\tilde{t}1}^q \rightarrow \xi_{\tilde{\tau}1}^q, 
\xi_{\tilde{t}2}^q \rightarrow \xi_{\tilde{\tau}2}^q, m_{\tilde{t}_1} \rightarrow m_{\tilde{\tau}_1}, 
 m_{\tilde{t}_2} \rightarrow m_{\tilde{\tau}_2} \right], 
\end{eqnarray}  
where $C_m^i, C_{mn}^i \equiv   C_m, C_{mn}(0, 0, \hat{s}, i)$ and $\xi_{\tilde{t}i}^q$, 
$\xi_{\tilde{b}i}^q$, $\xi_{\tilde{\tau}i}^q$ $(i=1,2)$ are the  coupling constants of quadratic vertexes $h_0-h_0-\tilde{t}-  
\tilde{t}$, $h_0-h_0-\tilde{b}-  
\tilde{b}$, and $h_0-h_0-\tilde{\tau}-  
\tilde{\tau}$, respectively,  which are given by 
\begin{eqnarray}  
\left( \begin{array}{cc}  
\xi_{\tilde{t}1}^q & \xi_{\tilde{t}3}^q\\  
\xi_{\tilde{t}3}^q & \xi_{\tilde{t}2}^q  
\end{array}\right)={i g^2\over 2} R^{\tilde{t}} A_{\tilde{t}}^q (R^{\tilde{t}})^T  
\end{eqnarray} 
with 
\begin{eqnarray} 
A_{\tilde{t}}^q= \left(\begin{array}{cc}  
{((1/2)-(2/3) \sin^2\theta_w)\cos(2 \alpha)\over \cos^2\theta_w}-{m_t^2 \cos^2\alpha\over m_{w}^2 \sin^2\beta}  
&0\\  
 0&  
 (2/3)\tan^2\theta_w \cos(2 \alpha)- {m_t^2\cos^2\alpha \over m_{w}^2 \sin^2\beta}  
 \end{array} \right), 
\end{eqnarray}  
\begin{eqnarray}  
\left( \begin{array}{cc}  
\xi_{\tilde{b}1}^q & \xi_{\tilde{b}3}^q\\  
\xi_{\tilde{b}3}^q & \xi_{\tilde{b}2}^q  
\end{array}\right)={i g^2\over 2} R^{\tilde{b}} A_{\tilde{b}}^q (R^{\tilde{b}})^T  
\end{eqnarray} 
with 
\begin{eqnarray} 
A_{\tilde{b}}^q= \left(\begin{array}{cc}  
{((-1/2)+(1/3) \sin^2\theta_w)\cos(2 \alpha)\over \cos^2\theta_w}-{m_b^2 \sin^2\alpha\over m_{w}^2 \cos^2\beta}  
&0\\  
 0&  
 (-1/3)\tan^2\theta_w \cos(2 \alpha)- {m_b^2\sin^2\alpha\over m_{w}^2 \cos^2\beta}  
 \end{array} \right), 
\end{eqnarray}  
and 
\begin{eqnarray}  
\left( \begin{array}{cc}  
\xi_{\tilde{\tau}1}^q & \xi_{\tilde{\tau}3}^q\\  
\xi_{\tilde{\tau}3}^q & \xi_{\tilde{\tau}2}^q  
\end{array}\right)={i g^2\over 2} \left(\begin{array}{cc}  
{((-1/2)+\sin^2\theta_w) \cos(2 \alpha )\over \cos^2\theta_w} 
&0\\  
 0&  
{-\sin^2\theta_w \cos(2 \alpha)\over \cos^2\theta_w} 
 \end{array} \right). 
\end{eqnarray}

The form factor $f_{1}^{(8)}$ is given by 
\begin{eqnarray}  
f_{1}^{(8)}=f_1^{(8, 2HDM)}+f_1^{(8, sfermions)},  
\end{eqnarray}  
with 
\begin{eqnarray}  
f_1^{(8,2HDM)}&=&{e^2 g^2 \over 32 \cos\theta_w^2 \pi^2}   
\left( 4 \cos\theta_w^2 + B_0(\hat{s}, m_{w}^2, m_{w}^2)   
      (-7 \cos\theta_w^2 + \sin\theta_w^2 \cos (2 \alpha) \cos (2 \beta) \right. \nonumber \\  
&-&   
        \cos\theta_w^2 \sin (2 \alpha) \sin (2 \beta)) +   
   B_0(\hat{s}, m_{H^\pm}^2, m_{H^\pm}^2) (-\cos\theta_w^2 \nonumber \\  
&-& \left. \sin\theta_w^2 \cos (2 \alpha) \cos (2 \beta) +   
        \cos\theta_w^2 \sin (2 \alpha) \sin (2 \beta))\right),  
\end{eqnarray}  
\begin{eqnarray}  
f_1^{(8, sfermions)}&=&  
-N_c \left[{i  e^2 e_t^2\over 8 \pi^2}\left(\xi_{\tilde{t}2}^q B_0( \hat{s}, m_{\tilde{t}_2}^2, m_{\tilde{t}_2}^2) +  
   \xi_{\tilde{t}1}^q B_0( \hat{s}, m_{\tilde{t}_1}^2, m_{\tilde{t}_1}^2)  \right) \right] \nonumber \\  
&-& N_c \left[e_t \rightarrow e_b, \xi_{\tilde{t}1}^q \rightarrow \xi_{\tilde{b}1}^q, 
\xi_{\tilde{t}2}^q \rightarrow \xi_{\tilde{b}2}^q, m_{\tilde{t}_1} \rightarrow m_{\tilde{b}_1}, 
 m_{\tilde{t}_2} \rightarrow m_{\tilde{b}_2} \right] \nonumber \\ 
&-& \left[e_t \rightarrow e_{\tau}, \xi_{\tilde{t}1}^q \rightarrow \xi_{\tilde{\tau}1}^q, 
\xi_{\tilde{t}2}^q \rightarrow \xi_{\tilde{\tau}2}^q, m_{\tilde{t}_1} \rightarrow m_{\tilde{\tau}_1}, 
 m_{\tilde{t}_2} \rightarrow m_{\tilde{\tau}_2} \right]. 
\end{eqnarray}

The form factor $f_{1}^{(9)}$ is given by 
\begin{eqnarray}  
f_{1}^{(9)}=f_1^{(9, 2HDM)}+f_1^{(9, sfermions)},  
\end{eqnarray}  
with 
\begin{eqnarray}  
f_1^{(9, 2HDM)}&=& {e^2 g^2 m_{z} m_{w} \over 32 \cos\theta_w \pi^2 (-m_{H}^2 + \hat{s})}  
\left( 4  \cos( \alpha - \beta)  -   
      2 B_0( \hat{s}, m_{H^\pm}^2, m_{H^\pm}^2)  \cos( \alpha - \beta) \right. \nonumber \\   
&-&   
      6 B_0( \hat{s}, m_{w}^2, m_{w}^2)  \cos( \alpha - \beta)   
 +  
B_0( \hat{s}, m_{H^\pm}^2, m_{H^\pm}^2)  \cos( 2 \beta)  \cos( \alpha + \beta) \nonumber \\  
 &-& \left.   
    B_0( \hat{s}, m_{w}^2, m_{w}^2)  \cos( 2 \beta)  \cos( \alpha + \beta) )   
    (-(\cos( 2 \alpha)  \cos( \alpha + \beta) ) + 2 \sin( 2 \alpha)  \sin( \alpha + \beta)\right)   
\nonumber \\  
  &+&  
{3 e^2  g^2 m_{z} m_{w} \cos( 2 \alpha)  \sin( \alpha + \beta) \over  32 \cos\theta_w \pi^2 (m_{h0}^2 - \hat{s})}  
    \left( 4 \sin( \alpha - \beta)  -   
      2 B_0( \hat{s}, m_{H^\pm}^2, m_{H^\pm}^2)  \sin( \alpha - \beta) \right. \nonumber \\   
&-&   
      6 B_0( \hat{s}, m_{w}^2, m_{w}^2)  \sin( \alpha - \beta) +  
B_0( \hat{s}, m_{H^\pm}^2, m_{H^\pm}^2)  \cos( 2 \beta)  \sin( \alpha + \beta) \nonumber \\  
 & - & \left.    
     B_0( \hat{s}, m_{w}^2, m_{w}^2)  \cos( 2 \beta)  \sin( \alpha + \beta) ) \right),  
\end{eqnarray}  
\begin{eqnarray}  
f_1^{(9, sfermions)}&=&-N_c \left[{2 \eta_2 e^2 e_t^2 \over 8 \pi^2 (m_{H}^2 - \hat{s})} \left( \xi_{\tilde{t}5} B_0( \hat{s}, m_{\tilde{t}_2}^2, m_{\tilde{t}_2}^2) + 
\xi_{\tilde{t}4} B_0( \hat{s}, m_{\tilde{t}_1}^2, m_{\tilde{t}_1}^2) \right) \right.\nonumber \\  
&+&\left.  
{2 \eta_1  e^2 e_t^2 \over 8 \pi^2 (m_{h0}^2 - \hat{s})}  \left(  
\xi_{\tilde{t}2} B_0( \hat{s}, m_{\tilde{t}_2}^2, m_{\tilde{t}_2}^2) + 
\xi_{\tilde{t}1} B_0( \hat{s}, m_{\tilde{t}_1}^2, m_{\tilde{t}_1}^2)  \right) \right]\nonumber \\  
&-&  
N_c\left[e_t \rightarrow e_b, \xi_{\tilde{t}1} \rightarrow \xi_{\tilde{b}1}, 
\xi_{\tilde{t}2}\rightarrow\xi_{\tilde{b}2},\xi_{\tilde{t}4}\rightarrow\xi_{\tilde{b}4},  
\xi_{\tilde{t}5}\rightarrow\xi_{\tilde{b}5}, 
m_{\tilde{t}_1}\rightarrow m_{\tilde{b}_1}, m_{\tilde{t}_2}\rightarrow m_{\tilde{b}_2} 
 \right]\nonumber \\ 
&-&\left[e_t \rightarrow e_{\tau}, \xi_{\tilde{t}1} \rightarrow \xi_{\tilde{\tau}1}, 
\xi_{\tilde{t}2}\rightarrow\xi_{\tilde{\tau}2},\xi_{\tilde{t}4}\rightarrow\xi_{\tilde{\tau}4}, 
\xi_{\tilde{t}5}\rightarrow\xi_{\tilde{\tau}5}, 
m_{\tilde{t}_1}\rightarrow m_{\tilde{\tau}_1}, m_{\tilde{t}_2}\rightarrow m_{\tilde{\tau}_2} 
\right]. 
\end{eqnarray}  
  
The form factor $f_{1}^{(10)}$ is given by 
\begin{eqnarray}  
f_{1}^{(10)}&=&e^2 g^2 \cos (\alpha - \beta)^2/(32 \pi^2)   
      ( (m_{h0}^2 - \hat{t}) (C_1^2 +   
     C_2^2) +   
  C_{00}^2  
+ 2 C_{00}^3) \nonumber \\  
 &+&   
 e^2 g^2 \sin (\alpha - \beta)^2)/(32 \pi^2)   
((m_{h0}^2 - \hat{t}) (C_1^1  
 +   
  C_2^1) +   
   3 C_{00}^1) \nonumber \\  
 &+&e^2 g^2 /(32 \cos\theta_w^2 \pi^2) (  
 -\cos\theta_w^2 B_0(0, m_{w}^2, m_{w}^2)  \nonumber \\  
&+&   
   C_0^1 \sin (\alpha - \beta)   
      (-3 \cos\theta_w m_{w}^2 \sin (\alpha - \beta) \nonumber \\  
&+& \cos\theta_w \hat{t} \sin (\alpha - \beta) -   
        m_{w} m_{z} \cos (2 \beta) \sin (\alpha + \beta)) \nonumber \\  
&+&   
   \cos (\alpha - \beta) C_0^3   
      (-(\cos\theta_w m_{H^\pm}^2 \cos (\alpha - \beta)) \nonumber \\  
&+& \cos\theta_w m_{w}^2 \cos (\alpha - \beta) +   
        \cos\theta_w \hat{t} \cos (\alpha - \beta) \nonumber \\  
&-& m_{w} m_{z} \sin (2 \beta) \sin (\alpha + \beta))  
), 
\end{eqnarray}  
where $C_m^i, C_{mn}^i \equiv   C_m, C_{mn}(m_{h0}^2, 0, \hat{t}, i)$ and $C_0^i \equiv   C_0(0,m_{h0}^2,\hat{t}, i)$.  
  
The form factor $f_{1}^{(11)}$ is given by 
\begin{eqnarray}  
f_{1}^{(11)}&=&e^2 g^2 \cos (\alpha - \beta)^2/(32 \pi^2)   
( (m_{h0}^2 - \hat{t})    
     (C_1^{2} +   
     C_2^{2})+   
  C_{00}^{2}  
   + 2 C_{00}^{3} )  
\nonumber \\  
&+&   
 e^2 g^2 \sin (\alpha - \beta)^2)/(32 \pi^2)   
 ((m_{h0}^2 - \hat{t}) (C_1^1  
 +   
   C_2^1) +   
   3 C_{00}^1 )\nonumber\\  
&+& e^2 g^2/(32 \cos\theta_w \pi^2)  
(-\cos\theta_w^2 B_0(0, m_{w}^2, m_{w}^2)  
       \nonumber \\  
	   &+&   
   C_0^1 \sin (\alpha - \beta)   
      (-3 \cos\theta_w m_{w}^2 \sin (\alpha - \beta)\nonumber \\  
	  &+& \cos\theta_w \hat{t} \sin (\alpha - \beta) -   
        m_{w} m_{z} \cos (2 \beta) \sin (\alpha + \beta))\nonumber \\  
&+&   
   \cos (\alpha - \beta) C_0^4   
      (-(\cos\theta_w m_{H^\pm}^2 \cos (\alpha - \beta)) \nonumber\\  
	  &+& \cos\theta_w m_{w}^2 \cos (\alpha - \beta) +   
        \cos\theta_w \hat{t} \cos (\alpha - \beta) \nonumber \\  
&-& m_{w} m_{z} \sin (2 \beta) \sin (\alpha + \beta))),  
\end{eqnarray}  
where $ C_m^i, C_{mn}^i \equiv   C_m, C_{mn}(m_{h0}^2, 0, \hat{t}, i)$ and $C_0^i \equiv   C_0(0,m_{h0}^2,\hat{t},i)$.

 The form factor $f_{1}^{(12)}$ is given by 
\begin{eqnarray}  
f_{1}^{(12)}&=&f_{1}^{(12, 2HDM)}+f_{1}^{(12, sfermions)},  
\end{eqnarray}  
with 
\begin{eqnarray}  
f_{1}^{(12, 2HDM)}&=&e^2 g^2 \cos (\alpha - \beta)^2/(32 \pi^2)   
(4 B_0(m_{h0}^2, m_{H^\pm}^2, m_{w}^2) -   
   2 (\hat{t} + \hat{u})   
      C_1^7 \nonumber \\  
&+&   
 (8 m_{h0}^2 - \hat{s})   
    C_1^8+   
   2 \hat{s} C_2^7 +   
   2 \hat{s} C_2^8  
   - 2 C_{00}^8) \nonumber \\   
&+&   
 e^2 g^2 \sin (\alpha - \beta)^2)/(32 \pi^2)   
 (4 B_0(m_{h0}^2, m_{w}^2, m_{w}^2) +   
   (4 m_{h0}^2 +\hat{s}) C_1^1+   
   4 \hat{s} C_2^1-  
2 C_{00}^1)\nonumber \\  
 &+&  
 e^2 g^2/(16 \cos\theta_w^2 \pi^2)   
(- C_0^6   
      (-2 \cos\theta_w m_{w} \sin (\alpha - \beta) \nonumber \\  
 &+& m_{z} \cos (2 \beta) \sin (\alpha + \beta))^2)  
     + C_0^1   
      (5 \cos\theta_w^2 m_{h0}^2 \sin (\alpha - \beta)^2 \nonumber \\  
&-&   
        10 \cos\theta_w^2 m_{w}^2 \sin (\alpha - \beta)^2 - 2 \cos\theta_w^2 \hat{s} \sin (\alpha - \beta)^2 \nonumber \\  
&-&   
        m_{z}^2 \cos (2 \beta)^2 \sin (\alpha + \beta)^2) +   
   C_0^8   
      (4 \cos\theta_w^2 m_{h0}^2 \cos (\alpha - \beta)^2 \nonumber \\  
&-& \cos\theta_w^2 \hat{s} \cos (\alpha - \beta)^2 +   
        2 \cos\theta_w m_{w} m_{z} \cos (\alpha - \beta) \sin (2 \beta) \sin (\alpha + \beta) \nonumber \\  
&-&   
        m_{z}^2 \sin (2 \beta)^2 \sin (\alpha + \beta)^2) +   
   C_0^7   
      (\cos\theta_w^2 m_{H^\pm}^2 \cos (\alpha - \beta)^2 \nonumber \\  
&+& \cos\theta_w^2 m_{h0}^2 \cos (\alpha - \beta)^2 -   
        \cos\theta_w^2 m_{w}^2 \cos (\alpha - \beta)^2 \nonumber \\  
&-& \cos\theta_w^2 \hat{s} \cos (\alpha - \beta)^2 +   
        2 \cos\theta_w m_{w} m_{z} \cos (\alpha - \beta) \sin (2 \beta) \sin (\alpha + \beta)\nonumber \\  
& - &  
        m_{z}^2 \sin (2 \beta)^2 \sin (\alpha + \beta)^2)  
),  
\end{eqnarray}  
\begin{eqnarray}  
f_{1}^{(12, sfermions)}&=&N_c \left[{ e^2 e_t^2 \over 4 \pi^2} \left(\xi_{\tilde{t}3}^2 C_0^{18}+  
\xi_{\tilde{t}2}^2 C_0^{17}+  
\xi_{\tilde{t}1}^2 C_0^{16}+  
\xi_{\tilde{t}3}^2 C_0^{19}\right)\right] \nonumber \\  
&+& N_c\left[e_t \rightarrow e_b, \xi_{\tilde{t}1} \rightarrow \xi_{\tilde{b}1}, 
\xi_{\tilde{t}2}\rightarrow\xi_{\tilde{b}2},\xi_{\tilde{t}3}\rightarrow\xi_{\tilde{b}3}, 
m_{\tilde{t}_1}\rightarrow m_{\tilde{b}_1}, m_{\tilde{t}_2}\rightarrow m_{\tilde{b}_2} 
 \right]\nonumber \\ 
&+&\left[e_t \rightarrow e_{\tau}, \xi_{\tilde{t}1} \rightarrow \xi_{\tilde{\tau}1}, 
\xi_{\tilde{t}2}\rightarrow\xi_{\tilde{\tau}2},\xi_{\tilde{t}3}\rightarrow\xi_{\tilde{\tau}3}, 
m_{\tilde{t}_1}\rightarrow m_{\tilde{\tau}_1}, m_{\tilde{t}_2}\rightarrow m_{\tilde{\tau}_2} 
\right],  
\end{eqnarray}  
where $C_m^i, C_{mn}^i \equiv   C_m, C_{mn}(m_{h0}^2, m_{h0}^2, \hat{s}, i)$.

\section*{Appendix B}  
The form factor $f_2^{(1)}$ is given by  
\begin{eqnarray}  
f_{2}^{(1)}&=&f_2^{(1, 2HDM)}+f_{2}^{(1, charginos)}+f_2^{(1,sfermions)},  
\end{eqnarray}  
with 
\begin{eqnarray}  
f_2^{(1, 2HDM)}&=&f_2^{(1, fermions)}+f_2^{(1,bosons)}. 
\end{eqnarray}  
 
Here $f_2^{(1, fermions)}$, $f_2^{(1,bosons)}$,  $f_{2}^{(1, charginos)}$ and $f_2^{(1,sfermions)}$ are 
\begin{eqnarray}  
&&f_2^{(1, fermions)}=\nonumber \\  
&&{e^2 g^2 m_t^2 \cos (\alpha)^2 csc(\beta)^2\over 6 m_{w}^2 \pi^2}   
    \left[ -8 m_t^2 D_0^7 -   
   4  (m_{h0}^2 + 8 m_t^2)   
      D_1^7  
     - (16 m_t^2 + \hat{s} + 2 \hat{t})   
      D_2^7\right. \nonumber \\  
 &- & \left. 4 D_{00}^7  
     - 4 (m_{h0}^2 + 4 m_t^2) (D_{11}^7+ D_{13}^7)  
    + 2 (-4 m_{h0}^2 - 16 m_t^2 - \hat{t} + \hat{u})   
      D_{12}^7   
      - (8 m_t^2 + \hat{s} + 2 \hat{t})   
      D_{22}^7 \right] \nonumber \\  
&+&  
{e^2 g^2 m_b^2 \sin (\alpha)^2 sec(\beta)^2\over 24 m_{w}^2 \pi^2}\left[m_t\rightarrow m_b\right] ,  
\end{eqnarray}  
\begin{eqnarray}  
&&f_2^{(1, bosons)}=\nonumber \\  
&&{e^2 g^2 \cos (\alpha - \beta)^2\over 16 \pi^2}   
      \left( 10 D_{00}^4- 16 D_{00}^5+ 12 D_{001}^4 - 32 D_{001}^5 +   
  6 D_{002}^4 - 16 D_{002}^5 \right. \nonumber \\  
&+&   
   (3 m_{h0}^2 + 2 \hat{t} + \hat{u}) ( D_{111}^4+ 3  D_{113}^4)  
+ 4 (-4 m_{h0}^2 + \hat{s}) (D_{111}^5 + 3  D_{113}^5)  
  + 2 (4 m_{h0}^2 + 4 \hat{t} + \hat{u}) (D_{112}^4+D_{123}^4)\nonumber \\  
    &-& 8 (3 m_{h0}^2 + 2 \hat{t} + \hat{u}) ( D_{112}^5 + D_{123}^5)   
    +   
 (7 m_{h0}^2 + 10 \hat{t} + \hat{u})  D_{122}^4  
     - 4 (6 m_{h0}^2 + 5 \hat{t} \nonumber\\  
	 &+& \hat{u})  D_{122}^5   
+\left.   
 (m_{h0}^2 + 2 \hat{t})  D_{222}^4  
    - 4 (m_{h0}^2 + \hat{t})  D_{222}^5\right) +  
 {e^2 g^2 \sin (\alpha - \beta)^2\over 16 \pi^2}  
\left( -6 D_{00}^1  
 -   
   20 D_{001}^1-   
   10 D_{002}^1 \right. \nonumber \\  
&+&   
   (-13 m_{h0}^2 + 4 \hat{s} + 2 \hat{t} + \hat{u}) ( D_{111}^1+ 3 D_{113}^1) -   
  2 (8 m_{h0}^2 + 4 \hat{t} + 3 \hat{u}) (D_{112}^1 +D_{123}^1) \nonumber \\  
&-&   
   (17 m_{h0}^2 + 10 \hat{t} + 3 \hat{u}) D_{122}^1   
-  
   (3 m_{h0}^2   
+\left. 2 \hat{t}) D_{222}^1 \right)   
 -   
   {e^2 g^2 C_0^1\over 8 \pi^2} +   
   {e^2 g^2\over 8 \cos\theta_w^2 \pi^2} \left(-2 \cos\theta_w m_{w} \sin (\alpha - \beta)   
\right. \nonumber \\  
&+& \left.  m_{z} \cos (2 \beta) \sin (\alpha + \beta) \right)^2  
(D_0^6  
+ 4 D_1^6 +   
2 D_2^6 +  
2 D_{11}^6   
+ 4 D_{12}^6 +   
2 D_{13}^6  
 + D_{22}^6)\nonumber \\  
 &+&{e^2 g^2\over 32 \cos\theta_w^2 \pi^2}   
 \left( 2 D_2^1   
      (-5 \cos\theta_w^2 m_{h0}^2 \sin (\alpha - \beta)^2 +   
        48 \cos\theta_w^2 m_{w}^2 \sin (\alpha - \beta)^2\right. \nonumber \\  
& - &11 \cos\theta_w^2 \hat{t} \sin (\alpha - \beta)^2 -   
        8 \cos\theta_w^2 \hat{u} \sin (\alpha - \beta)^2 +   
        4 m_{z}^2 \cos (2 \beta)^2 \sin (\alpha + \beta)^2) \nonumber \\  
&+&   
   D_0^1   
      (48 \cos\theta_w^2 m_{w}^2 \sin (\alpha - \beta)^2 - 7 \cos\theta_w^2 \hat{t} \sin (\alpha - \beta)^2 -   
        8 \cos\theta_w^2 \hat{u} \sin (\alpha - \beta)^2 \nonumber \\  
&-&   
        6 \cos\theta_w m_{w} m_{z} \cos (2 \beta) \sin (\alpha - \beta) \sin (\alpha + \beta) +   
        4 m_{z}^2 \cos (2 \beta)^2 \sin (\alpha + \beta)^2) \nonumber \\  
&+&   
  4 D_{12}^1   
      (-(\cos\theta_w^2 m_{h0}^2 \sin (\alpha - \beta)^2) +   
        48 \cos\theta_w^2 m_{w}^2 \sin (\alpha - \beta)^2 - 21 \cos\theta_w^2 \hat{t} \sin (\alpha - \beta)^2 \nonumber \\  
&-&   
        17 \cos\theta_w^2 \hat{u} \sin (\alpha - \beta)^2 +   
        6 \cos\theta_w m_{w} m_{z} \cos (2 \beta) \sin (\alpha - \beta) \sin (\alpha + \beta) +   
        4 m_{z}^2 \cos (2 \beta)^2 \sin (\alpha + \beta)^2) \nonumber \\  
&+&   
  D_{22}^1   
      (-16 \cos\theta_w^2 m_{h0}^2 \sin (\alpha - \beta)^2 +   
        48 \cos\theta_w^2 m_{w}^2 \sin (\alpha - \beta)^2 + 8 \cos\theta_w^2 \hat{s} \sin (\alpha - \beta)^2 \nonumber \\  
&-&   
        15 \cos\theta_w^2 \hat{t} \sin (\alpha - \beta)^2 - 8 \cos\theta_w^2 \hat{u} \sin (\alpha - \beta)^2 +   
        6 \cos\theta_w m_{w} m_{z} \cos (2 \beta) \sin (\alpha - \beta) \sin (\alpha + \beta) \nonumber \\  
&+&   
        4 m_{z}^2 \cos (2 \beta)^2 \sin (\alpha + \beta)^2) +   
  2 (D_{11}^1+ D_{13}^1)   
      (-34 \cos\theta_w^2 m_{h0}^2 \sin (\alpha - \beta)^2 \nonumber \\  
&+&   
        48 \cos\theta_w^2 m_{w}^2 \sin (\alpha - \beta)^2 + 16 \cos\theta_w^2 \hat{s} \sin (\alpha - \beta)^2 -   
        3 \cos\theta_w^2 \hat{t} \sin (\alpha - \beta)^2 - 2 \cos\theta_w^2 \hat{u} \sin (\alpha - \beta)^2 \nonumber \\  
&+&   
        6 \cos\theta_w m_{w} m_{z} \cos (2 \beta) \sin (\alpha - \beta) \sin (\alpha + \beta) +   
        4 m_{z}^2 \cos (2 \beta)^2 \sin (\alpha + \beta)^2) \nonumber \\  
&+&   
  2 D_1^1   
      (-49 \cos\theta_w^2 m_{h0}^2 \sin (\alpha - \beta)^2 +   
        96 \cos\theta_w^2 m_{w}^2 \sin (\alpha - \beta)^2 + 20 \cos\theta_w^2 \hat{s} \sin (\alpha - \beta)^2 \nonumber \\  
&+&   
        \cos\theta_w^2 \hat{u} \sin (\alpha - \beta)^2 +   
        8 m_{z}^2 \cos (2 \beta)^2 \sin (\alpha + \beta)^2) \nonumber \\  
&+&   
   8 (D_{11}^5+ D_{13}^5)   
      (-12 \cos\theta_w^2 m_{h0}^2 \cos (\alpha - \beta)^2 + 4 \cos\theta_w^2 \hat{s} \cos (\alpha - \beta)^2 \nonumber \\  
&-&   
        2 \cos\theta_w m_{w} m_{z} \cos (\alpha - \beta) \sin (2 \beta) \sin (\alpha + \beta) +   
        m_{z}^2 \sin (2 \beta)^2 \sin (\alpha + \beta)^2) \nonumber \\  
&+&   
   16 D_{12}^5   
      (-4 \cos\theta_w^2 m_{h0}^2 \cos (\alpha - \beta)^2 - 5 \cos\theta_w^2 \hat{t} \cos (\alpha - \beta)^2 -   
        3 \cos\theta_w^2 \hat{u} \cos (\alpha - \beta)^2 \nonumber \\  
&-&   
        2 \cos\theta_w m_{w} m_{z} \cos (\alpha - \beta) \sin (2 \beta) \sin (\alpha + \beta) +   
        m_{z}^2 \sin (2 \beta)^2 \sin (\alpha + \beta)^2) \nonumber \\  
&+&   
  4 D_{22}^5   
      (-4 \cos\theta_w^2 m_{h0}^2 \cos (\alpha - \beta)^2 - 6 \cos\theta_w^2 \hat{t} \cos (\alpha - \beta)^2 -   
        2 \cos\theta_w^2 \hat{u} \cos (\alpha - \beta)^2 \nonumber \\  
&-&   
        2 \cos\theta_w m_{w} m_{z} \cos (\alpha - \beta) \sin (2 \beta) \sin (\alpha + \beta) +   
        m_{z}^2 \sin (2 \beta)^2 \sin (\alpha + \beta)^2) \nonumber \\  
&+&   
   8 D_2^5   
      (-(\cos\theta_w^2 m_{h0}^2 \cos (\alpha - \beta)^2) - 3 \cos\theta_w^2 \hat{t} \cos (\alpha - \beta)^2 -   
        2 \cos\theta_w^2 \hat{u} \cos (\alpha - \beta)^2 \nonumber \\  
&-&   
        2 \cos\theta_w m_{w} m_{z} \cos (\alpha - \beta) \sin (2 \beta) \sin (\alpha + \beta) +   
        m_{z}^2 \sin (2 \beta)^2 \sin (\alpha + \beta)^2) \nonumber \\  
&+&   
   4 D_0^5   
      (-2 \cos\theta_w^2 \hat{t} \cos (\alpha - \beta)^2 - 2 \cos\theta_w^2 \hat{u} \cos (\alpha - \beta)^2 -   
        2 \cos\theta_w m_{w} m_{z} \cos (\alpha - \beta) \sin (2 \beta) \sin (\alpha + \beta) \nonumber \\  
&+&   
        m_{z}^2 \sin (2 \beta)^2 \sin (\alpha + \beta)^2) +   
8 D_1^5   
      (-12 \cos\theta_w^2 m_{h0}^2 \cos (\alpha - \beta)^2 \nonumber \\  
&+& 5 \cos\theta_w^2 \hat{s} \cos (\alpha - \beta)^2 -   
        4 \cos\theta_w m_{w} m_{z} \cos (\alpha - \beta) \sin (2 \beta) \sin (\alpha + \beta) +   
        2 m_{z}^2 \sin (2 \beta)^2 \sin (\alpha + \beta)^2) \nonumber \\  
&+&   
   D_0^4   
      (-11 \cos\theta_w^2 m_{H^\pm}^2 \cos (\alpha - \beta)^2 +   
        11 \cos\theta_w^2 m_{w}^2 \cos (\alpha - \beta)^2 + \cos\theta_w^2 \hat{t} \cos (\alpha - \beta)^2 \nonumber \\  
&-&   
        14 \cos\theta_w m_{w} m_{z} \cos (\alpha - \beta) \sin (2 \beta) \sin (\alpha + \beta) +   
        4 m_{z}^2 \sin (2 \beta)^2 \sin (\alpha + \beta)^2) \nonumber \\  
&+&   
   2 D_2^4   
      (-3 \cos\theta_w^2 m_{H^\pm}^2 \cos (\alpha - \beta)^2 -   
        \cos\theta_w^2 m_{h0}^2 \cos (\alpha - \beta)^2 + 3 \cos\theta_w^2 m_{w}^2 \cos (\alpha - \beta)^2 \nonumber \\  
&+&   
        \cos\theta_w^2 \hat{t} \cos (\alpha - \beta)^2 -   
       8 \cos\theta_w m_{w} m_{z} \cos (\alpha - \beta) \sin (2 \beta) \sin (\alpha + \beta) \nonumber \\  
&+&   
        4 m_{z}^2 \sin (2 \beta)^2 \sin (\alpha + \beta)^2) +   
  D_{22}^4   
      (\cos\theta_w^2 m_{H^\pm}^2 \cos (\alpha - \beta)^2 - \cos\theta_w^2 m_{w}^2 \cos (\alpha - \beta)^2 \nonumber \\  
&+&   
        8 \cos\theta_w^2 \hat{s} \cos (\alpha - \beta)^2 + 9 \cos\theta_w^2 \hat{t} \cos (\alpha - \beta)^2 -   
        2 \cos\theta_w m_{w} m_{z} \cos (\alpha - \beta) \sin (2 \beta) \sin (\alpha + \beta) \nonumber \\  
&+&   
        4 m_{z}^2 \sin (2 \beta)^2 \sin (\alpha + \beta)^2) +   
   4 D_{12}^4   
      (\cos\theta_w^2 m_{H^\pm}^2 \cos (\alpha - \beta)^2 \nonumber \\  
&+&   
        15 \cos\theta_w^2 m_{h0}^2 \cos (\alpha - \beta)^2 - \cos\theta_w^2 m_{w}^2 \cos (\alpha - \beta)^2 -   
        \cos\theta_w^2 \hat{t} \cos (\alpha - \beta)^2\nonumber \\  
& - &5 \cos\theta_w^2 \hat{u} \cos (\alpha - \beta)^2 -   
        2 \cos\theta_w m_{w} m_{z} \cos (\alpha - \beta) \sin (2 \beta) \sin (\alpha + \beta) +   
        4 m_{z}^2 \sin (2 \beta)^2 \sin (\alpha + \beta)^2) \nonumber \\  
&+&   
  2 (D_{11}^4+ D_{13}^4)   
      (\cos\theta_w^2 m_{H^\pm}^2 \cos (\alpha - \beta)^2 +   
        14 \cos\theta_w^2 m_{h0}^2 \cos (\alpha - \beta)^2 - \cos\theta_w^2 m_{w}^2 \cos (\alpha - \beta)^2 \nonumber \\  
&-&   
        3 \cos\theta_w^2 \hat{t} \cos (\alpha - \beta)^2 - 2 \cos\theta_w^2 \hat{u} \cos (\alpha - \beta)^2 -   
        2 \cos\theta_w m_{w} m_{z} \cos (\alpha - \beta) \sin (2 \beta) \sin (\alpha + \beta)\nonumber\\  
		&+&   
        4 m_{z}^2 \sin (2 \beta)^2 \sin (\alpha + \beta)^2)   
+   
  2 D_1^4   
      (-6 \cos\theta_w^2 m_{H^\pm}^2 \cos (\alpha - \beta)^2\nonumber \\  
& -& \cos\theta_w^2 m_{h0}^2 \cos (\alpha - \beta)^2 +   
        6 \cos\theta_w^2 m_{w}^2 \cos (\alpha - \beta)^2  
 + \cos\theta_w^2 \hat{u} \cos (\alpha - \beta)^2 \nonumber \\  
&- &\left.  
        16 \cos\theta_w m_{w} m_{z} \cos (\alpha - \beta) \sin (2 \beta) \sin (\alpha + \beta) +   
        8 m_{z}^2 \sin (2 \beta)^2 \sin (\alpha + \beta)^2)  
\right),  
\end{eqnarray}  
\begin{eqnarray}  
&&f_2^{(1, charginos)}=\nonumber \\  
&&\left[{e^2\over 2 \pi^2} \left(4 m_{\tilde{\chi}_1^\pm} (a_3^2 m_{\tilde{\chi}_1^\pm} - b_{3}^2 m_{\tilde{\chi}_1^\pm} + a_3^2 m_{\tilde{\chi}_2^\pm} + b_{3}^2 m_{\tilde{\chi}_2^\pm})   
     D_0^{12} \right.\right. \nonumber \\  
&+&   
    4 (a_3^2 m_{h0}^2 - b_{3}^2 m_{h0}^2 + 3 a_3^2 m_{\tilde{\chi}_1^\pm}^2 - 3 b_{3}^2 m_{\tilde{\chi}_1^\pm}^2 +   
       4 a_3^2 m_{\tilde{\chi}_1^\pm} m_{\tilde{\chi}_2^\pm} + 4 b_{3}^2 m_{\tilde{\chi}_1^\pm} m_{\tilde{\chi}_2^\pm} + a_3^2 m_{\tilde{\chi}_2^\pm}^2 - b_{3}^2 m_{\tilde{\chi}_2^\pm}^2)   
     D_1^{12} \nonumber \\  
&+&   
    (6 a_3^2 m_{\tilde{\chi}_1^\pm}^2 - 6 b_{3}^2 m_{\tilde{\chi}_1^\pm}^2 + 8 a_3^2 m_{\tilde{\chi}_1^\pm} m_{\tilde{\chi}_2^\pm} + 8 b_{3}^2 m_{\tilde{\chi}_1^\pm} m_{\tilde{\chi}_2^\pm} +   
       2 a_3^2 m_{\tilde{\chi}_2^\pm}^2 - 2 b_{3}^2 m_{\tilde{\chi}_2^\pm}^2 + a_3^2 \hat{s} - b_{3}^2 \hat{s} \nonumber \\  
	   &+& 2 a_3^2 \hat{t} - 2 b_{3}^2 \hat{t})   
     D_2^{12}   
+   
    4 (a_3^2 - b_{3}^2) D_{00}^{12}+   
    4 (a_3^2 m_{h0}^2 - b_{3}^2 m_{h0}^2 + a_3^2 m_{\tilde{\chi}_1^\pm}^2 - b_{3}^2 m_{\tilde{\chi}_1^\pm}^2 +   
       2 a_3^2 m_{\tilde{\chi}_1^\pm} m_{\tilde{\chi}_2^\pm}\nonumber\\  
	   &+& 2 b_{3}^2 m_{\tilde{\chi}_1^\pm} m_{\tilde{\chi}_2^\pm} + a_3^2 m_{\tilde{\chi}_2^\pm}^2 - b_{3}^2 m_{\tilde{\chi}_2^\pm}^2)   
     D_{11}^{12} \nonumber \\  
&+&   
    2 (4 a_3^2 m_{h0}^2 - 4 b_{3}^2 m_{h0}^2 + 4 a_3^2 m_{\tilde{\chi}_1^\pm}^2 - 4 b_{3}^2 m_{\tilde{\chi}_1^\pm}^2 +   
       8 a_3^2 m_{\tilde{\chi}_1^\pm} m_{\tilde{\chi}_2^\pm}\nonumber\\  
	   &+& 8 b_{3}^2 m_{\tilde{\chi}_1^\pm} m_{\tilde{\chi}_2^\pm} + 4 a_3^2 m_{\tilde{\chi}_2^\pm}^2 - 4 b_{3}^2 m_{\tilde{\chi}_2^\pm}^2 +   
       a_3^2 \hat{t} - b_{3}^2 \hat{t} - a_3^2 \hat{u} + b_{3}^2 \hat{u}) D_{12}^{12} \nonumber \\  
&+&   
    4 (a_3^2 m_{h0}^2 - b_{3}^2 m_{h0}^2 + a_3^2 m_{\tilde{\chi}_1^\pm}^2 - b_{3}^2 m_{\tilde{\chi}_1^\pm}^2 +   
       2 a_3^2 m_{\tilde{\chi}_1^\pm} m_{\tilde{\chi}_2^\pm} + 2 b_{3}^2 m_{\tilde{\chi}_1^\pm} m_{\tilde{\chi}_2^\pm} + a_3^2 m_{\tilde{\chi}_2^\pm}^2 - b_{3}^2 m_{\tilde{\chi}_2^\pm}^2)   
     D_{13}^{12}\nonumber \\  
& + &  
    (2 a_3^2 m_{\tilde{\chi}_1^\pm}^2 - 2 b_{3}^2 m_{\tilde{\chi}_1^\pm}^2 + 4 a_3^2 m_{\tilde{\chi}_1^\pm} m_{\tilde{\chi}_2^\pm} + 4 b_{3}^2 m_{\tilde{\chi}_1^\pm} m_{\tilde{\chi}_2^\pm} +   
       2 a_3^2 m_{\tilde{\chi}_2^\pm}^2 - 2 b_{3}^2 m_{\tilde{\chi}_2^\pm}^2 \nonumber \\  
&+&\left.\left.  a_3^2 \hat{s} - b_{3}^2 \hat{s} + 2 a_3^2 \hat{t} - 2 b_{3}^2 \hat{t})   
     D_{22}^{12}\right) \right]  
+  
\left[a_3\rightarrow a_2, b_{3}\rightarrow b_{2}, m_{\tilde{\chi}_1^\pm}\rightarrow m_{\tilde{\chi}_2^\pm}\right]\nonumber \\  
&+&  
\left[a_3\rightarrow a_1, b_{3}\rightarrow b_{1}, m_{\tilde{\chi}_2^\pm}\rightarrow m_{\tilde{\chi}_1^\pm}\right]\nonumber   
+  
\left[m_{\tilde{\chi}_1^\pm}\rightarrow m_{\tilde{\chi}_2^\pm}, m_{\tilde{\chi}_2^\pm}\rightarrow m_{\tilde{\chi}_1^\pm}\right],  
\end{eqnarray}  
\begin{eqnarray}  
&&f_2^{(1, sfermions)}=\nonumber \\  
&&-N_c \left[{ e^2 e_t^2\over 2 \pi^2}\left(  
\xi_{\tilde{t}3}^2 ( D_0^{18} + 4 D_1^{18}+   
       2 D_2^{18} +   
       2 D_{11}^{18}+   
       4 D_{12}^{18}+   
       2 D_{13}^{18}\right. \right.\nonumber \\  
& +&\left.\left.   
       D_{22}^{18})+  
\xi_{\tilde{t}2}^2 (18\rightarrow 15)  
+\xi_{\tilde{t}1}^2 (18\rightarrow 14)  
+\xi_{\tilde{t}3}^2 (18\rightarrow 19) \right)\right] \nonumber \\  
&-& N_c\left[e_t \rightarrow e_b, \xi_{\tilde{t}1} \rightarrow \xi_{\tilde{b}1}, 
\xi_{\tilde{t}2}\rightarrow\xi_{\tilde{b}2},\xi_{\tilde{t}3}\rightarrow\xi_{\tilde{b}3}, 
m_{\tilde{t}_1}\rightarrow m_{\tilde{b}_1}, m_{\tilde{t}_2}\rightarrow m_{\tilde{b}_2} 
 \right]\nonumber \\ 
&-&\left[e_t \rightarrow e_{\tau}, \xi_{\tilde{t}1} \rightarrow \xi_{\tilde{\tau}1}, 
\xi_{\tilde{t}2}\rightarrow\xi_{\tilde{\tau}2},\xi_{\tilde{t}3}\rightarrow\xi_{\tilde{\tau}3}, 
m_{\tilde{t}_1}\rightarrow m_{\tilde{\tau}_1}, m_{\tilde{t}_2}\rightarrow m_{\tilde{\tau}_2} 
\right],  
\end{eqnarray}  
where $C_m^i, C_{mn}^i\equiv   C_m, C_{mn}(0, 0, \hat{s},i)$ and $D_x^i
\equiv    D_x(m_{h0}^2, 0, 0, m_{h0}^2, \hat{t}, \hat{s}, i)$.  
  
The form factor $f_{2}^{(2)}$ is given by 
\begin{eqnarray}  
f_{2}^{(2)}&=&f_2^{(2, 2HDM)}+f_{2}^{(2, charginos)}+f_2^{(2, sfermions)},  
\end{eqnarray}  
with 
\begin{eqnarray}  
f_2^{(2, 2HDM)}&=&f_2^{(2, fermions)}+f_2^{(2,bosons)}. 
\end{eqnarray}  
Here $f_2^{(2, fermions)}$, $f_2^{(2,bosons)}$,  $f_{2}^{(2, charginos)}$ and  
$f_2^{(2, sfermions)}$ are 
\begin{eqnarray}  
&&f_2^{(2, fermions)}=\nonumber \\  
&&{e^2 g^2 m_t^2 \cos (\alpha)^2 csc(\beta)^2\over 6 m_{w}^2 \pi^2}   
\left[ 2 C_0^9  
 - (8 m_t^2 + \hat{s})   
      D_2^7  
 \right. \nonumber \\  
&-& 8 m_t^2 D_3^7- 4 D_{00}^7  
     + (\hat{t} - \hat{u})  (D_{12}^7+ D_{13}^7)  
     - (8 m_t^2 + \hat{s} + 2 \hat{u})  D_{22}^7\nonumber\\  
&+&  \left. (-4 m_{h0}^2 - 16 m_t^2 + \hat{t} - \hat{u})  D_{23}^7  
 -   
   2 (m_{h0}^2 + 4 m_t^2) D_{33}^7 \right]  
\nonumber \\  
&+&{e^2 g^2 m_b^2 \sin (\alpha)^2 sec(\beta)^2\over 24 m_{w}^2 \pi^2}\left[m_t\rightarrow m_b\right],  
\end{eqnarray}  
\begin{eqnarray}  
&&f_2^{(2, bosons)}=\nonumber \\  
&&{e^2 g^2 \cos (\alpha - \beta)^2\over 16 \pi^2}   
\left( 4 C_0^5   
   - 4 C_1^2 +   
  4 C_1^3+  
    2 (-\hat{t} + \hat{u})    
     D_1^3 \right. \nonumber \\  
&-&   
4 C_2^2 +   
   4 C_2^3  
+4 D_{00}^2  
 + 4 D_{00}^3  
    + (-\hat{t} + \hat{u}) ( D_{12}^2+D_{12}^3+ D_{13}^2+ D_{13}^3)  
+ 8 (D_{002}^2- D_{002}^3+ D_{003}^2 - D_{003}^3)\nonumber \\  
&+&   
   (-\hat{t} + \hat{u}) ( D_{122}^2- D_{122}^3 +2 D_{123}^2-2 D_{123}^3+D_{133}^2-D_{133}^3)  
    +   
   2 (m_{h0}^2 + \hat{u}) ( D_{222}^2 - D_{222}^3)\nonumber \\  
   & +&\left.   
   (6 m_{h0}^2 + \hat{t} + 5 \hat{u}) ( D_{223}^2 - D_{223}^3)  
    +   
   2 (3 m_{h0}^2 + \hat{t} + 2 \hat{u}) ( D_{233}^2 - D_{233}^3)   
   +   
   (4 m_{h0}^2 - \hat{s}) ( D_{333}^2-D_{333}^3\right) \nonumber \\  
 &+&  
 {e^2 g^2 \sin (\alpha - \beta)^2\over 8 \pi^2}   
  \left( 2 C_0^1 +   
 (-\hat{t} + \hat{u}) (D_1^1+D_{12}^1+D_{13}^1)  
 +   
   4  D_{00}^1\right) \nonumber \\  
 &+ &  
 { e^2 g^2 \over 16 \cos\theta_w^2 \pi^2} \left(   
2 (-2 \cos\theta_w m_{w} \sin (\alpha - \beta) + m_{z} \cos (2 \beta) \sin (\alpha + \beta))^2   
 (D_2^6+ D_3^6+D_{22}^6+ 2 D_{23}^6+ D_{33}^6) \right. \nonumber \\  
&+&  
 2 D_0^1 \sin (\alpha - \beta)   
(3 \cos\theta_w m_{h0}^2 \sin (\alpha - \beta) -   
        2 \cos\theta_w \hat{t} \sin (\alpha - \beta)\nonumber \\  
& - &\cos\theta_w \hat{u} \sin (\alpha - \beta) +   
        2 m_{w} m_{z} \cos (2 \beta) \sin (\alpha + \beta)) \nonumber \\  
&+ &  
   2 \cos (\alpha - \beta) D_0^3   
(3 \cos\theta_w m_{h0}^2 \cos (\alpha - \beta) -  
       2 \cos\theta_w \hat{t} \cos (\alpha - \beta) \nonumber \\  
&-& \cos\theta_w \hat{u} \cos (\alpha - \beta) +  
        2 m_{w} m_{z} \sin (2 \beta) \sin (\alpha + \beta))\nonumber \\  
 &+& 2 D_2^1   
      (4 \cos\theta_w^2 m_{h0}^2 \sin (\alpha - \beta)^2 +   
        12 \cos\theta_w^2 m_{w}^2 \sin (\alpha - \beta)^2 \nonumber \\  
&-& 4 \cos\theta_w^2 \hat{t} \sin (\alpha - \beta)^2 +   
        2 \cos\theta_w m_{w} m_{z} \cos (2 \beta) \sin (\alpha - \beta) \sin (\alpha + \beta) \nonumber \\  
&+&   
        m_{z}^2 \cos (2 \beta)^2 \sin (\alpha + \beta)^2) +   
   2 D_{33}^1   
      (4 \cos\theta_w^2 m_{h0}^2 \sin (\alpha - \beta)^2 \nonumber \\  
&+&   
        12 \cos\theta_w^2 m_{w}^2 \sin (\alpha - \beta)^2 - 3 \cos\theta_w^2 \hat{t} \sin (\alpha - \beta)^2 -   
        3 \cos\theta_w^2 \hat{u} \sin (\alpha - \beta)^2\nonumber \\  
& +&   
        2 \cos\theta_w m_{w} m_{z} \cos (2 \beta) \sin (\alpha - \beta) \sin (\alpha + \beta) +   
        m_{z}^2 \cos (2 \beta)^2 \sin (\alpha + \beta)^2) \nonumber \\  
&+&   
   2 D_{22}^1   
      (4 \cos\theta_w^2 m_{h0}^2 \sin (\alpha - \beta)^2 +   
        12 \cos\theta_w^2 m_{w}^2 \sin (\alpha - \beta)^2 - 4 \cos\theta_w^2 \hat{t} \sin (\alpha - \beta)^2 \nonumber \\  
&- &  
        2 \cos\theta_w^2 \hat{u} \sin (\alpha - \beta)^2 +   
        2 \cos\theta_w m_{w} m_{z} \cos (2 \beta) \sin (\alpha - \beta) \sin (\alpha + \beta) +   
        m_{z}^2 \cos (2 \beta)^2 \sin (\alpha + \beta)^2) \nonumber \\  
&+&   
  2 D_3^1   
      (6 \cos\theta_w^2 m_{h0}^2 \sin (\alpha - \beta)^2 +   
        12 \cos\theta_w^2 m_{w}^2 \sin (\alpha - \beta)^2 - 4 \cos\theta_w^2 \hat{t} \sin (\alpha - \beta)^2\nonumber \\  
& -&   
        2 \cos\theta_w^2 \hat{u} \sin (\alpha - \beta)^2 +   
        2 \cos\theta_w m_{w} m_{z} \cos (2 \beta) \sin (\alpha - \beta) \sin (\alpha + \beta) +   
        m_{z}^2 \cos (2 \beta)^2 \sin (\alpha + \beta)^2) \nonumber \\  
&+&   
  D_{23}^1   
      (6 \cos\theta_w^2 m_{h0}^2 \sin (\alpha - \beta)^2 +   
        48 \cos\theta_w^2 m_{w}^2 \sin (\alpha - \beta)^2 + 5 \cos\theta_w^2 \hat{s} \sin (\alpha - \beta)^2 \nonumber \\  
&-&   
        9 \cos\theta_w^2 \hat{t} \sin (\alpha - \beta)^2 - 5 \cos\theta_w^2 \hat{u} \sin (\alpha - \beta)^2 +   
        8 \cos\theta_w m_{w} m_{z} \cos (2 \beta) \sin (\alpha - \beta) \sin (\alpha + \beta)   
\nonumber  
\\ &+&   
        4 m_{z}^2 \cos (2 \beta)^2 \sin (\alpha + \beta)^2) +   
   (D_{2}^2 +D_3^2)   
      (-4 \cos\theta_w^2 m_{H^\pm}^2 \cos (\alpha - \beta)^2\nonumber \\  
& - &  
        \cos\theta_w^2 m_{h0}^2 \cos (\alpha - \beta)^2 + 4 \cos\theta_w^2 m_{w}^2 \cos (\alpha - \beta)^2 +   
        \cos\theta_w^2 \hat{u} \cos (\alpha - \beta)^2 \nonumber \\  
&- &  
        6 \cos\theta_w m_{w} m_{z} \cos (\alpha - \beta) \sin (2 \beta) \sin (\alpha + \beta) +   
        2 m_{z}^2 \sin (2 \beta)^2 \sin (\alpha + \beta)^2) \nonumber \\  
&+&   
   D_{33}^3   
      (5 \cos\theta_w^2 m_{h0}^2 \cos (\alpha - \beta)^2 - 5 \cos\theta_w^2 \hat{t} \cos (\alpha - \beta)^2 -   
        4 \cos\theta_w^2 \hat{u} \cos (\alpha - \beta)^2\nonumber \\  
& - &  
        2 \cos\theta_w m_{w} m_{z} \cos (\alpha - \beta) \sin (2 \beta) \sin (\alpha + \beta) +   
        2 m_{z}^2 \sin (2 \beta)^2 \sin (\alpha + \beta)^2)\nonumber \\  
&+&   
   D_{22}^2   
      (7 \cos\theta_w^2 m_{h0}^2 \cos (\alpha - \beta)^2 - 4 \cos\theta_w^2 \hat{t} \cos (\alpha - \beta)^2 -   
        3 \cos\theta_w^2 \hat{u} \cos (\alpha - \beta)^2 \nonumber \\  
&-&  
        2 \cos\theta_w m_{w} m_{z} \cos (\alpha - \beta) \sin (2 \beta) \sin (\alpha + \beta) +   
        2 m_{z}^2 \sin (2 \beta)^2 \sin (\alpha + \beta)^2) \nonumber \\  
&+&   
   D_{33}^2   
      (3 \cos\theta_w^2 m_{h0}^2 \cos (\alpha - \beta)^2 - \cos\theta_w^2 \hat{t} \cos (\alpha - \beta)^2 -   
        2 \cos\theta_w^2 \hat{u} \cos (\alpha - \beta)^2 \nonumber \\  
&-&   
        2 \cos\theta_w m_{w} m_{z} \cos (\alpha - \beta) \sin (2 \beta) \sin (\alpha + \beta) +   
        2 m_{z}^2 \sin (2 \beta)^2 \sin (\alpha + \beta)^2) \nonumber \\  
&+&   
   D_{22}^3   
      (\cos\theta_w^2 m_{h0}^2 \cos (\alpha - \beta)^2 - 4 \cos\theta_w^2 \hat{t} \cos (\alpha - \beta)^2 -   
        \cos\theta_w^2 \hat{u} \cos (\alpha - \beta)^2 \nonumber \\  
&-&   
        2 \cos\theta_w m_{w} m_{z} \cos (\alpha - \beta) \sin (2 \beta) \sin (\alpha + \beta) +   
        2 m_{z}^2 \sin (2 \beta)^2 \sin (\alpha + \beta)^2) \nonumber \\  
&+&   
  D_3^3   
      (13 \cos\theta_w^2 m_{h0}^2 \cos (\alpha - \beta)^2 - 8 \cos\theta_w^2 \hat{t} \cos (\alpha - \beta)^2 -   
        5 \cos\theta_w^2 \hat{u} \cos (\alpha - \beta)^2 \nonumber \\  
&+&   
        2 \cos\theta_w m_{w} m_{z} \cos (\alpha - \beta) \sin (2 \beta) \sin (\alpha + \beta) +   
        2 m_{z}^2 \sin (2 \beta)^2 \sin (\alpha + \beta)^2) \nonumber \\  
&+&   
  D_2^3   
      (9 \cos\theta_w^2 m_{h0}^2 \cos (\alpha - \beta)^2 - 8 \cos\theta_w^2 \hat{t} \cos (\alpha - \beta)^2 -   
        \cos\theta_w^2 \hat{u} \cos (\alpha - \beta)^2 \nonumber \\  
&+&   
        2 \cos\theta_w m_{w} m_{z} \cos (\alpha - \beta) \sin (2 \beta) \sin (\alpha + \beta) +   
        2 m_{z}^2 \sin (2 \beta)^2 \sin (\alpha + \beta)^2) \nonumber \\  
&+&   
   D_{23}^2   
      (5 \cos\theta_w^2 \hat{s} \cos (\alpha - \beta)^2 -   
        4 \cos\theta_w m_{w} m_{z} \cos (\alpha - \beta) \sin (2 \beta) \sin (\alpha + \beta) +   
        4 m_{z}^2 \sin (2 \beta)^2 \sin (\alpha + \beta)^2)\nonumber \\  
& +&   
  D_{23}^3   
      (6 \cos\theta_w^2 m_{h0}^2 \cos (\alpha - \beta)^2 - 9 \cos\theta_w^2 \hat{t} \cos (\alpha - \beta)^2 -   
        5 \cos\theta_w^2 \hat{u} \cos (\alpha - \beta)^2 \nonumber \\  
&-& \left.   
        4 \cos\theta_w m_{w} m_{z} \cos (\alpha - \beta) \sin (2 \beta) \sin (\alpha + \beta) +   
        4 m_{z}^2 \sin (2 \beta)^2 \sin (\alpha + \beta)^2)  
\right),  
\end{eqnarray}  
\begin{eqnarray}  
&&f_2^{(2, charginos)}=\nonumber\\  
&&\left[{e^2\over 2 \pi^2 }\left((4 a_3^2 m_{\tilde{\chi}_1^\pm}^2 - 4 b_{3}^2 m_{\tilde{\chi}_1^\pm}^2 + 4 a_3^2 m_{\tilde{\chi}_1^\pm} m_{\tilde{\chi}_2^\pm} + 4 b_{3}^2 m_{\tilde{\chi}_1^\pm} m_{\tilde{\chi}_2^\pm} +   
       a_3^2 \hat{s} - b_{3}^2 \hat{s}) D_2^{10}\right. \right. \nonumber \\  
&+&   
    4 m_{\tilde{\chi}_1^\pm} (a_3^2 m_{\tilde{\chi}_1^\pm} - b_{3}^2 m_{\tilde{\chi}_1^\pm} + a_3^2 m_{\tilde{\chi}_2^\pm} + b_{3}^2 m_{\tilde{\chi}_2^\pm}) D_3^{10} +   
   (a_3^2 - b_{3}^2) (-2  C_0^{13}+ 4 D_{00}^{10}+ (-\hat{t} + \hat{u}) (D_{12}^{10}\nonumber \\  
&+& D_{13}^{10}))  
   (2 a_3^2 m_{\tilde{\chi}_1^\pm}^2 - 2 b_{3}^2 m_{\tilde{\chi}_1^\pm}^2 + 4 a_3^2 m_{\tilde{\chi}_1^\pm} m_{\tilde{\chi}_2^\pm} + 4 b_{3}^2 m_{\tilde{\chi}_1^\pm} m_{\tilde{\chi}_2^\pm} +   
       2 a_3^2 m_{\tilde{\chi}_2^\pm}^2 - 2 b_{3}^2 m_{\tilde{\chi}_2^\pm}^2 + a_3^2 \hat{s} - b_{3}^2 \hat{s} \nonumber \\  
&+& 2 a_3^2 \hat{u} - 2 b_{3}^2 \hat{u})   
D_{22}^{10} +   
    (4 a_3^2 m_{h0}^2 - 4 b_{3}^2 m_{h0}^2 + 4 a_3^2 m_{\tilde{\chi}_1^\pm}^2 - 4 b_{3}^2 m_{\tilde{\chi}_1^\pm}^2 +   
       8 a_3^2 m_{\tilde{\chi}_1^\pm} m_{\tilde{\chi}_2^\pm}\nonumber\\  
	   &+& 8 b_{3}^2 m_{\tilde{\chi}_1^\pm} m_{\tilde{\chi}_2^\pm} + 4 a_3^2 m_{\tilde{\chi}_2^\pm}^2 - 4 b_{3}^2 m_{\tilde{\chi}_2^\pm}^2 -   
       a_3^2 \hat{t} + b_{3}^2 \hat{t} + a_3^2 \hat{u} - b_{3}^2 \hat{u}) D_{23}^{10}\nonumber \\  
&+& \left. \left.  
    2 (a_3^2 m_{h0}^2 - b_{3}^2 m_{h0}^2 + a_3^2 m_{\tilde{\chi}_1^\pm}^2 - b_{3}^2 m_{\tilde{\chi}_1^\pm}^2 + 2 a_3^2 m_{\tilde{\chi}_1^\pm} m_{\tilde{\chi}_2^\pm} +   
       2 b_{3}^2 m_{\tilde{\chi}_1^\pm} m_{\tilde{\chi}_2^\pm} + a_3^2 m_{\tilde{\chi}_2^\pm}^2 - b_{3}^2 m_{\tilde{\chi}_2^\pm}^2) D_{33}^{10}\right)  
\right]\nonumber \\  
&+&  
\left[a_3\rightarrow a_2, b_{3}\rightarrow b_{2}, m_{\tilde{\chi}_1^\pm}\rightarrow m_{\tilde{\chi}_2^\pm}\right]+  
\left[a_3\rightarrow a_1, b_{3}\rightarrow b_{1}, m_{\tilde{\chi}_2^\pm}\rightarrow m_{\tilde{\chi}_1^\pm}\right]\nonumber \\  
&+&  
\left[m_{\tilde{\chi}_1^\pm}\rightarrow m_{\tilde{\chi}_2^\pm}, m_{\tilde{\chi}_2^\pm}\rightarrow m_{\tilde{\chi}_1^\pm}\right], 
\end{eqnarray}  
\begin{eqnarray}  
&&f_2^{(2,sfermions)}=\nonumber \\  
&&-N_c \left[{e^2 e_t^2\over 2 \pi^2}\left(  
\xi_{\tilde{t}3}^2 (D_2^{16} +  
      D_3^{16} +  
      D_{22}^{16} +   
      2 D_{23}^{16} +   
      D_{33}^{16}) \right.\right. \nonumber \\  
&-&\left.\left.  
\xi_{\tilde{t}2}^2 (16\rightarrow 15)  
-\xi_{\tilde{t}1}^2 (16\rightarrow 14)  
-\xi_{\tilde{t}3}^2 (16\rightarrow 17)\right)\right] \nonumber \\  
&-& N_c\left[e_t \rightarrow e_b, \xi_{\tilde{t}1} \rightarrow \xi_{\tilde{b}1}, 
\xi_{\tilde{t}2}\rightarrow\xi_{\tilde{b}2},\xi_{\tilde{t}3}\rightarrow\xi_{\tilde{b}3}, 
m_{\tilde{t}_1}\rightarrow m_{\tilde{b}_1}, m_{\tilde{t}_2}\rightarrow m_{\tilde{b}_2} 
 \right]\nonumber \\ 
&-&\left[e_t \rightarrow e_{\tau}, \xi_{\tilde{t}1} \rightarrow \xi_{\tilde{\tau}1}, 
\xi_{\tilde{t}2}\rightarrow\xi_{\tilde{\tau}2},\xi_{\tilde{t}3}\rightarrow\xi_{\tilde{\tau}3}, 
m_{\tilde{t}_1}\rightarrow m_{\tilde{\tau}_1}, m_{\tilde{t}_2}\rightarrow m_{\tilde{\tau}_2} 
\right],  
\end{eqnarray}  
where $D_m^i, D_{mn}^i, D_{mnl}^i
\equiv    D_m, D_{mn}, D_{mnl}(0, m_{h0}^2, 0, m_{h0}^2, \hat{u}, \hat{t}, i)$,  
      $C_m^i, C_{mn}^i\equiv$ $C_m, C_{mn}(m_{h0}^2, 0, \hat{t}, i)$ and $C_0^i
      \equiv   C_0(0, m_{h0}^2,\hat{t}, i)$.

 The form factor $f_{2}^{(10)}$ is given by 
\begin{eqnarray}  
f_{2}^{(10)}&=&{e^2 g^2\cos (\alpha - \beta)^2\over 32 \pi^2}   
\left( 4 C_0^5 -   
  C_0^4 +   
   6 C_1^3  
     + 6 C_2^3 +   
   C_{11}^2 +  
  2 C_{11}^3 +   
   2 C_{12}^2 +  
 4 C_{12}^3 +   
   C_{22}^2 +  
2 C_{22}^3\right) \nonumber \\   
&+&  
 { 3 e^2 g^2\sin (\alpha - \beta)^2\over 32 \pi^2}  
\left( C_0^1+   
   2 C_1^1  
   + 2 C_2^1 +   
   C_{11}^1 +   
   2 C_{12}^1 +   
   C_{22}^1 \right),  
\end{eqnarray}  
where $C_m^i, C_{mn}^i \equiv    C_m, C_{mn}(m_{h0}^2, 0, \hat{t}, i)$ and  $C_0^i \equiv    C_0(0,m_{h0}^2,\hat{t}, i)$.  
  
The form factor $f_{2}^{(11)}$ is given by 
\begin{eqnarray}  
f_{2}^{(11)}&=&{e^2 g^2 \cos (\alpha - \beta)^2\over 32 \pi^2}   
\left( 4 C_0^5 -   
  C_0^4 +   
   6 C_1^3  
     + 6 C_2^3+   
   C_{11}^2  
   + 2 C_{11}^3 +   
  2 C_{12}^2  
    +4 C_{12}^3 +   
   C_{22}^2  
+2 C_{22}^3 \right) \nonumber \\  
&+ &  
 {3 e^2 g^2 \sin (\alpha - \beta)^2\over 32 \pi^2}  
\left(  
   C_0^1 +   
   2 C_1^1+   
   2 C_2^1  
 +   
   C_{11}^1 +   
   2 C_{12}^1+   
   C_{22}^1\right),  
\end{eqnarray}  
where $ C_m^i, C_{mn}^i \equiv   C_m, C_{mn}(m_{h0}^2, 0, \hat{t}, i)$ and $C_0^i \equiv   C_0(0,m_{h0}^2,\hat{t},i)$.

 The form factor $f_{2}^{(12)}$ is given by 
\begin{eqnarray}  
f_{2}^{(12)}={e^2 g^2 \cos (\alpha - \beta)^2\over 16 \pi^2}  
\left( -4 C_0^8 -   
   4 C_1^8-  
C_{11}^8 \right)-   
 {e^2 g^2 \sin (\alpha - \beta)^2\over 16 \pi^2}   
 \left(4 C_0^1 -   
  4 C_1^1-   
C_{11}^1\right),  
\end{eqnarray}  
where $C_m^i, C_{mn}^i \equiv   C_m, C_{mn}(m_{h0}^2, m_{h0}^2, \hat{s}, i)$.  
  
\section*{Appendix C}  
The form factor $f_3^{(1)}$ is given by  
\begin{eqnarray}  
f_{3}^{(1)}&=&\left[{i e^2 a_3 b_{3} m_{\tilde{\chi}_1^\pm} m_{\tilde{\chi}_2^\pm} D_0^{12}\over \pi^2}\right]+  
\left[a_3\rightarrow a_2,b_{3}\rightarrow b_{2}, m_{\tilde{\chi}_1^\pm}\rightarrow m_{\tilde{\chi}_2^\pm}\right]\nonumber \\  
&+&  
\left[a_3\rightarrow a_1,b_{3}\rightarrow b_{1}, m_{\tilde{\chi}_2^\pm}\rightarrow m_{\tilde{\chi}_1^\pm}\right]  
+  
\left[m_{\tilde{\chi}_1^\pm}\rightarrow m_{\tilde{\chi}_2^\pm},m_{\tilde{\chi}_2^\pm}\rightarrow m_{\tilde{\chi}_1^\pm}\right],  
\end{eqnarray}  
where  $D_0^i \equiv   D_0( m_{h0}^2, 0, 0, m_{h0}^2, \hat{t}, \hat{s}, i)$.  
  
The form factor $f_{3}^{(2)}$ is given by 
\begin{eqnarray}  
f_{3}^{(2)}&=&\left[{i e^2 a_3 b_{3} m_{\tilde{\chi}_1^\pm} m_{\tilde{\chi}_2^\pm} D_0^{10} \over \pi^2}\right]+  
\left[m_{\tilde{\chi}_1^\pm}\rightarrow m_{\tilde{\chi}_2^\pm},m_{\tilde{\chi}_2^\pm}\rightarrow m_{\tilde{\chi}_1^\pm}\right] \nonumber \\  
&+&  
\left[a_3\rightarrow a_2, b_{3}\rightarrow b_{2}, m_{\tilde{\chi}_1^\pm}\rightarrow m_{\tilde{\chi}_2^\pm}\right]+  
\left[a_3\rightarrow a_1, b_{3}\rightarrow b_{1}, m_{\tilde{\chi}_2^\pm}\rightarrow m_{\tilde{\chi}_1^\pm}\right],  
\end{eqnarray}  
where  
$D_0^i \equiv   D_0(0, m_{h0}^2, 0, m_{h0}^2, \hat{u}, \hat{t}, i)$.

 The form factor $f_{3}^{(5)}$ is given by 
\begin{eqnarray}  
f_{3}^{(5)}={-i e^2 \eta_2 (b_{5} m_{\tilde{\chi}_2^\pm} C_0^{11}+ b_{4} m_{\tilde{\chi}_1^\pm} C_0^{10}) \over 2 \pi^2 (m_{H}^2 - \hat{s})}+  
{-i e^2 \eta_1 (b_{2} m_{\tilde{\chi}_2^\pm} C_0^{11}+ b_{1} m_{\tilde{\chi}_1^\pm} C_0^{10})\over 2 \pi^2 (m_{h0}^2 - \hat{s})},  
\end{eqnarray}  
where $C_0^i\equiv   C_0(0, 0, \hat{s}, i)$.


%% file: feyn.tex
\begin{figure}
\epsfxsize=10 cm
\centerline{\epsffile{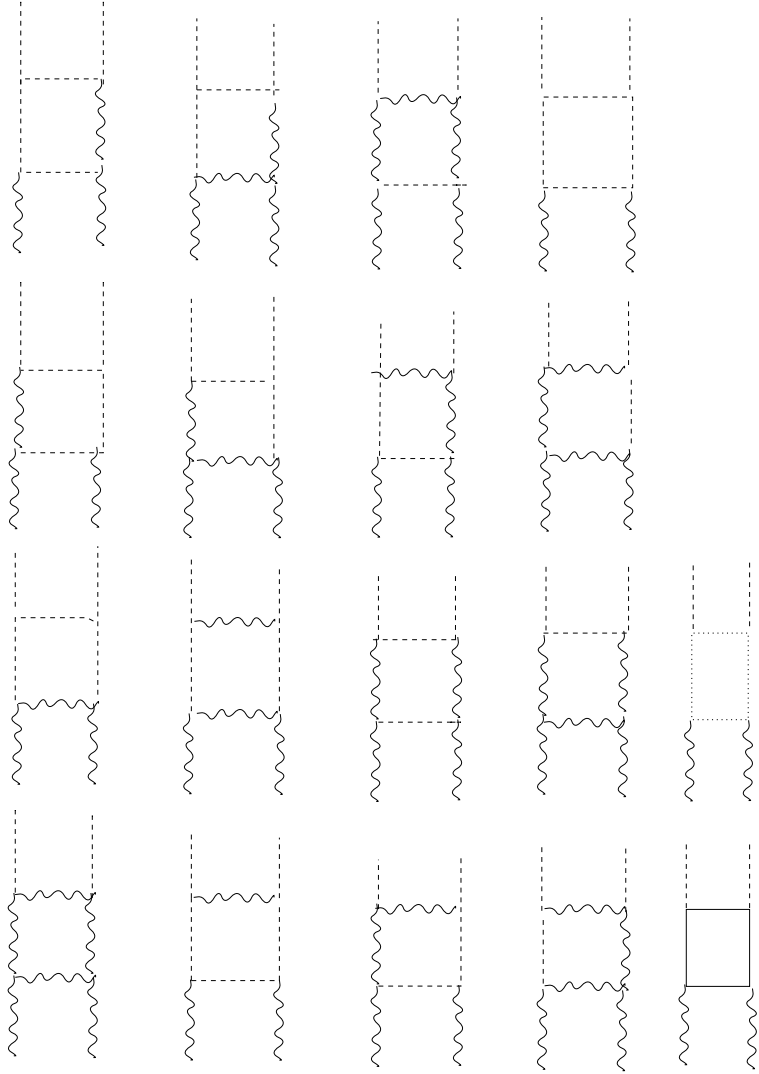}}
\caption{Feynman diagrams for the precess $\gamma\gamma\rightarrow h_0h_0$, 
where the solid line, wavy line, dashed line and dotted line represent 
the fermions (the top and the bottom quark as well as the charginos), 
the gauge bosons(the photon and the $W^\pm$ boson), 
the scalars (the charged Higgs bosons, the charged Goldstone bosons,
 the squarks and the sleptons) and the charged ghosts, respectively.}
\label{fig1}
\end{figure}

\begin{figure}
\epsfxsize=10 cm
\centerline{\epsffile{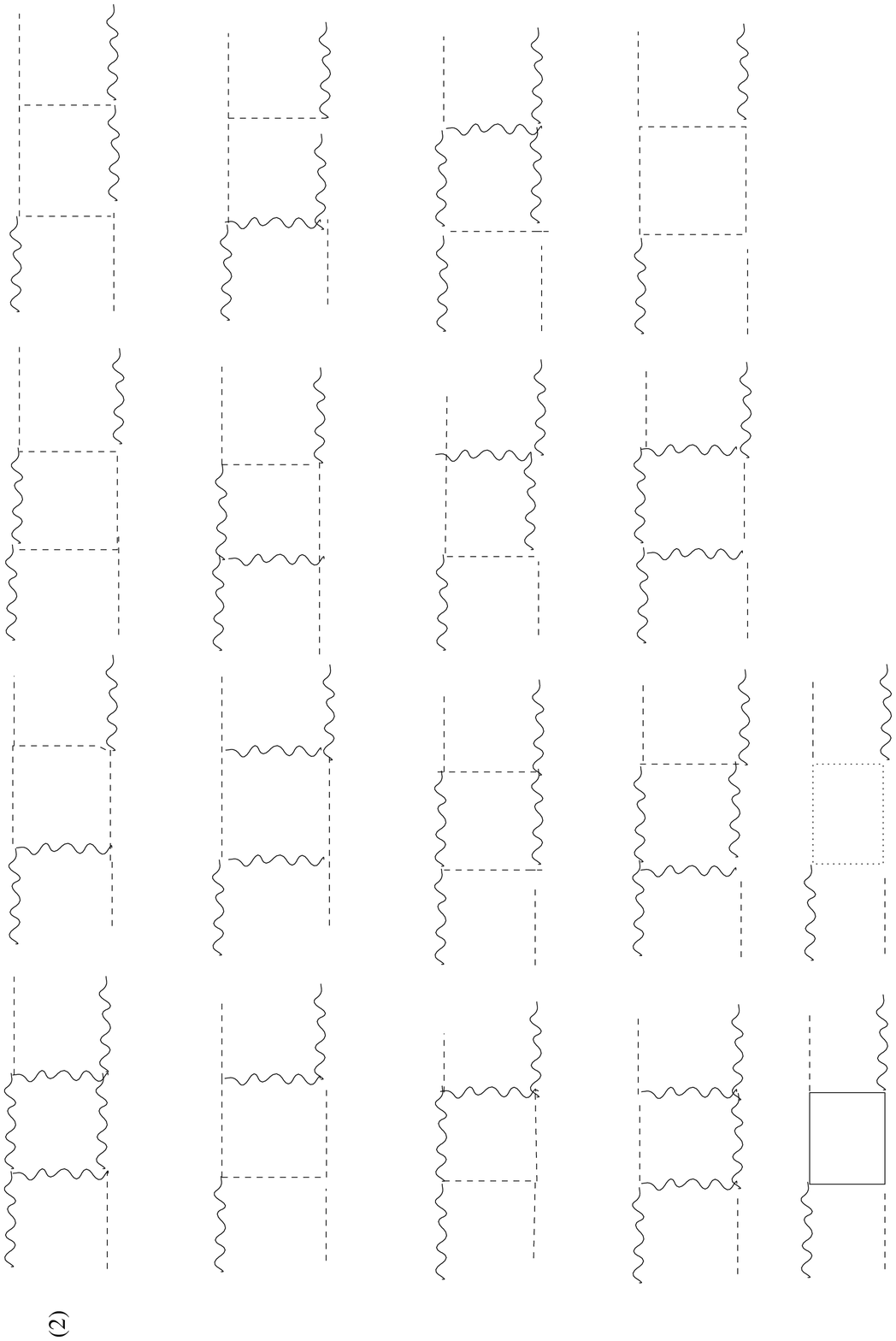}}
\caption{ Same with Figure \ref{fig1}}
\label{fig2}
\end{figure}

\begin{figure}
\epsfxsize=10 cm
\centerline{\epsffile{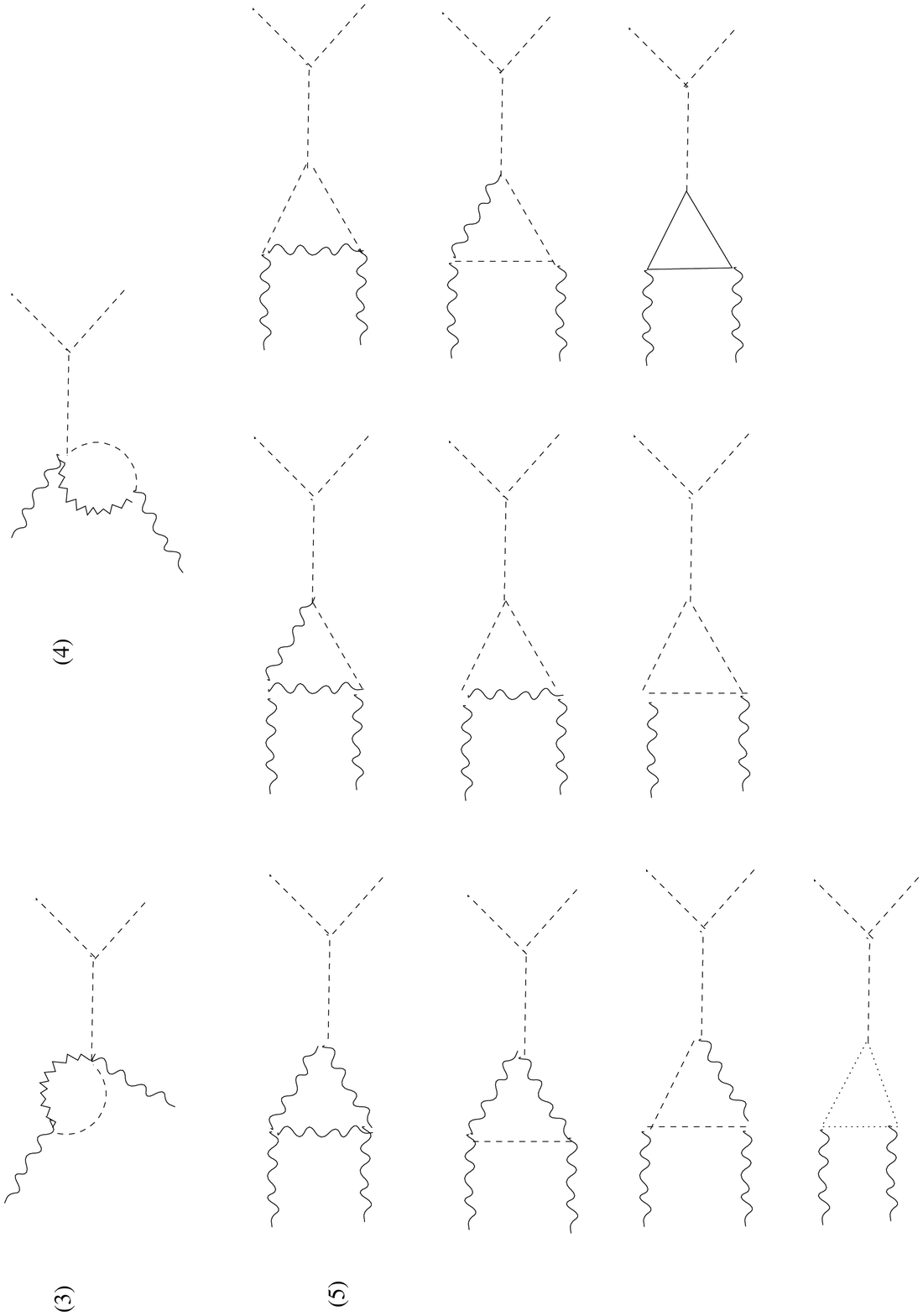}}
\caption{ Same with Figure \ref{fig1}}
\label{fig3}
\end{figure}

\begin{figure}
\epsfxsize=10 cm
\centerline{\epsffile{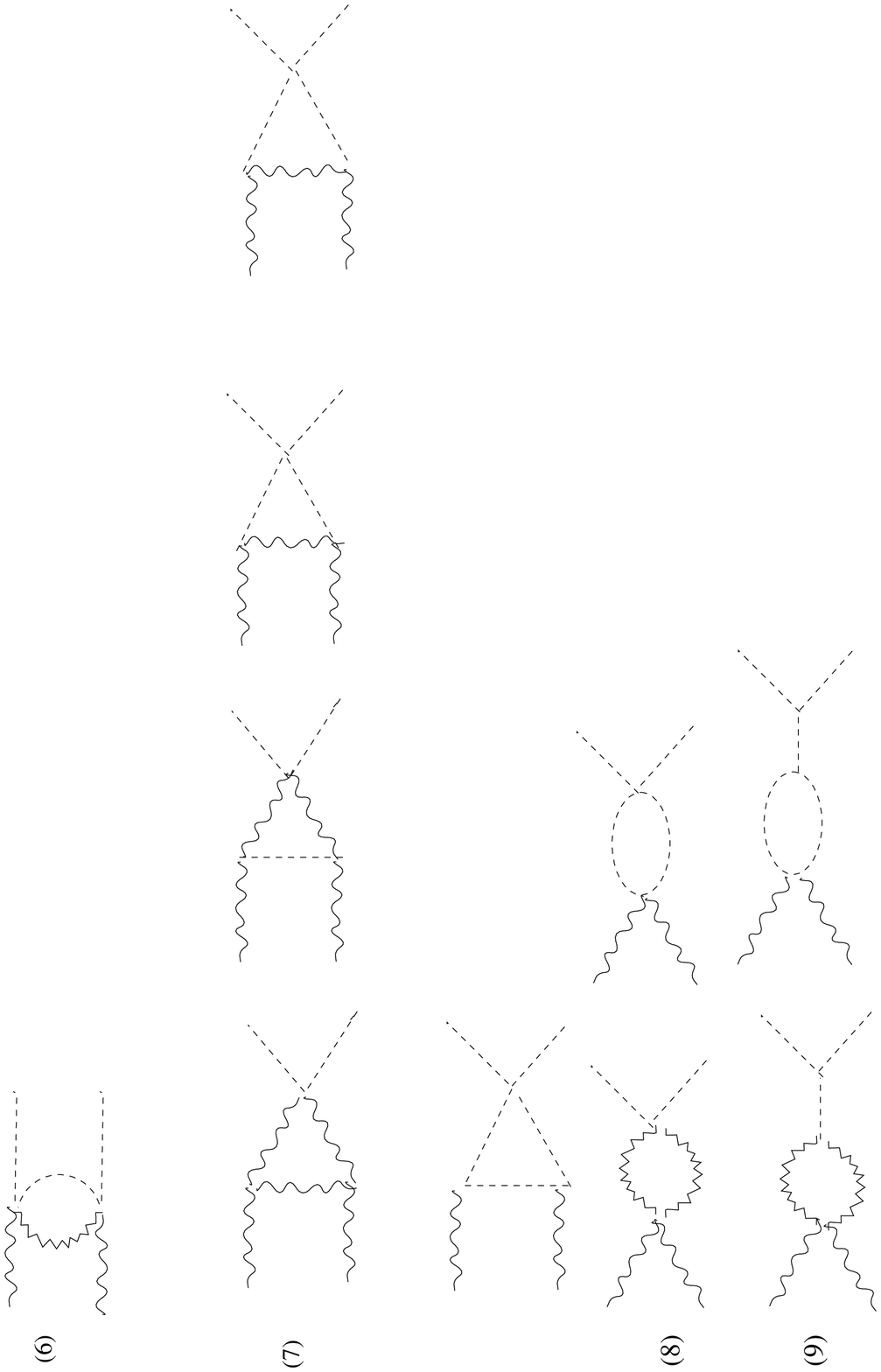}}
\caption{ Same with Figure \ref{fig1}}
\label{fig4}
\end{figure}

\begin{figure}
\epsfxsize=9 cm
\centerline{\epsffile{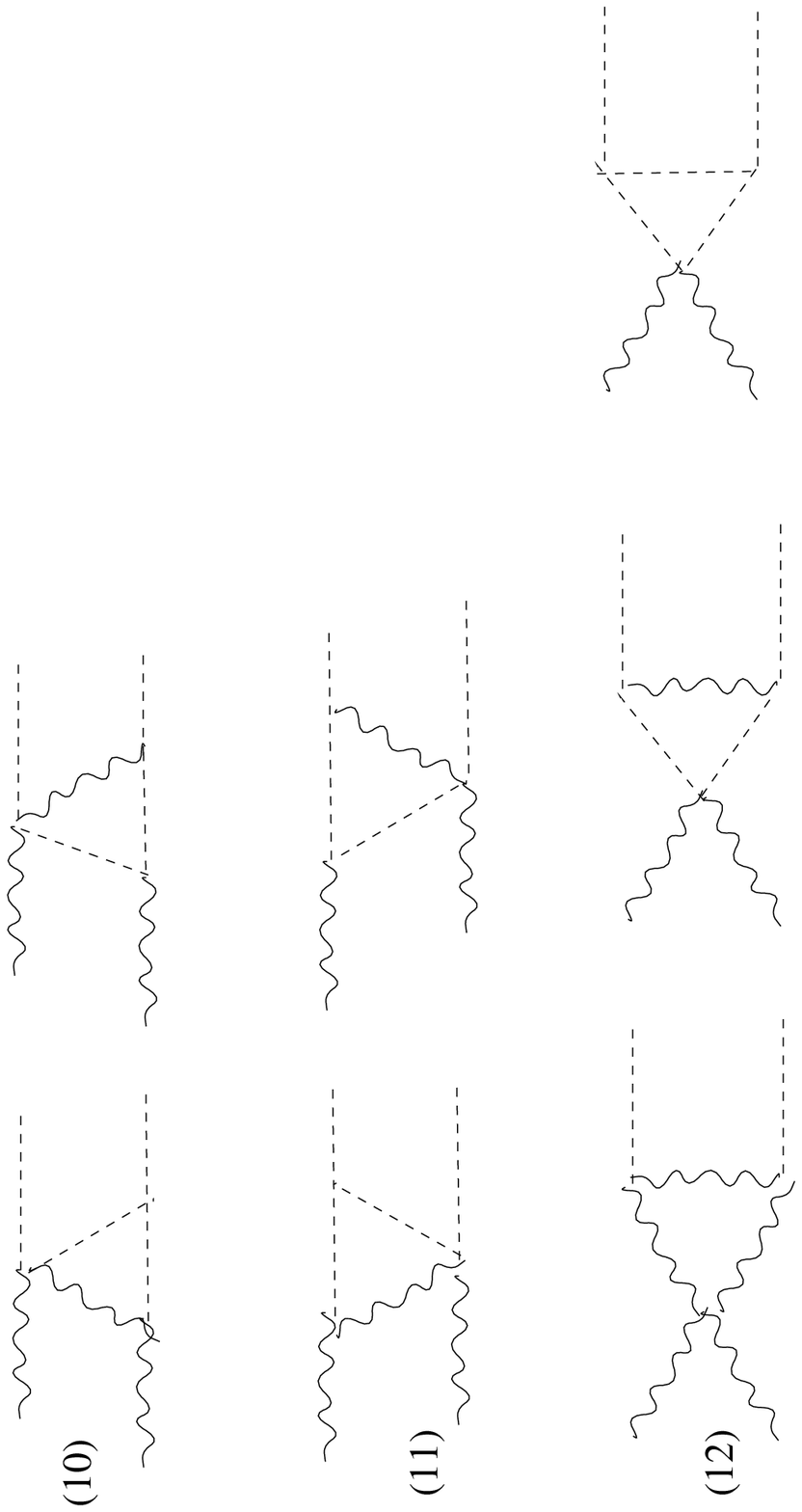}}
\caption{ Same with Figure \ref{fig1}}
\label{fig5}
\end{figure}


%% file: fig.tex
 
\begin{figure} 
\epsfxsize=10 cm 
\centerline{\epsffile{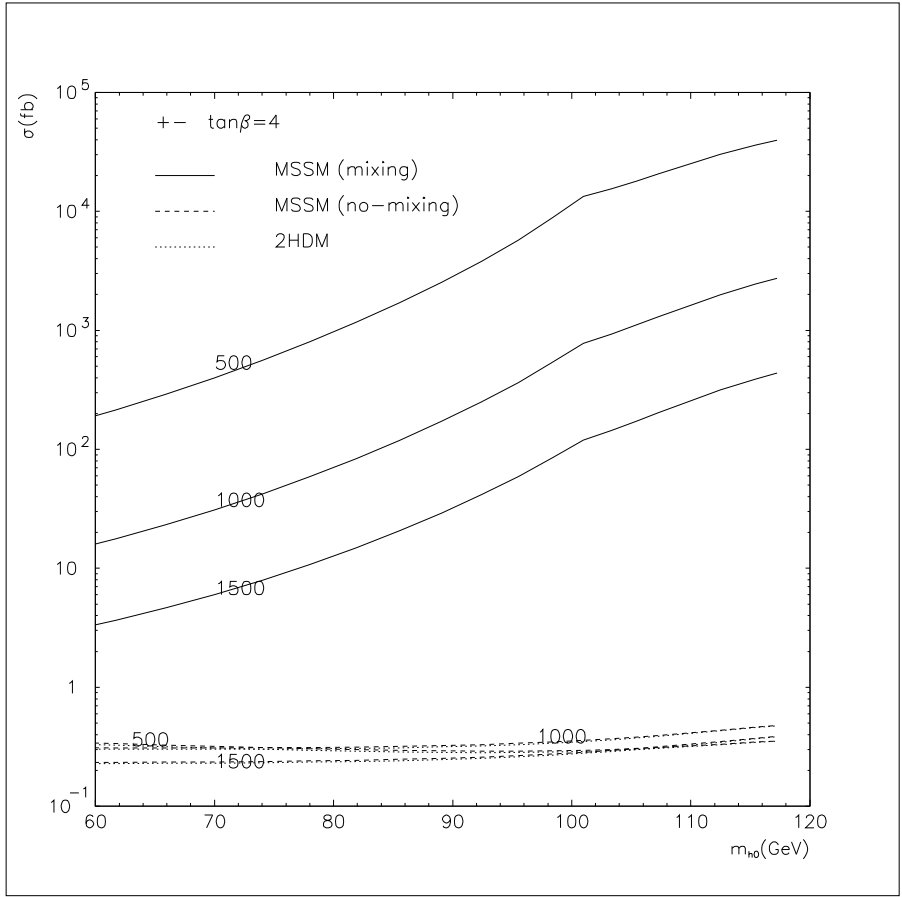}} 
\caption[]{
The cross sections for the process $\gamma\gamma
\rightarrow h_0h_0 $  
as 
a function  of the lightest Higgs boson mass 
for the opposite photon helicities 
$\lambda_1=-\lambda_2=1$ 
at $\sqrt{\hat{s}}=500, 1000, 1500 GeV$ and $\tan\beta=4$.  
The numbers on the curves are the 
phton-photon center-of-mass energy$\sqrt{\hat{s}}$. 
The dotted lines represent the 
cross sections of  the process in the 2HDM 
and the dotted lines - 
in the MSSM 
with the $A_t=\mu cot\beta$ and $A_b=\mu \tan\beta$ (there does not exist 
 mixing between 
the stops, we label it $no-mixing$),
 the solid lines -  in the MSSM  
with the stops 
mixing (we label it $mixing$). 
We have taken $m^2_{\tilde{Q}}=m^2_{\tilde{U}}=m^2_{\tilde{D}}=
(1 TeV)^2$, and when the stops exist mixing, we choose the value of 
$A_t$ which
make the mass of the lighter stop equal to $50 GeV$.
The other relevant SUSY 
parameters are 
$M=100 GeV$ and $\mu=-300 GeV$.
} 
\label{fig6} 
\end{figure}

\begin{figure} 
\epsfxsize=10 cm 
\centerline{\epsffile{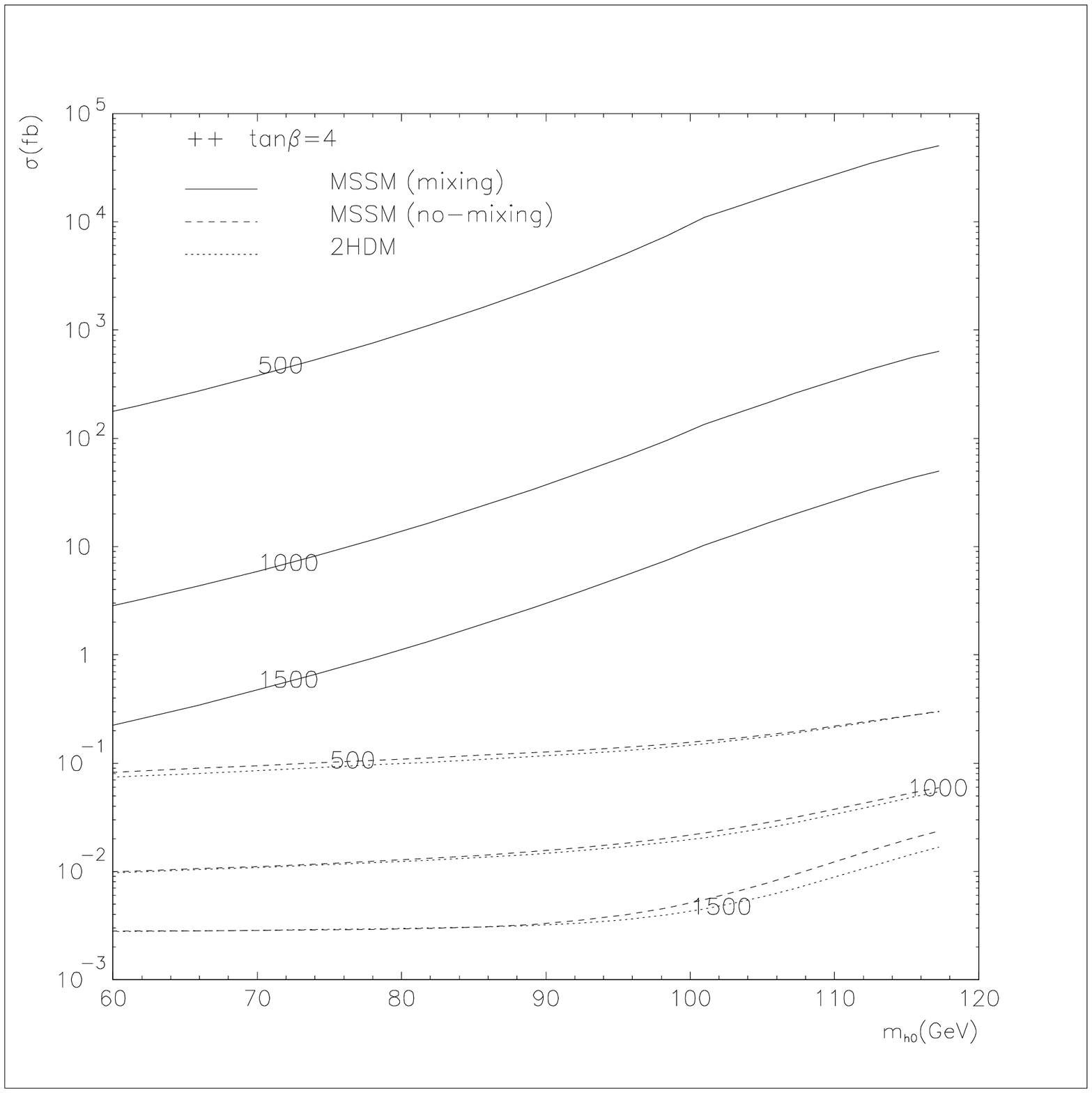}} 
\caption[]{Same with Figure \ref{fig6} but for the equal photon helicities 
$\lambda_1=\lambda_2=1$.} 
\label{fig7} 
\end{figure}

\begin{figure} 
\epsfxsize=10 cm 
\centerline{\epsffile{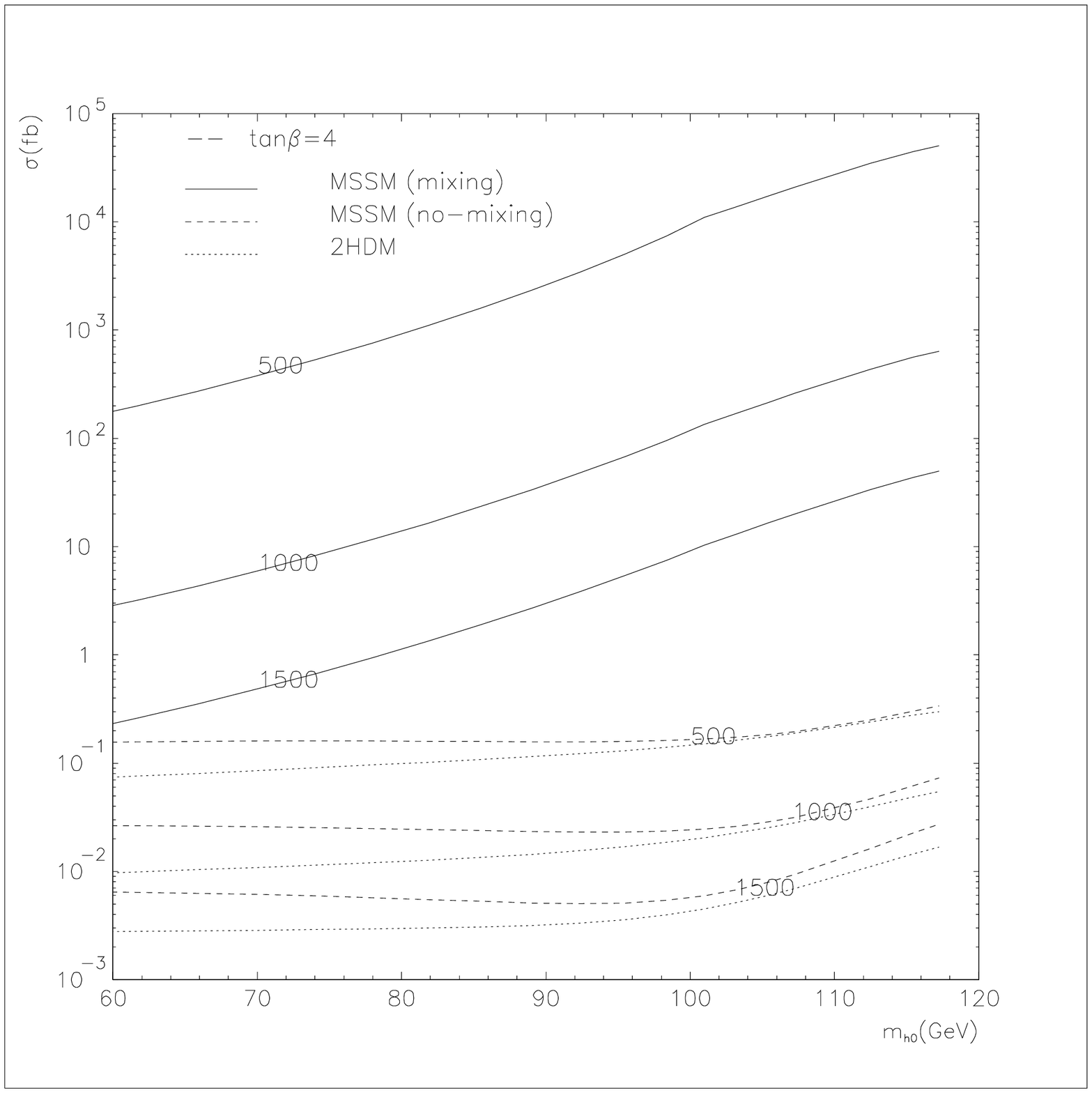}} 
\caption[]{Same with Figure \ref{fig6} but for  
the equal photon helicities 
$\lambda_1=\lambda_2=-1$.} 
\label{fig8} 
\end{figure}

\begin{figure} 
\epsfxsize=10 cm 
\centerline{\epsffile{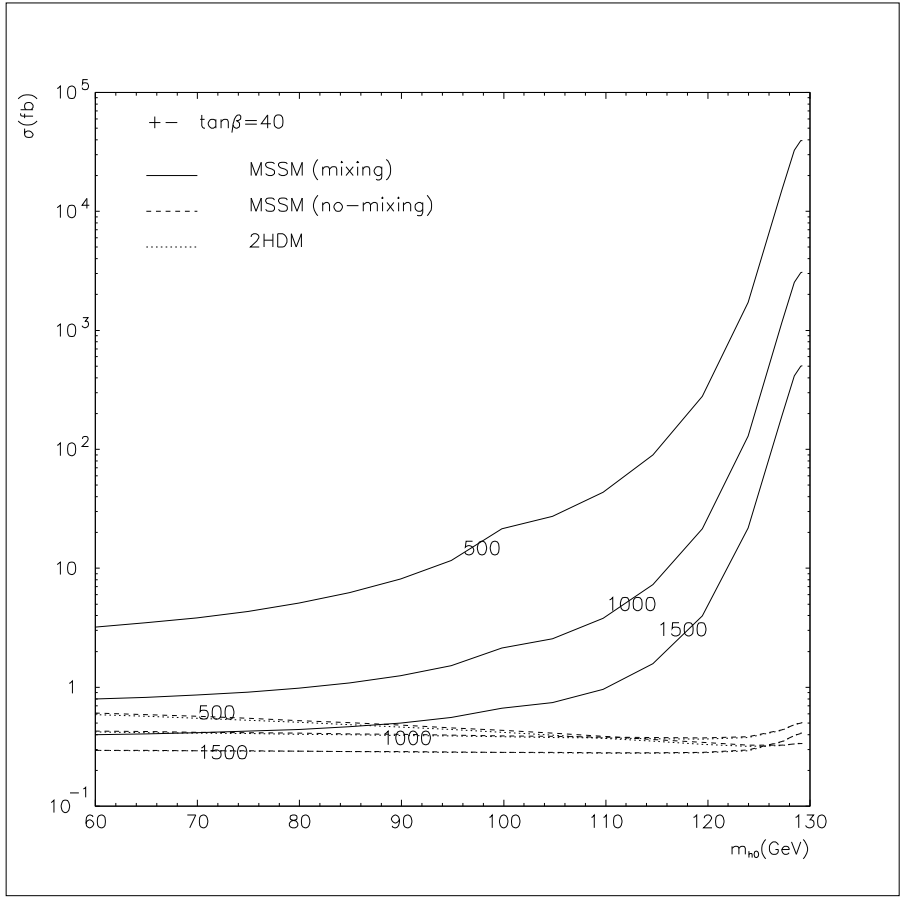}} 
\caption[]{
The cross sections for the process $\gamma\gamma
\rightarrow h_0h_0 $  
as 
a function  of the lightest Higgs boson mass 
for the opposite photon helicities 
$\lambda_1=-\lambda_2=1$ 
at $\sqrt{\hat{s}}=500, 1000, 1500 GeV$ and $\tan\beta=40$,  
the other parameters are the same with 
Figure \ref{fig6}. 
} 
\label{fig9} 
\end{figure}

\begin{figure} 
\epsfxsize=10 cm 
\centerline{\epsffile{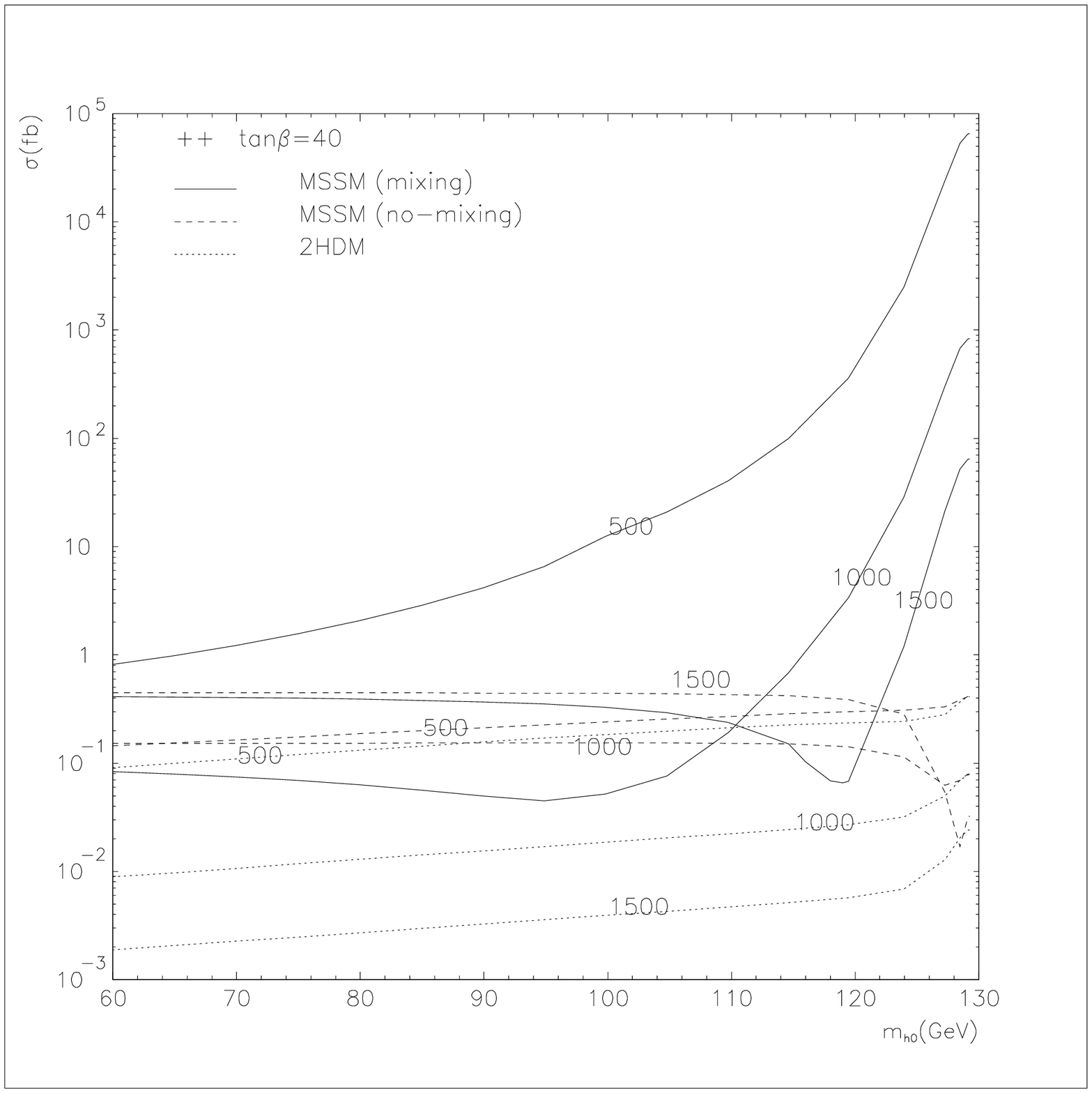}} 
\caption[]{Same with Figure \ref{fig9} but for  
the equal photon helicities 
$\lambda_1=\lambda_2= 1$.} 
\label{fig10} 
\end{figure} 
 

\begin{figure} 
\epsfxsize=10 cm 
\centerline{\epsffile{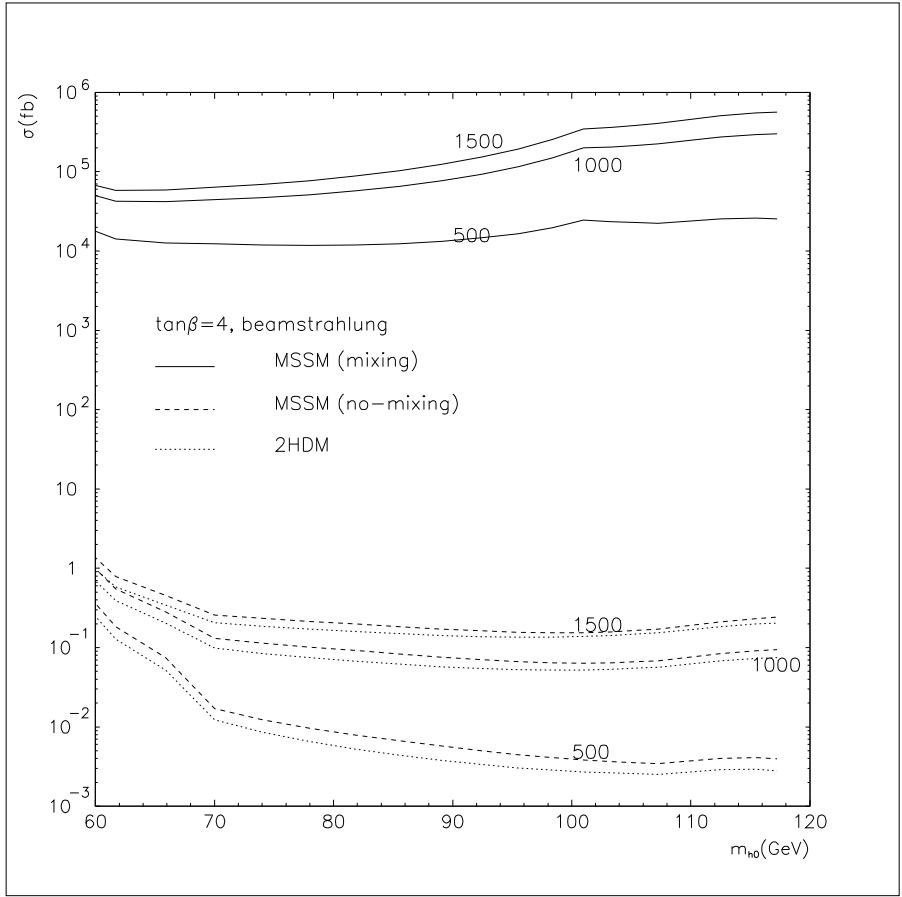}} 
\caption[]{ The total cross sections for the process $e^+e^-\rightarrow  
\gamma\gamma\rightarrow 
h_0h_0$ as a function of the lightest Higgs boson mass for the beamstrahlung photons at $\sqrt{s}=500, 1000$ and $1500 GeV$, where  
$\tan\beta=4$ 
and the other SUSY parameters are the same with 
Figure \ref{fig6}. 
The numbers on the curves are the $e^+e^-$ CMS energy.} 
\label{fig11} 
\end{figure}

\begin{figure} 
\epsfxsize=10 cm 
\centerline{\epsffile{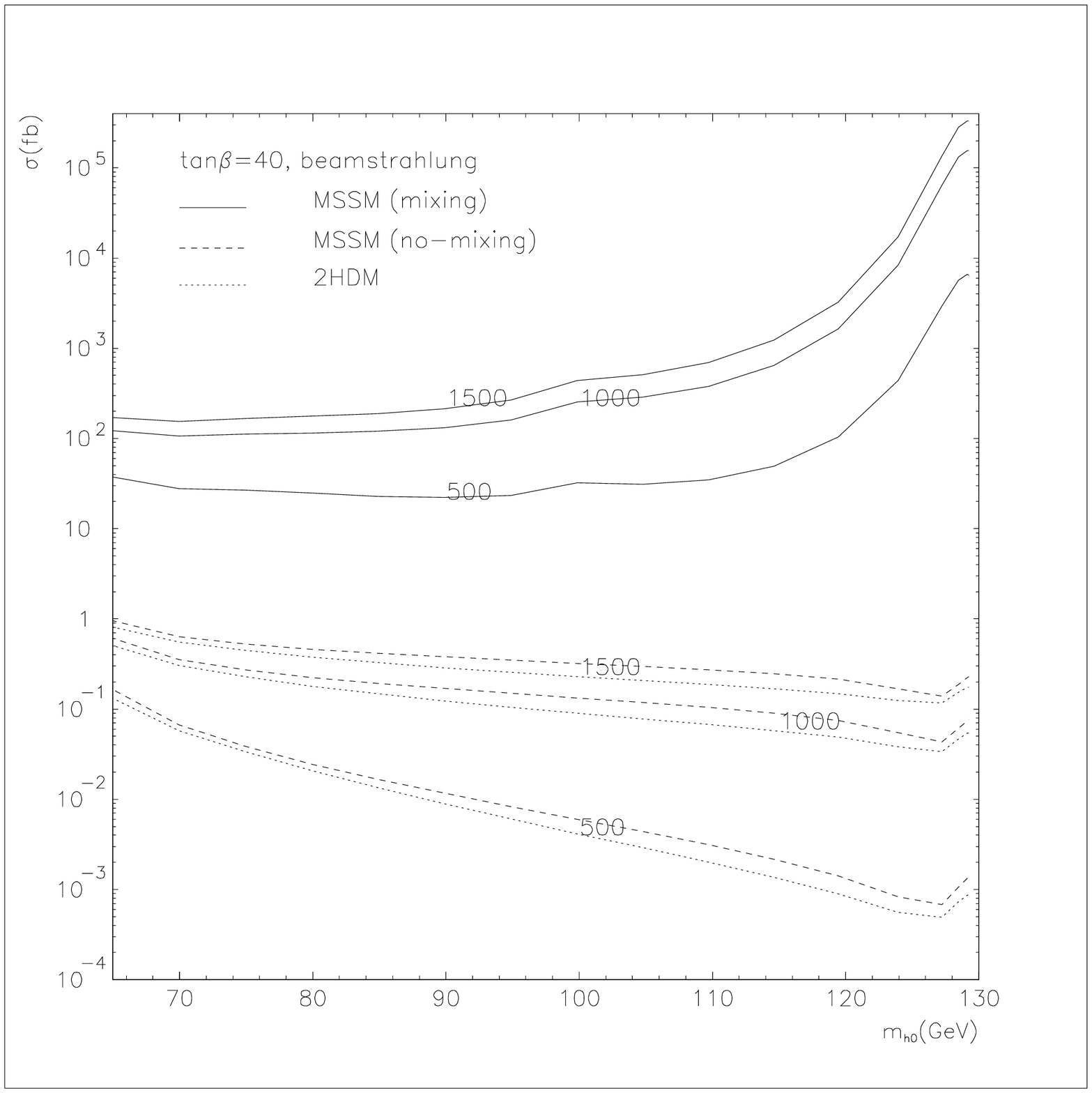}} 
\caption[]{Same with Figure \ref{fig11} but for $\tan\beta=40$.} 
\label{fig12}
\end{figure} 
 
\begin{figure} 
\epsfxsize=10 cm 
\centerline{\epsffile{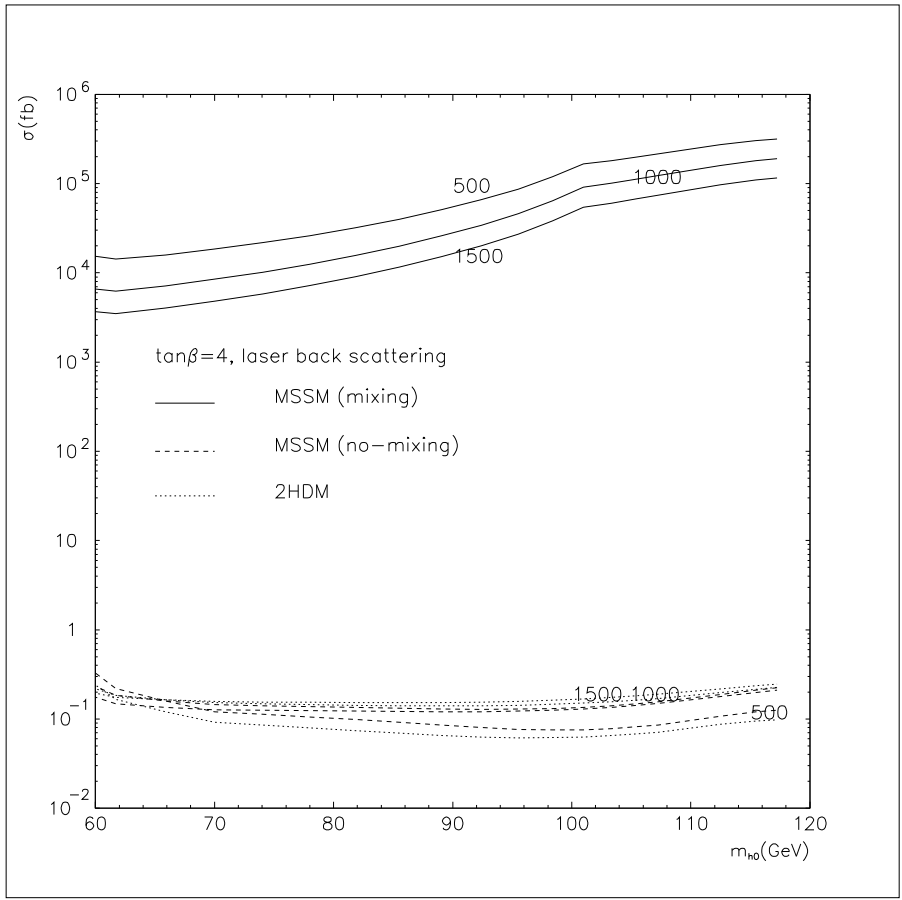}} 
\caption[]{Same with Figure \ref{fig11} but for 
laser back scattering photons.} 
\label{fig13} 
\end{figure} 
 
\begin{figure} 
\epsfxsize=10 cm 
\centerline{\epsffile{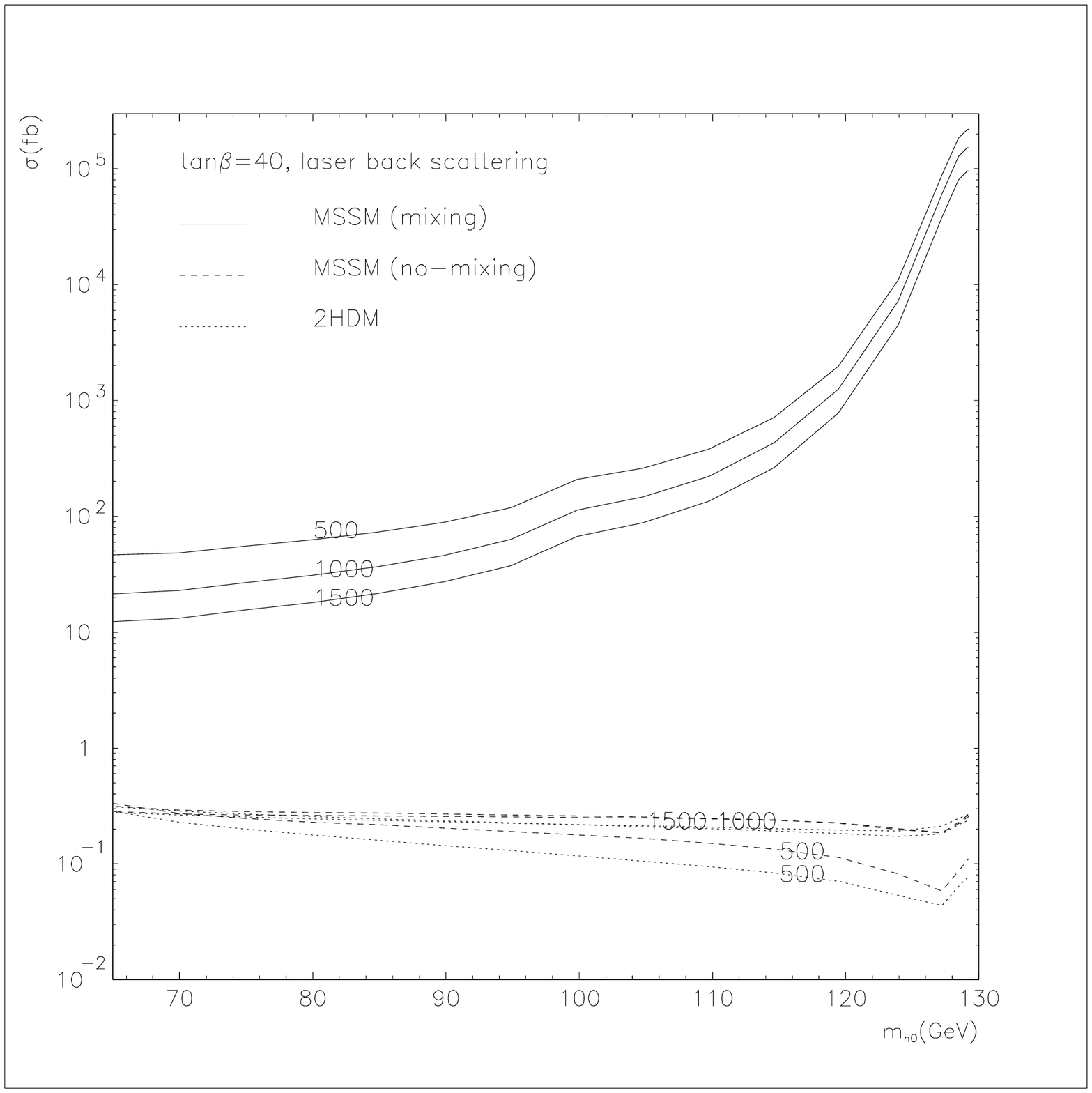}} 
\caption[]{Same with Figure \ref{fig13} but for $\tan\beta=40$.} 
\label{fig14} 
\end{figure}
